\documentclass[reprint,prd,floatfix,
superscriptaddress,nofootinbib,notitlepage]{revtex4-1}
\pdfoutput=1
\usepackage{graphicx,epsfig}
\usepackage{amsmath}
\usepackage{amssymb}
\usepackage{longtable}
\usepackage{mathtools}
\usepackage{multirow}
\usepackage{comment}
\usepackage{dcolumn}
\usepackage{physics}
\usepackage{bm}
\usepackage{amsfonts}
\usepackage{subfigure}
\usepackage{color}
\usepackage{xcolor}
\usepackage{relsize}
\usepackage{float}

\usepackage{tabularray}
\makeatletter
\newcommand\emailx[1]{%
\move@AF%
\def\@affil{{\normalfont\,#1\strut}{}}%
}%
\newcommand{\e}[1]{\times 10^{#1}}

\begin{document}

%%%%%%%%%%%%%%%%%%%%%%%%%%%%%%%%%%%%%%%%%%%%%%%%%%%%%%%%%%%%%%%%%%%%%%%%%%%
%%%%%%%%%%%%%%%%%%%%%%%%%%%%%%%%%%%%%%%%%%%%%%%%%%%%%%%%%%%%%%%%%%%%%%%%%%%
%%%%%%%%%%%%%%%%%%%%%%%%%%%%%%%%%%%%%%%%%%%%%%%%%%%%%%%%%%%%%%%%%%%%%%%%%%%
%%%%%%%%%%%%%%%%%%%%%%%%%%%%%%%%%%%%%%%%%%%%%%%%%%%%%%%%%%%%%%%%%%%%%%%%%%%
\title{Quasinormal modes of Proca and Maxwell fields in $d$-dimensional 
Schwarzschild-AdS black holes }
%%%%%%%%%%%%%%%%%%%%%%%%%%%%%%%%%%%%%%%%%%%%%%%%%%%%%%%%%%%%%%%%%%%%%%%%%%%
%%%%%%%%%%%%%%%%%%%%%%%%%%%%%%%%%%%%%%%%%%%%%%%%%%%%%%%%%%%%%%%%%%%%%%%%%%%
%%%%%%%%%%%%%%%%%%%%%%%%%%%%%%%%%%%%%%%%%%%%%%%%%%%%%%%%%%%%%%%%%%%%%%%%%%%
%%%%%%%%%%%%%%%%%%%%%%%%%%%%%%%%%%%%%%%%%%%%%%%%%%%%%%%%%%%%%%%%%%%%%%%%%%%

\author{David C. Lopes}
\email{david.d.lopes@tecnico.ulisboa.pt}
\affiliation{Centre for Astrophysics Research, 
Department of Physics, Astronomy and Mathematics,
University of Hertfordshire,
Hatfield AL10 9AB, United Kingdom, \&\\
Centro de Astrof\'{\i}sica e Gravita\c c\~ao  - CENTRA,
Departamento de F\'{\i}sica, Instituto Superior T\'ecnico - IST,
Universidade de Lisboa - UL, Avenida Rovisco Pais 1, 1049-001
Lisboa, Portugal \vskip 0.3cm}

\author{Tiago V. Fernandes}
\email{tiago.vasques.fernandes@tecnico.ulisboa.pt}
\affiliation{Centro de Astrof\'{\i}sica e Gravita\c c\~ao  - CENTRA,
Departamento de F\'{\i}sica, Instituto Superior T\'ecnico - IST,
Universidade de Lisboa - UL, Avenida Rovisco Pais 1, 1049-001
Lisboa, Portugal \vskip 0.3cm}

\author{Jos\'e P. S. Lemos}
\email{joselemos@ist.utl.pt}
\affiliation{
Center for Astrophysics and Space Science,
School of Mathematical and Physical Sciences,\\
Macquarie University, Sydney, New South Wales 2109, Australia, \&\\
Centro de Astrof\'{\i}sica e Gravita\c c\~ao  - CENTRA,
Departamento de F\'{\i}sica,\\ Instituto Superior T\'ecnico - IST,
Universidade de Lisboa - UL, \\Av. Rovisco Pais 1, 1049-001
Lisboa, Portugal
\vskip 0.3cm}

\begin{abstract}

Proca and Maxwell fields in $d$-dimensional Schwarzschild
black holes with anti-de Sitter (AdS) asymptotics are investigated
through their linear perturbations and associated quasinormal modes
(QNMs) with Dirichlet boundary conditions at infinity.  We begin by
reviewing how the Proca field equations in $d$-dimensional spherically
symmetric spacetimes reduce to three radial wave-like equations, one
fully decoupled equation and two mutually coupled equations. We also
demonstrate explicitly how the Maxwell equations emerge from the
zero-mass limit of the Proca system. Several analytical properties of
the corresponding QNM spectrum are examined.  To compute the
QNM frequencies, we employ two complementary numerical methods
particularly suited to asymptotically AdS spacetimes. Using these
techniques, we determine the QNMs modes of Proca field
perturbations in $4$, $5$, $6$, and $7$-dimensional Schwarzschild-AdS
backgrounds. As a new result, we find numerically that scalar-type
Maxwell perturbations in large $d \geq 5$ Schwarzschild-AdS black
holes exhibit purely imaginary low-frequency modes, analogous to those
found in vector-type gravitational perturbations. The presence of such
modes is especially relevant within the AdS/CFT correspondence, as
they correspond to the linearized hydrodynamic regime in the dual
conformal field theory.  We further analyze the influence of the Proca
mass on the QNM spectrum, also emphasizing how Maxwell modes
are recovered in the massless limit. The dependence of the spectrum on
the black hole radius is explored.  In addition, analytic expressions
for the QNM frequencies of vector-type and monopole Proca
perturbations, as well as Maxwell modes, are derived for small
$d$-dimensional Schwarzschild-AdS black holes by matching asymptotic
expansions using an intermediate region. These analytic results show
good agreement with the numerical findings, confirming, in particular,
the existence of purely imaginary low-frequency scalar-type Maxwell
modes in large $d \geq 5$ Schwarzschild-AdS spacetimes.
\centerline{}
\vskip 1.0cm
\end{abstract}


\centerline{}
\vskip 1.0cm

\maketitle

%%%%%%%%%%%%%%%%%%%%%%%%%%%%%%%%%%%%%%%%%%%%%%
\section{Introduction}\label{sec:intro}
%%%%%%%%%%%%%%%%%%%%%%%%%%%%%%%%%%%%%%%%%%%%%%

An important aspect of analyzing the linear stability of stationary
systems involves understanding how these systems respond when
subjected to small perturbations. Such analyses are
essential for determining whether a
system will return to its original
state after a slight deviation, evolve toward a new equilibrium
configuration, or become unstable over time when perturbed
from its initial state. In the context of black hole spacetimes, the
response to a small external perturbation is typically dominated, at
intermediate times, by the quasinormal mode (QNM) ringing phase,
during which the perturbations can be described in terms of a
discrete, countable set of QNM frequencies and their associated
eigenfunctions. These frequencies are generally complex numbers.
The
real part of the frequency
corresponds to the actual oscillation frequency, and the
imaginary part determines whether the perturbation decays or grows
in time. The imaginary component of the QNM frequencies, in
particular, provides critical information about the linear stability
of the perturbed spacetime. Specifically, if all QNMs are
damped,
which means that their amplitudes decrease in time,
the
spacetime is considered linearly stable under the class of
perturbations being investigated. On the other hand, if any mode
exhibits growth with time, it signals an instability within
the system.
Beyond their role in stability analysis, these
modes also carry
information about how perturbations evolve in
curved spacetimes. They are closely related to observable phenomena,
for instance, the gravitational wave signals produced when black holes
respond to dynamical perturbations.

To
provide an overview on QNMs, we focus 
in three  classes of black hole
spacetimes. First, asymptotically flat black hole
spacetimes, in which the geometry approaches that of Minkowski
space at large distances, representing isolated gravitational systems.
Second, asymptotically de Sitter (dS) black hole
spacetimes, which describe black holes
in universes with a positive cosmological
constant. Third, asymptotically anti-de
Sitter (AdS) black hole spacetimes, characterized by a negative cosmological
constant, which play an important role in
several areas of physics,
particularly in supergravity, holography, and the
AdS/CFT correspondence, where QNMs are closely related to relaxation
processes in the dual conformal field theory.

We  first turn our attention to black holes in
asymptotically flat spacetimes, i.e., black hole
spacetime where the cosmological constant
is set to zero. The
Schwarzschild black hole, as the simplest spherically symmetric
solution of general relativity, provided the natural starting point
for investigations of black hole stability under linear perturbations.
The stability against
gravitational 
perturbations was  tested in
several early works, notably by Regge and Wheeler,
Zerilli, and Vishveshwara
\cite{regge_wheeler,zerilli,vishveshwara}
that laid the
basis for the development of black hole perturbation theory.
Extending the analysis to Maxwell electromagnetic perturbations,
Reissner-Nordstr\"om black holes, i.e., spherical black holes with
electric charge, were investigated in \cite{maxwell_4d_sch1972}.  It
was shown that the equations governing perturbations in such a black
hole background could be reduced to a single, wave-like equation,
simplifying the analysis.
For rotating, uncharged black holes in general relativity, namely Kerr
black holes, it was shown in \cite{teukolsky} that both
electromagnetic and gravitational perturbations can be separated 
analytically. This work also addressed the stability of rotating black
holes and the phenomenon of superradiant scattering.
Related to perturbations are 
the QNMs, i.e., the
damped oscillations that dominate the late-time response of black
holes to perturbations. The QNMs
of the Schwarzschild black hole were
computed for the first time in \cite{chandra_detweiler1975}, and their
interpretation has  become central in
black hole and
gravitational wave physics.
A systematic treatment of massless field perturbations with various
spin values, including scalar, electromagnetic, and gravitational,
was
provided by Chandrasekhar
\cite{chandra_mathematical_BH1983}, further developing  the
framework of linear perturbation theory in black hole spacetimes.
The situation becomes more complicated in the
presence of massive vector fields, known as Proca fields. Unlike the
massless case, where gauge freedom simplifies the analysis, the Proca
field's mass term explicitly breaks gauge invariance. This breaking
introduces an additional propagating degree of freedom, leading to
coupling between the different components of the field. As a result,
the perturbation equations become more intricate and, in general, do
not decouple for arbitrary values of the angular momentum quantum
number, even in the static black hole background
\cite{galtsov1984}. This complexity poses
challenges to attempts at an analytical treatment and
obliges to use numerical methods for further understanding.
For scalar massive fields, an initial treatment can be found in
\cite{simone_will1992}.
A comprehensive review of the subject, encompassing not only
QNMs of black holes but also those of relativistic stars,
is provided in \cite{k1999}, where
the
theoretical
framework is discussed and applications are given.
 The case of massless
vector fields, specifically Maxwell fields, and their perturbations on
static black hole spacetimes was explored in \cite{maxwell_d2001},
and for Proca fields 
it was demonstrated in \cite{pawl} that
monopole, i.e., zero angular momentum, Proca perturbations can be
reduced to a decoupled wave-like equation, a simplification 
that does not extend to perturbations with general angular
momentum, where the coupling between different modes persists.
Further studies of massive scalar field perturbations were performed
in \cite{zhidenko_massive_scalar2005,zhidenko_massive_scalar2006},
providing insights into their dynamics and into the QNM
spectrum. The phenomenon of superradiant instabilities through scalar
fields was investigated in \cite{cdly2004}, which analyzed the
conditions under which energy extraction from black holes becomes
possible.
In another study, Proca field perturbations and their
associated QNMs were analyzed in detail in
\cite{konoplya_massive2006}, where a comprehensive examination of
the spectrum of such massive vector fields in black hole
spacetimes was done.
Additionally, the behavior of massless vector fields in
higher-dimensional, $d \geq 4$, black holes was analyzed in
\cite{ortega2006}, where it was shown that the perturbation equations
reduce to two decoupled wave-like equations, one governing a
scalar-type degree of freedom and the other governing the remaining $d
- 3$ vector-type degrees of freedom.
For a more extensive discussion of these topics and further
developments, the reader is referred to the review in
\cite{phd_zhidenko2009}.
In \cite{dolan2012}, the dynamics of a massive vector, i.e.,
a Proca field in the
4-dimensional 
Schwarzschild spacetime was explored, with a particular focus on
QNMs and bound states for higher multipole
moments. This study marked an advancement in understanding
the spectrum of Proca fields in spherically symmetric black hole
backgrounds beyond the monopole case.
A subsequent and more comprehensive analysis of Proca field
perturbations was carried out
for $d$-dimensional
black holes in \cite{herdeiro_proca_eqs2012}. There,
the authors showed that the field equations can be systematically
decomposed into a set of three radial wave-like equations,
two coupled equations governing the scalar-type modes of
the Proca field, and 
one fully
decoupled equation describing the $d-3$ vector-type degrees of
freedom. This decomposition provides a more tractable
framework for analyzing the field dynamics, although the coupled
nature of the scalar-type sector still presents computational
challenges.
A further review and discussion of perturbations,
particularly in relation to massive fields and their stability
properties, is presented in \cite{pani2013}.
The case of rotating black holes was tackled in \cite{dolan2018,
percivaldolan2020}, where the behavior of massive Proca fields in the
Kerr geometry was studied. These works identified conditions under
which superradiant instabilities may arise, contributing to the
broader investigation of massive field instabilities in rotating
spacetimes.
More recent work in \cite{garcia2022} extended the analysis to
scenarios involving vector fields that are nonminimally coupled to
curvature, examining how such couplings influence the QNM
spectrum and may modify the stability of black hole
solutions.
Finally, in \cite{dolan2024}, superradiant instabilities were
investigated in the context of charged regular black holes. This study
treated energy extraction mechanisms in regular black
hole spacetimes considering both the regularity of the metric and
the presence of an electromagnetic field, showing how such
features alter the superradiant phenomenon.

We next mention black holes in asymptotically dS
spacetimes, i.e., black hole spacetimes where the cosmological
constant is set to be positive.  The presence of a cosmological
horizon in these geometries introduces new features absent in
asymptotically flat spacetimes, such as a finite volume between the
event and cosmological horizons, and the setting of boundary
conditions relevant for the analysis.
These differences influence the behavior of QNMs and imply the
need for an appropriate treatment.
The first analysis of QNMs and the stability of black holes in dS
space was carried out in \cite{mellormoss}, where insights into the
characteristic decay of fields in the presence of an event and a
cosmological horizon were given. 
Further developments were achieved in the context of near-extremal dS
black holes, where the event and cosmological horizons become nearly
coincident, demonstrating that an analytical approach is possible to
test the stability of the spacetimes \cite{cardosolemosl2003desitter}.
Other results on the QNMs of Schwarzschild-dS black holes were
obtained in \cite{zhidenko20024desitter, konoplyazhidenko2022}. These
works employed analytical and numerical techniques to compute the QNM
spectra for scalar, electromagnetic, and gravitational perturbations
and showed that increasing the cosmological constant lowers the real
part of the QNM frequencies and modifies the damping rates.
The effects of massive scalar fields propagating in Schwarzschild-dS
black hole backgrounds were investigated in \cite{konoplya2024}. In
particular, the interplay between the scalar field mass and the
cosmological constant leads to the appearance of an effective
potential that alters the stability of the spacetime and the
propagation of waves.

Finally, we turn our view onto
black holes in asymptotically AdS
spacetimes, i.e., black hole spacetimes where the cosmological
constant is set to be negative,
which the subject of our main interest. 
The study of
QNMs in AdS spacetimes began with the analysis of
scalar perturbations in Schwarzschild-AdS black holes in
\cite{hubeny2000}, followed by investigations of scalar,
electromagnetic, and Weyl perturbations in BTZ black holes in
\cite{cardoso_lemos2001btz}.
Subsequent work extended the analysis to electromagnetic and
gravitational perturbations of Schwarzschild-AdS black holes in
\cite{cardoso_lemos2001}, and to planar, cylindrical, and toroidal
black holes in \cite{cardoso_lemos2001cylindrical}. Further
developments in \cite{bertik2003}
analyzed Reissner-Nordstr\"om-AdS
spacetimes and computed its quasinormal modes,
and in
\cite{cardoso_lemos_konoplya2003}
the QNM spectrum of Schwarzschild-AdS black holes
was shown to exhibit distinct
asymptotic regimes where frequencies scale with the black hole size
and approach evenly spaced patterns determined by the AdS boundary
conditions.
A general framework for studying perturbations and QNMs
in arbitrary spacetime dimensions was developed in
\cite{ishibashi_kodama12003, ishibashi_kodama_stability2003,
ishibashi_kodama2004}, a method that also encompasses zero and
positive cosmological constant cases. The dynamics and wave equations
of various fields in AdS spacetimes were analyzed in \cite{wald2004}.
Tensorial, i.e., gravitational perturbations of planar black holes were
studied in \cite{zanchin2006}, while scalar field perturbations and
instabilities in Kerr-AdS spacetimes were reported in
\cite{cardoso_dias_yoshida2006}. QNMs of AdS black holes
in the context of expanding plasmas were explored in
\cite{gubser2007}, and the high-wavelength eikonal regime was
examined in \cite{siopsis2007}.
The stability of Reissner-Nordstr\"om-AdS black holes
in
general relativity in four and
higher dimensions, was investigated through perturbation analysis in
\cite{konoplya_stability2008,cardoso_breit_wigner2009}.
Finally, a broad spectrum of perturbations, including
Maxwell fields, and the emergence of superradiant instabilities in AdS
black hole spacetimes were studied in
\cite{wang_herdeiro_matching2014, herdeiro_maxwell2015}.
Proca field perturbations in extremal and near-extremal static black
hole spacetimes were  investigated in \cite{ueda2018}
and the
QNMs of Proca fields in four-dimensional
Schwarzschild-AdS spacetime were further analyzed in
\cite{tiago_sads2022}, with a focus on the decoupling of the 
degrees of freedom. 
Normal modes of Proca perturbations in pure AdS spacetime
were explored in \cite{fhlc2023}.
QNMs
in cylindrical AdS black holes were further
studied in \cite{kailin2023},
while a comprehensive analysis of various boundary conditions in AdS
black hole backgrounds was presented in \cite{kinoshita2024}. In
\cite{lopes2024}, the normal modes of Proca fields in pure AdS
spacetimes for general spacetime dimensions $d\geq4$ were derived. This
work showed that the Proca equations reduce to a system of three
radial wave-like equations, namely,
two coupled equations describing the
scalar-type modes and
a single decoupled equation governing the
$d-3$ vector-type mode.
Finally, the QNM spectrum of a Proca field in five-dimensional
Schwarzschild-AdS spacetime was obtained using the isomonodromy method
in \cite{bfl2025}.

Taken together, previous studies, some of which we mentioned above,
have built a consistent picture of black hole perturbations in
asymptotically flat, de Sitter, and Anti-de Sitter spacetimes. This
provides an effective framework where different boundary conditions and
geometric setups can be compared on the same footing. In particular,
the QNMs in black hole spacetimes play an important
role in describing how
perturbations evolve and decay in time and how
the spacetime reacts to external and internal excitations.
These results have been important for understanding several physical
and observational aspects, including
stability, energy dissipation, and gravitational wave emission.
In the AdS case, they also connect
naturally with holographic descriptions.
In this work, we focus on QNMs of Proca fields,
specifically, in massive vector
perturbations, in higher-dimensional, spherically symmetric,
asymptotically AdS black hole spacetimes. The
field's mass and the presence of extra dimensions can 
affect the spectrum. Our main motivation is to better understand how
mass and spin affect the QNM spectrum and to explore the effect of
higher dimensions on the dynamics of black hole spacetimes.
Let us consider these topics more carefully.
First, higher-dimensional spacetimes play an important role in various
theoretical settings, including string theory and braneworld
scenarios, which allow for extra spatial dimensions beyond the usual
four. These approaches have motivated the study of new black hole
solutions in higher dimensions \cite{lemos1997}, which can show a
different behavior from the four-dimensional ones.
Second, asymptotically AdS spacetimes are interesting not just from a
theoretical point of view but also mathematically. They show up in 
theories like supergravity and also play a role in the AdS/CFT
correspondence, which links gravity in AdS space to a conformal field
theory on its boundary. For a more detailed discussion 
physical and mathematical significance
of AdS spacetimes, see \cite{sokolowski}.
Third, the study of massive spin-1 fields is highly relevant both
within and beyond the standard model of particle physics. In
particular, massive vector fields, such as the Proca field, have
gained increasing interest in the context of dark matter models,
including scenarios involving light bosonic candidates such as axions
and fuzzy cold dark matter. These fields interact
predominantly through gravity, making it essential to understand their
behavior in curved spacetime, especially in regions
near the horizon of black holes.
For an overview and classification of
dark matter candidates, see \cite{bertonehooper}.
Fourth and finally, given the growing theoretical and phenomenological
interest in massive vector fields, a comprehensive study of the QNMs
associated with Proca perturbations in higher-dimensional
asymptotically AdS black hole spacetimes is both opportune and
appropriate. Such a study is essentially absent from the
literature, leaving a gap that we attempt to fill here.
Therefore, our research is focused on providing a systematic and
detailed investigation of QNMs of Proca fields in higher-dimensional,
spherically symmetric AdS black holes, dealing with the QNM
dynamical
properties and the stability behavior in these geometries.

The paper is organized as follows.  In Sec.~\ref{sec:proca_field}, we
introduce the equations describing a Proca field minimally coupled to
a general relativistic $d$-dimensional curved spacetime.
In Sec.~\ref{sec:proca_in_sads}, we review how the Proca field
equations in $d$-dimensional spherically symmetric spacetimes reduce
to three radial wave-like equations,
two equation coupled between themselves, and another
completely decoupled.  We also carefully study how the
Maxwell equations arise from the zero mass limit of the Proca
equations.
In Sec.~\ref{sec:analytical}, we investigate some analytical
properties of the QNMs of Proca and Maxwell field
perturbations.
In Sec.~\ref{sec:numerics_integration}, we detail the numerical
techniques employed to compute the QNM 
frequencies. Two methods are used.
The shooting method, which involves numerically integrating the
perturbation equations. This approach is well-established in the
literature, particularly for asymptotically flat spacetimes, see,
e.g., \cite{chandra_detweiler1975,pani2013,phd_zhidenko2009}, and has
also been adapted for asymptotically AdS backgrounds
\cite{konoplya_stability2008}.
And the
Horowitz-Hubeny numerical method, specifically designed for asymptotically
AdS spacetimes, which is known for its efficiency and reliability in this
context, see \cite{hubeny2000,tiago_sads2022}.
These complementary methods allow us to determine
with accuracy
the QNM spectra across
different spacetime geometry parameters.
In
Sec.~\ref{sec:num_results}, we compute the  Proca and Maxwell 
QNM frequencies in $4,5,6,7$-dimensional
Schwarzschild-AdS spacetime.
In particular, we find numerically that scalar-type
Maxwell perturbations in $d\geq 5$ large
Schwarzschild-AdS spacetimes exhibit purely imaginary low frequency
modes, which are similar to those found for vector-type gravitational
perturbations in \cite{cardoso_lemos2001,
cardoso_lemos2001cylindrical,cardoso_lemos_konoplya2003,
zanchin2006}.
The existence of such modes is relevant in
the AdS/CFT correspondence, as these correspond to the linearized
hydrodynamic regime in the CFT side
\cite{siopsis2007,gubser2007}.
Furthermore, the effect of the mass of
the Proca field on the QNMs is studied, with an emphasis on how
the Maxwell modes can be obtained from the massless limit of the Proca
modes.
Additionally, we investigate the dependence of the
QNMs on the black hole radius.
In Sec.~\ref{sec:matching}, we obtain analytic expressions for the
QNMs of vector-type and monopole Proca perturbations, as
well as the Maxwell modes, for small $d$-dimensional Schwarzschild-AdS
black holes.  We do this by matching asymptotic expansions of the
solution in an intermediate region, see e.g.,
\cite{cdly2004,dolan2012} for asymptotically flat spacetimes and
\cite{cardoso_dias_yoshida2006,cardoso_breit_wigner2009,
wang_herdeiro_matching2014} for asymptotically AdS.
The analytical results are  compared with the numerical
ones, and the comparison shows good agreement in the relevant
regimes.
In
Sec.~\ref{sec:conclusions}, we conclude emphasizing that
we have obtained the QNMs of Proca
and Maxwell perturbations in
higher-dimensional Schwarzschild-AdS spacetime with Dirichlet boundary
conditions, thus generalizing the study made in
\cite{tiago_sads2022}.

\vskip 0.2cm

We use Planck units throughout the paper, i.e., $\hbar = 1$, $G=1$, and
$c=1$. Moreover, spacetime indices are denoted by Greek letters, e.g.,
$\mu$, $\nu$, run from $0$ to $d-1$, where $0$ is the time index, and
$1$ to $d-1$ specify the spatial indices.

\vfill

%%%%%%%%%%%%%%%%%%%%%%%%%%%%%%%%%%%%%%%%%%%%%%%%%%%%%%%%%%%%%%%%%
\section{Proca and Maxwell fields
in curved spacetime}\label{sec:proca_field}
%%%%%%%%%%%%%%%%%%%%%%%%%%%%%%%%%%%%%%%%%%%%%%%%%%%%%%%%%%%%%%%%%

The dynamics of a Proca field minimally coupled to a $d$-dimensional 
gravitational background with negative cosmological constant is encoded in 
the Einstein-Proca action, $S = S_{\rm E} + S_{\rm P}$, where 
$S_{\rm E} = \frac{1}{16\pi}
\int d^{d}x \sqrt{-g}\,(R -2\Lambda)$ is the 
Einstein action, 
$S_{\rm P} = -\int d^{d}x \sqrt{-g}\left(\frac{1}{2} \mu^{2}
A_{\mu}A^{\mu} +\frac{1}{4}F_{\mu \nu}F^{\mu \nu} \right)$ is the Proca 
field action, $g$ is the determinant of the metric 
$g_{\mu\nu}$, ${R} = R_{\mu \nu}g^{\mu \nu}$ is the Ricci scalar, 
defined as the trace of the Ricci tensor $R_{\mu \nu}$ calculated in terms 
of the metric itself and its first and second derivatives, $\Lambda = -
\frac{(d-1)(d-2)}{2l^2}$ is the cosmological constant, with $l$ being
the characteristic AdS length, $A_\mu$ is the Proca field with mass
$\mu$, and $F_{\mu \nu} \equiv \nabla_{\mu}A_\nu-\nabla_{\nu}A_\mu$ is
the Proca field strength, with $\nabla_\mu$ representing the covariant 
derivative with respect to the $x^\mu$ coordinate.
For zero field mass, $\mu=0$, the action $S$
turns into the Einstein-Maxwell action.
The field equations for the metric and for the Proca field are
obtained by applying 
the variational principle to the action $S$. 
Variation
of the action $S$ with respect to
the metric  gives the Einstein equation,
\begin{align}
\label{eq:efe}
&G_{\mu \nu} -\frac{(d-1)(d-2)}{2l^2} g_{\mu \nu} =
8\pi T_{\mu \nu}\,,\nonumber \\
&T_{\mu \nu}
\hskip -0.12cm
\equiv
\hskip -0.08cm
g^{\alpha \beta}
\hskip -0.06cm
F_{\mu \alpha}
\hskip -0.02cm
F_{\nu \beta}
\hskip -0.05cm
+
\hskip -0.05cm
\mu^2
\hskip -0.02cm
A_{\mu}A_{\nu}
\hskip -0.07cm
-
\hskip -0.03cm
g_{\mu \nu}
\hskip -0.11cm
\left(
\hskip -0.05cm
\frac{1}{4}F_{\alpha \beta}
\hskip -0.02cm
F^{\alpha \beta}
\hskip -0.05cm
+
\hskip -0.05cm
\frac{\mu^2}{2}A_{\alpha}A^{\alpha}
\hskip -0.05cm
\right)
\hskip -0.09cm
,
\end{align}
where $G_{\mu \nu}= {R}_{\mu \nu}-\frac{1}{2}g_{\mu \nu}{R}$ is the
Einstein tensor and $T_{\mu \nu}$ is the Proca stress-energy tensor.
Moreover, the Einstein tensor obeys the Bianchi identity $\nabla^\nu
G_{\mu \nu} =0$, which in turn implies the conservation law for $T_{\mu
\nu}$, i.e., $\nabla^\nu T_{\mu \nu} = 0$.
The Proca field equation
can be obtained  by
varying the action $S$ with respect to the Proca field $A_\mu$
to give
\begin{equation}\label{eq:proca_equation}
\nabla_\nu F^{\mu \nu}+ \mu^2A^\mu = 0\,.
\end{equation}
Due to $F_{\mu \nu}$ being an antisymmetric tensor, one can calculate
the divergence of
the Proca field equation,
 to obtain
 an
 equation for $A_\mu$, i.e., $\nabla^{\mu}A_{\mu}=0$.
For $\mu^2 \neq 0$ this equation is 
the Bianchi
identity for the field $A_\mu$.
In this nonzero field mass case, $A_\mu$ is a physical field with
only $d-1$
dynamical degrees of freedom, as one component
of the field can always be
determined from the others precisely by
the Proca Bianchi identity $\nabla^{\mu}A_{\mu}=0$.
For $\mu^2=0$,
$A_\mu$ reduces to the Maxwell field
obeying the field equation $\nabla_\nu F^{\mu \nu} = 0$,
and
the Bianchi identity $\nabla^{\mu}A_{\mu}= 0$
ceases to exist, as it 
no
longer follows from the field equation. In turn,
$A_\mu$ becomes invariant under
the gauge transformation $ A^\mu \longrightarrow A^\mu+\partial^\mu h
$, where $h$ is an arbitrary scalar field,
and what was 
the Bianchi identity for $A_{\mu}$
becomes the Lorenz gauge condition of the
Maxwell field, which does not completely fix the gauge, as it is still
gauge-invariant if $\nabla_\mu \nabla^\mu h = 0$, i.e., if $h$ obeys
the massless Klein-Gordon equation. Thus, in the massless case,
$A_\mu$ only has $d-2$ dynamical degrees of freedom.

The full dynamics of a Proca field propagating in curved spacetime can
only be captured by solving Eqs.~\eqref{eq:efe}
and~\eqref{eq:proca_equation} simultaneously for both $g_{\mu \nu}$
and $A_\mu$, taking into account
the Proca stress-energy tensor
$T_{\mu \nu}$
defined in Eq.~\eqref{eq:efe}.  Here,
we will consider the Schwarzschild-AdS solution of the vacuum Einstein
field equations with negative cosmological constant and the trivial
$A_\mu = 0$ solution, and then take into account linear perturbations
of the Proca field around $A_\mu = 0$. Since $T_{\mu \nu}$ defined
in 
Eq.~\eqref{eq:efe} is of second order in $A_\mu$,
perturbations in the Proca field do not induce a curvature
perturbation on $g_{\mu \nu}$ in first order.  Thus, in first order,
the metric is still described by the background metric as a solution
to vacuum Einstein field equations in Eq.~\eqref{eq:efe} with $A_\mu =
0$, while the perturbations of the Proca field are described by the
Proca field equations Eq.~\eqref{eq:proca_equation}, with a fixed
background metric. In the same way, the Maxwell field can be treated
as a perturbation on a background that is not itself perturbed,
either by taking the massless limit of the Proca field
with care or by
considering the Maxwell field from the start.

%%%%%%%%%%%%%%%%%%%%%%%%%%%%%%%%%%%%%%%%%%%%%%%%%%%%%%%%%%%%%%%
\section{Linear Proca and Maxwell field perturbations in Schwarzschild-AdS}
\label{sec:proca_in_sads}
%%%%%%%%%%%%%%%%%%%%%%%%%%%%%%%%%%%%%%%%%%%%%%%%%%%%%%%%%%%%%%%

%%%%%%%%%%%%%%%%%%%%%%%%%%%%%%%%%%%%%%%%%%%%%%%%%%%%%%%%%%%%%%%%
\subsection{Proca equations in spherically symmetric backgrounds:
The radial equations in Schwarzschild-AdS}
%%%%%%%%%%%%%%%%%%%%%%%%%%%%%%%%%%%%%%%%%%%%%%%%%%%%%%%%%%%%%%%%

\subsubsection{Proca equations in spherically symmetric backgrounds}

We are interested in solving Eq.~\eqref{eq:proca_equation} in a
$d$-dimensional Schwarzschild-AdS spacetime, which is a
static spherically
symmetric vacuum solution to the Einstein field equations with
negative cosmological constant described by
Eq.~\eqref{eq:efe}.
The coordinates $x^\mu$ used are the time coordinate
$x^0\equiv t$, the radial coordinate $x^1\equiv r$, and
the angular polar coordinates 
$x^i\equiv \theta^i$, with $i=2,3,..., d-1$.
For a metric $g_{\mu \nu}$, the line element
$ds^2$ is generically given by
$ds^2= g_{\mu \nu} dx^\mu dx^\nu$, and for 
$d$-dimensional Schwarzschild-AdS
spacetime the line element
in Schwarzschild spherical coordinates
is given by
\begin{align}
ds^2= -f(r)\,dt^2+\frac{dr^2}{f(r)}+r^2
d\Omega_{d-2}^2 \,,
\label{eq:background}
\end{align} 
where $f(r)$ is
\begin{align}
f(r) = 1 + \frac{r^2}{l^2}- \left(1+
\frac{r_{\rm h}^2}{l^2}\right)
\left(\frac{r_{\rm h}^2}{l^2}\right)^{d-3} 
\,,
\label{eq:f}
\end{align} 
$d\Omega_{d-2}^2$ is the line element of the $(d-2)$-sphere in polar
coordinates $\theta^i$, with $i=2,3,..., d-1$, given by
$d\Omega^2_{d-2} = (d\theta^2)^2 + \sum_{i=3}^{d-1} \prod^{i-1}_{k=2}
\sin^2 (\theta^k) (d\theta^i)^2$, and $r_{\rm h}$ is the event horizon
radius, the only positive root of $f(r)$.

In spherically symmetric spacetimes, the Proca equations
Eq.~\eqref{eq:proca_equation} reduce to a set of radial wave-like
equations
 \cite{herdeiro_proca_eqs2012,ueda2018,lopes2024}, using the
decomposition given
in~\cite{ishibashi_kodama12003}. This decomposition of the
Proca field consists in expanding its components according to their
tensorial behavior on the $(d-2)$-sphere, using the appropriate
spherical harmonics. More concretely, we assume the following ansatz
for the Proca field \cite{lopes2024}
\begin{align}\label{eq:proca_expansion}
&A_\mu dx^\mu = r^{1-\frac{d}{2}}\sum_{\vec{k}_s} 
\biggl(u_{0 \vec{k}_s}(t,r)dt
+  \frac{u_{1\vec{k}_s}(t,r)}{f(r)}dr\biggr)Y_{\vec{k}_s}
\nonumber\\
&+  r^{1-\frac{d}{2}}\sum_{\vec{k}_s}
\biggl[\frac{r u_{2\vec{k}_s}(t,r)}{\ell(\ell+d-3)} 
\hat{\nabla}_i Y_{\vec{k}_s} 
d\theta^i\biggr]\nonumber \\
& + r^{2-\frac{d}{2}}\sum_{\vec{k}_v} u_{3\vec{k}_v}(t,r) Y_{\vec
{k}_vi} d\theta^i \,, 
\end{align}
where $Y_{\vec{k}_s}$ and $Y_{\vec{k}_vi}$ are, respectively, the
scalar and vector spherical harmonics on the $(d-2)$-sphere,
$\hat{\nabla}_i$ denotes the covariant derivative associated to the
$(d-2)$-sphere, the functions $u_{0 \vec{k}_s}$, $u_{1\vec{k}_s}$, and
$u_{2\vec{k}_s}$ are the functions that describe the time and radial
dependence of the scalar sector of the Proca field, while
$u_{3\vec{k}_v}$ describes the time and radial dependence of the
vector sector of the Proca field, and $\vec{k}_s$, $\vec{k}_v$ in the
subscript are vectors containing the angular momentum number, $\ell$,
and the $d-3$ azimuthal numbers associated to each
harmonic. Substituting Eq.~\eqref{eq:proca_expansion} in
Eq.~\eqref{eq:proca_equation}, the Proca field equations completely
separate in two different sectors, namely,
the scalar-type sector, whose
components behave as scalars on the sphere, covering  two 
degrees of freedom of the Proca field,
and 
the vector-type sector, whose
components behave as vectors on the sphere, covering
the other $d-3$ degrees of
freedom of the Proca field.

The scalar-type modes of the Proca field perturbations are coupled, 
and are described by the following system of 
wave-like equations for each harmonic vector $\vec{k}_s$
\begin{align}
&\mathcal{D}_{\ell} u_0
+\frac{(d-4)f}{r^2}\left(1-f+\frac{rf'}{2}\right)u_0\notag\\
&+f'\left(\partial_t u_1-\partial_{r_*} u_0\right) = 0 \,,
\label{eq:u0}\\
&\mathcal{D}_{\ell} u_1 +
f\left(\frac{d-4}{r^2}-\frac{2(d-3)}{r^2}f+\frac{(d-3)}{r}
f'\right)u_1\notag\\
&-\frac{f}{r}\left(f'-\frac{2f}{r}\right)u_2 = 0 \,,
\label{eq:u1}\\
&\mathcal{D}_{\ell} u_2 + f\frac{(d-4)}{r^2}
u_2+\frac{2f\ell(\ell+d-3)}{r^2}u_1 = 0 \,,
\label{eq:u2}
\end{align}
where $\mathcal{D}_\ell$ is the operator defined by
\begin{align}
\label{eq:Doperator}
\mathcal{D}_\ell= &-\partial^2_t +\partial^2_{r_*}-\notag\\
&\begin{aligned}[b]f
\biggl(&\frac{(\ell+1)(\ell+d-4)}{r^2}+
\frac{(d-4)(d-6)}{4r^2}f\\&
+\frac{(d-4)}{2r}f' + \mu^2 \biggr)\,,\end{aligned}
\end{align}
with $dr_*$ being defined through $dr_* = \frac{dr}{f(r)}$,
$\partial_t$ and $\partial_{r_*}$
denoting partial derivatives in relation to $t$ and 
$r_*$, respectively, a dash meaning derivative
with respect to $r$, 
 and where we
have suppressed the harmonic vectors in the subscript 
of $u_0$, $u_1$ and $u_2$. The Bianchi identity,
for $A_{\mu}$, i.e., 
$\nabla^{\mu}A_{\mu}= 0$,
relates
the scalar sector 
functions through
\begin{align}
\partial_t u_0- \partial_{r_*} u_1 =
\frac{f}{r}\left(\frac{d-2}{2}u_1-u_2\right) \,.
\label{eq:u_bianchi}
\end{align}
On the other hand, the vector-type modes are all
decoupled and described by the same wave-like equation,
\begin{equation}\label{eq:uv}
\mathcal{D}_\ell u_3=0\,,
\end{equation}
where we  have suppressed the harmonic indices in $u_3$.

\subsubsection{Proca radial equations in Schwarzschild-AdS}

The QNMs of Proca field perturbations in
Schwarzschild-AdS are dynamical solutions of
Eqs.~\eqref{eq:u0}-\eqref{eq:uv} for $u_0$, $u_1$, $u_2$ and $u_3$,
with the appropriate boundary conditions specified at the event
horizon and at spatial infinity. By disregarding the static solution
of $u_0$ to Eq.~\eqref{eq:u0}, one can choose $u_0$ to be completely
determined from $u_1$ and $u_2$ through Eq.~\eqref{eq:u_bianchi}, so
that the physical solutions are described solely by
Eqs.~\eqref{eq:u1}-\eqref{eq:u2} for $u_1$ and $u_2$, and by
Eq.~\eqref{eq:uv} for $u_3$.  By further assuming a time dependence of
the form
\begin{equation}\label{eq:timed}
u(r,t)=u(r)e^{-i\omega t}\,,
\end{equation}
the wave-like equations become
Schr\"{o}dinger-like equations for the mode frequencies $\omega$,
with associated effective potentials which, alongside with the
boundary conditions, determine the dynamics of the perturbations of
the Proca field.

For the scalar-type modes, the perturbations are ruled by the coupled
system of equations
\begin{equation}\label{eq:sch_coupled}
\partial^2_{r_*} \boldsymbol{u}+ (\omega^2
\boldsymbol{I}-\boldsymbol{V_{\rm s}}) \boldsymbol{u} = 0 \,,
\end{equation}
where $\boldsymbol{u}=(u_1 \;\; u_2)^T$, $\boldsymbol{I}$ is the
identity matrix and $\boldsymbol{V_{\rm s}}$ can be read from
Eqs.~\eqref{eq:u1} and~\eqref{eq:u2} as
\begin{gather}\label{eq:scalarpotential_sads}
\boldsymbol{V_{\rm s}}=
\begin{pmatrix}
V_{s 1 1}  &  V_{s 1 2}  \\  
V_{s 2 1} &  V_{s 2 2}
\end{pmatrix}\,,
\end{gather}
with the following components
\begin{align}
V_{s 1 1}=&f
\left(\frac{(d-4)(d-2)+4\mu^2l^2}{4l^2}\right. \nonumber\\
&+\left.\frac{4\ell(\ell+d-3)+d(d-2)}{4r^2}\right. \nonumber\\
&-\left.\frac{3(d-2)^2}{4r_{\rm h}^2}
\left(1+\frac{r_{\rm h}^2}{l^2}\right)
\left(\frac{r_{\rm h}}{r}\right)^{d-1}\right)\,,\nonumber\\
V_{s 1 2}=&f
\left(-\frac{2}{r^2}+\frac{d-1}{4r_{\rm h}^2}
\left(1+\frac{r_{\rm h}^2}{l^2}\right)
\left(\frac{r_{\rm h}}{r}\right)^{d-1}\right)\,,\nonumber\\
V_{s 2 1}=&-f\frac{2\ell(\ell+d-3)}{r^2}\,,\nonumber\\
V_{s 2 2}=&f
\left(\frac{(d-4)(d-2)+4\mu^2l^2}{4l^2}\right.\nonumber\\
&+\left.\frac{4\ell(\ell+d-3)+(d-4)(d-6)}{4r^2}\right.\nonumber\\
&-\left.\frac{d(d-4)}{4r_{\rm h}^2}\left(1+
\frac{r_{\rm h}^2}{l^2}\right)\left(\frac{r_{\rm h}}{r}
\right)^{d-1}\right)\,.
\label{eq:matrixpotential}
\end{align}
The monopole mode, $\ell = 0$, is described uniquely by
Eq.~\eqref{eq:u1} with $u_2=0$, and so by the one-dimensional
Sch\"{o}dinger-like with potential $V_{s11}$.

For vector-type perturbations, the radial equation in
Eq.~\eqref{eq:uv} can be written as
\begin{equation}\label{eq:sch_u3}
    \partial^2_{r_*} u_3 + (\omega^2-V_{\rm v})u_3 = 0\,,
\end{equation}
with the potential
\begin{align}
V_{\rm v} = f
\biggl(&\frac{(d-4)(d-2)+4\mu^2l^2}{4 l^2}\notag\\
&+\frac{(2\ell+d-4)(2\ell+d-2)}{4 r^2}\notag\\
&+\frac{d(d-4)}{4r_{\rm h}^2}\left(1+
\frac{r_{\rm h}^2}{l^2}\right)
\left(\frac{r_{\rm h}}{r}\right)^{d-1}
 \biggr) \,.
\label{eq:vectorpotential_sads}
\end{align}

%%%%%%%%%%%%%%%%%%%%%%%%%%%%%%%%%%%%%%%%%%%%%%%%%%%%%%%%%%%%%%%%
\subsection{Maxwell
equations in spherically symmetric backgrounds:
The radial equations in Schwarzschild-AdS}
%%%%%%%%%%%%%%%%%%%%%%%%%%%%%%%%%%%%%%%%%%%%%%%%%%%%%%%%%%%%%%%%

The Maxwell field equations can be obtained by carefully taking the
massless limit of the Proca field equations. As explained in
Section~\ref{sec:proca_field}, the equations are now invariant under
a gauge transformation, and the Maxwell field carries pure gauge
degrees of freedom that need to be eliminated.

The scalar-type sector contains pure-gauge degrees of freedom, and one
needs to find a suitable gauge-invariant combination of scalar-type
variables to describe the unique scalar-type physical degree of
freedom. This is achieved by defining the gauge-invariant variable
\begin{align}
u_{1 2} = \frac{u_1}{r} -\frac{r^{\frac{d}{2}-2}}{\ell(\ell+d-3)}
\partial_{r_*} \left(r^{2-\frac{d}{2}}u_2\right)\,,
\label{eq:u12}
\end{align}
which yields the
Maxwell equation for the physical scalar-type degree of freedom,
\begin{equation}\label{eq:sch_u12}
\partial^2_{r_*} u_{1 2} + (\omega^2-V_{\rm sm})u_{12} = 0\,,
\end{equation}
with 
\begin{align}
V_{\rm sm} = f\biggl(
&\frac{(d-4)(d-6)}{4 l^2}\notag\\
&+\frac{(2\ell+d-4)(2\ell+d-2)}{4 r^2}\notag\\
&-\frac{(3d-8)(d-4)}{4r_{\rm h}^2}\left(1+
\frac{r_{\rm h}^2}{l^2}\right)\left(\frac{r_{\rm h}}{r}\right)^
{d-1}\biggr)\label{eq:scalarpotentialmaxwell_sads}\,,
\end{align}
where the subscript m stands for Maxwell to distinguish from
the Proca case.
In order to understand what happens when taking the massless limit of
the Proca scalar-type sector,
i.e., to understand
Eqs.~\eqref{eq:u12}-\eqref{eq:scalarpotentialmaxwell_sads},
it is useful to write Eq.~\eqref{eq:u2}
as
\begin{equation}\label{eq:rw1}
\mathcal{D}^{s=0}_\ell \left(ru_2\right) = -2\ell(\ell+d-3)f u_{12}\,,
\end{equation}
where $\mathcal{D}^{s=0}_\ell$ is the operator ruling the dynamics of
a massive scalar field propagating in a spherically symmetric
background which is given by \cite{hubeny2000}
\begin{align}\label{eq:operator_scalarfield}
\mathcal{D}_{\ell}^{s=0}=&-\partial_t^2+\partial^2_{r_*}\notag\\
&\begin{aligned}[b]
-f\biggl(&\frac{\ell(\ell+d-3)}{r^2}+\frac{(d-2)(d-4)}{4r^2}f\\
&+\frac{d-2}{2r}f'+\mu^2\biggr)
\,.\end{aligned}
\end{align}
Now, by applying the operator of Eq.~\eqref{eq:Doperator} to
Eq.~\eqref{eq:rw1} and using  Eqs.~\eqref{eq:u1}-\eqref{eq:u2} and
the definition of $u_{12}$ given in
Eq.~\eqref{eq:u12} to reduce the order of the derivatives, one
obtains the decoupled quartic order equation for $u_2$ as
\begin{equation}\label{eq:u2dec}
\mathcal{D}^{s=1}_\ell(f^{-1}\mathcal{D}^{s=0}_\ell (ru_2)) =
2 \mu^2 f f' u_2 \,,
\end{equation}
with $\mathcal{D}_{\ell}^{s=1}$
being given by 
\begin{align}\label{eq:Ds1operator}
\mathcal{D}_{\ell}^{s=1}=&-\partial_t^2+\partial^2_{r_*}\notag\\
&\begin{aligned}[b]
-f\biggl(&\frac{\ell(\ell+d-3)}{r^2}+\frac{(d-2)(d-4)}{4r^2}f\\\
&-\frac{d-4}{2r}f' + \mu^2\biggr)\,.
\end{aligned} 
\end{align}
The operator $\mathcal{D}_{\ell}^{s=1}$ of Eq.~\eqref{eq:Ds1operator}
is the $d$-dimensional generalization of the
four-dimensional operator appearing in \cite{dolan2012}, and
$\mathcal{D}_{\ell}^{s=1}+\mu^2f$
being the operator  ruling the dynamics of
the scalar sector of the Maxwell field in $d$ dimensions.
The decoupled equation Eq.~\eqref{eq:u2dec} shows that the presence of
a nonzero mass $\mu$ makes the degrees of freedom of the function
$u_2$ physical.
On the other hand,
if one performs the massless limit $\mu^2=0$,
Eq.~\eqref{eq:u2dec} becomes
\begin{equation}\label{eq:rw2}
\mathcal{D}_{\ell}^{s=1}|_{\mu=0}
\left(f^{-1}\mathcal{D}_{\ell}^{s=0}|_{\mu=0}
\left(ru_2\right)\right) 
= 0\,,
\end{equation}
where
$\mathcal{D}_{\ell}^{s=0}|_{\mu=0}$ is the operator of
Eq.~\eqref{eq:operator_scalarfield} with $\mu=0$ and
$\mathcal{D}_{\ell}^{s=1}|_{\mu=0}$ is the operator of
Eq.~\eqref{eq:Ds1operator} with $\mu=0$.  Thus, in this
limit, the equations completely factorize. This was found
for four dimensions
in
\cite{dolan2012,tiago_sads2022}. One can interpret the
factorization by Eq.~\eqref{eq:rw2} in the following way.  The
solution to $\mathcal{D}_{l}^{s=0}\left(ru_2\right) = 0$ describes the
pure-gauge degree of freedom of the Maxwell field, i.e., $u_{12} = 0$,
see Eq.~\eqref{eq:rw1}.
The physical degree of freedom is described by the function
$u_{12}$ and it obeys the Maxwell scalar-type equation
$\mathcal{D}^{s=1}_\ell u_{1 2} = 0$, see again Eq.~\eqref{eq:rw1}.
When one puts $\partial_t^2=-\omega^2$ this latter equation yields
Eqs.~\eqref{eq:u12}-\eqref{eq:scalarpotentialmaxwell_sads}.

The vector-type sector,
on the other hand, can be shown to be gauge
invariant, meaning that vector-type modes are  ruled by
Eq.~\eqref{eq:sch_u3} after setting the mass $\mu$ to zero, i.e.,
$\mu=0$, in the potential of Eq.~\eqref{eq:vectorpotential_sads}.
Thus, from Eqs.~\eqref{eq:sch_u3} and \eqref{eq:vectorpotential_sads}
we write
\begin{equation}\label{eq:sch_u3m}
    \partial^2_{r_*} u_3 + (\omega^2-V_{\rm vm})u_3 = 0\,,
\end{equation}
with the potential
\begin{align}
V_{\rm vm} = f
\biggl(&\frac{(d-4)(d-2)}{4 l^2}\notag\\
&+\frac{(2\ell+d-4)(2\ell+d-2)}{4 r^2}\notag\\
&+\frac{d(d-4)}{4r_{\rm h}^2}\left(1+
\frac{r_{\rm h}^2}{l^2}\right)
\left(\frac{r_{\rm h}}{r}\right)^{d-1}
 \biggr) \,,
\label{eq:vectorpotential_sadsm}
\end{align}
where again the subscript m stands for Maxwell.

%%%%%%%%%%%%%%%%%%%%%%%%%%%%%%%%%%%%%
\section{Analytical properties of Proca and Maxwell quasinormal modes
\label{sec:analytical}}
%%%%%%%%%%%%%%%%%%%%%%%%%%%%%%%%%%%%%

\subsection{Asymptotic behavior of the solutions and boundary conditions}

\subsubsection{Proca field perturbations}

%here

To obtain the QNM solutions in
Schwarzschild-AdS, one needs to impose appropriate boundary conditions
at the horizon and at spatial infinity.

At the horizon, we have
$f(r_{\rm h})=0$, and all the potentials vanish, meaning that one
should have at first
order $u \sim e^{\pm i \omega r_{*}}$ near the horizon. 
The only classically allowed boundary condition is then
that there are only ingoing waves, i.e., 
$u \sim e^{-i \omega r_*}$ which in terms of the coordinate
$r$ can be written as
\begin{equation}\label{eq:bc_horizon}
u_k(r) =\alpha^{r=r_{\rm h}}_k 
\left(r-r_{\rm h}\right)^{-\frac{i \omega}{f'(r_{\rm h})}},
\quad\quad 
\quad r\rightarrow r_{\rm h} \,,
\end{equation}
where $\alpha^{r=r_{\rm h}}_k$ is a constant,
with $k\in \{1,2,3\}$, and
we have expanded $f(r)=f'(r_{\rm h})(r-r_{\rm
h})+\mathcal{O}\left((r-r_{\rm h})^2\right)$.

The suitable boundary conditions at spatial infinity in pure AdS were
determined in \cite{wald2004}, see also \cite{lopes2024}.
To address this
question, one must study how the Proca field behaves at spatial
infinity.  Since the effect of the black hole can be neglected
asymptotically, the behavior of the Proca field near spatial infinity
is essentially the same as the one found for pure AdS. Namely, near
$r=+\infty$, all Proca solutions behave as
$u_k = \alpha^{r=\infty}_k r^{-
\frac{1}{2}\left(1 + \sqrt{(d-3)^2 + 4\mu^2l^2}\right)}  
+ \beta^{r=\infty}_k r^{-\frac{1}{2}\left(1 - \sqrt{(d-3)^2 +
4\mu^2l^2}\right)}$
where $\alpha^{r=\infty}_k$ and $\beta^{r=\infty}_k$ are constants,
with $k\in \{1,2,3\}$.
Now we have to impose conditions on the functions $u_k$
so that they behave well at infinity.
For $d=4$ and $0<\mu^2l^2<\frac{3}{4}$
the functions
$u_k$ are square-integrable in the sense of $\int {\bar u}_k u_k
dr_*$ for 
one-parameter family of boundary conditions that includes the Dirichlet
boundary condition, the latter meaning $u_k (r\rightarrow
\infty) = 0$ which implies 
$\beta^{r=\infty}_k=0$.
For $d=4$ and $\frac{3}{4}\leq\mu^2l^2<\infty$
the functions
$u_k$ are square-integrable only if one imposes 
a Dirichlet
boundary condition $\beta^{r=\infty}_k=0$.
For $d>4$, the functions
$u_k$ are square-integrable
also only for $\beta^{r=\infty}_k=0$, which corresponds to the Dirichlet
boundary condition, see \cite{wald2004}  for these conditions.
In any case, throughout this work, we assume  Dirichlet
boundary conditions at spatial infinity, i.e., $u_k (r\rightarrow
\infty) = 0$, and so $\beta^{r=\infty}_k=0$.
Thus,
\begin{align}\label{eq:infbehaviorPROCA}
u_k = \alpha^{r=\infty}_k r^{-
\frac{1}{2}\left(1 + \sqrt{(d-3)^2 + 4\mu^2l^2}\right)}  ,
\quad r\rightarrow \infty,
\end{align}
gives the behavior of the function $u_k$ at infinity.

\subsubsection{Maxwell field perturbations}

The boundary condition at the horizon for Maxwell field perturbations
is \begin{equation}\label{eq:bc_horizonMaxwell}
u_i(r) =\alpha^{r=r_{\rm h}}_i
\left(r-r_{\rm h}\right)^{-\frac{i \omega}{f'(r_{\rm h})}},
\quad\quad 
\quad r\rightarrow r_{\rm h} \,,
\end{equation}
where $\alpha^{r=r_{\rm h}}_i$ is a constant
with  $i\in \{12,3\}$. Note that Eq.~\eqref{eq:bc_horizonMaxwell}
for Maxwell 
is analogous to Eq.~\eqref{eq:bc_horizon} for Proca,
the difference is in the running of the subscripts $\scriptsize i$
and $\scriptsize k$.

On the other hand, the behavior of the
Maxwell field near spatial infinity 
is different from  Proca. Near $r=+\infty$,
the Maxwell solutions behave
accordingly with the dimension $d$ \cite{wald2004,lopes2024}. 
Regarding the scalar-type sector, for $d=4$ the behavior is
$u_{12} = 
\alpha_{12}^{r=\infty}r^{-1}
+
\beta_{12}^{r=\infty}$,
for $d=5$ the behavior is
$u_{12} = 
\alpha_{12}^{r=\infty}r^{-\frac12}
+
\beta_{12}^{r=\infty}r^{-\frac12}\ln r
$,
for $d\geq6$ the behavior is
$u_{12} = \alpha_{12}^{r=\infty}r^{2-\frac{d}{2}}
+
\beta_{12}^{r=\infty}r^{\frac{d}{2}-3}$. Regarding the 
vector-type sector, for $d\geq4$ the behavior is
$u_3 = 
\alpha_{3}^{r=\infty}r^{1-\frac{d}{2}}
+
\beta_{3}^{r=\infty}r^{\frac{d}{2}-2}$
where $\alpha_i^{r=\infty}$ and $\beta_i^{r=\infty}$ are constants,
with $i\in \{12,3\}$.
For $d=4,5,6$, a whole one-parameter family of boundary
conditions can be imposed. Nonetheless, we assume
boundary conditions that allow us to
connect with the Proca perturbations and so we set $\beta_i=0$.
For $d=4$, to connect with the Proca
perturbations, we impose the Dirichlet boundary condition,
which corresponds to setting $u_i (r\rightarrow
\infty) = 0$ and so $\beta_i^{r=\infty}=0$.
For
$d=5$ for the
scalar-type Maxwell perturbations,
we have $u_{12} (r\rightarrow \infty) = 0$
even for $\beta_i^{r=\infty}\neq0$. In this case, we impose the
Dirichlet-Neumann condition, that is, the vanishing of the
dominant logarithmic term, thus $\beta_i^{r=\infty}=0$
in $d=5$ as well.
For $d=6$, again to connect with the Proca
perturbations, we impose the Dirichlet boundary condition,
$u_i (r\rightarrow
\infty) = 0$ and so $\beta_i^{r=\infty}=0$.
For $d\geq 7$, the Dirichlet boundary condition
is the only one that guarantees well-defined dynamics for the Maxwell
field \cite{wald2004}.
Thus, after all this discussion,
the Maxwell field $u_i$ at infinity can be written as
\begin{align}%\label{eq:infbehaviourMAXWELL}
u_{12}& = \begin{cases}
\dfrac{\alpha_{12}^{r=\infty}}{r},
&\,r\to\infty,\quad\quad d=4\,,\nonumber\\
\\
\dfrac{\alpha_{12}^{r=\infty}}{r^{\frac{d}{2}-2}},
& \,r\to\infty,\quad\quad d\geq5\,,\nonumber\\
\end{cases}
\\\nonumber
\\\label{eq:infbehaviourMAXWELL2}
u_3 &= \quad
\dfrac{\alpha_{3}^{r=\infty}}{r^{\frac{d}{2}-1}} ,
\,\,\,r\to\infty,\quad\quad d\geq4
\,,
\end{align}
where the $\alpha_i^{r=\infty}$ are constants,
with $i\in \{12,3\}$.

\subsection{Linear stability of Schwarzschild-AdS}
\label{sec:stability}

\subsubsection{Method}

We now prove that, for the boundary conditions used, all the decoupled
Schr\"{o}dinger-like equations of Section~\ref{sec:proca_in_sads},
i.e., the monopole mode and  vector-type
modes of the Proca field, and the scalar
and vector-type
modes of the Maxwell field,
must
yield QNM frequencies $\omega$ with negative imaginary
part, which amounts to say that Schwarzschild-AdS is linearly stable
against such perturbations.
The approach we use for the proof is
well-known and can be found in,
e.g.,
\cite{hubeny2000,cardoso_lemos2001,ishibashi_kodama_stability2003,
zhidenko_massive_scalar2005,phd_zhidenko2009}. We consider a general coupled
system of Schr\"{o}dinger-like equations with a Hermitian potential
$\boldsymbol{V}$ with solutions $\boldsymbol{u}$,
\begin{align}\label{eq:schrodingergeneral}
\frac{d^2\boldsymbol{u}}{d r_*^2}
+\left( \omega^2- \boldsymbol{V}\right)
\boldsymbol{u}=0
\,.
\end{align}
Multiplying the system
given in Eq.~\eqref{eq:schrodingergeneral}
by $\boldsymbol{u}^\dagger$ and integrating in
the region of interest, one has an energy balance like equation
\begin{align}\label{eq:stability1}
&\left(\boldsymbol{u}^\dagger \frac{d \boldsymbol{u}}{d r_*}\right) 
\biggr\rvert_{r_*=-\infty}^{r_*=r_*^{(\infty)}} 
+ \omega^2 \int^{r_*^{(\infty)}}_{-\infty}\left|
\boldsymbol{u} \right|^2 dr_*\notag\\
&=\int^{r_*^{(\infty)}}_{-\infty} \left(\left|
\frac{d\boldsymbol{u}}{d r_*}\right|^2
+ \boldsymbol{u}^\dagger \boldsymbol{V} \boldsymbol{u} \right) dr_*
\,,
\end{align}
where the first term means the difference between the evaluation at 
spatial infinity and the evaluation at the horizon, in this order.
We can conveniently introduce an arbitrary operator
given by
\begin{equation}\label{eq:sdeformation1}
\boldsymbol{D}=\boldsymbol{I}\frac{d}{d r_*}+\boldsymbol{S}(r_*) \,,
\end{equation}
where $\boldsymbol{S}$ must be Hermitian.
Also, we use the boundary conditions at the horizon, i.e.,
Eq.~\eqref{eq:bc_horizon}
for Proca and Eq.~\eqref{eq:bc_horizonMaxwell}
for Maxwell, which sets one of the boundary terms to
$\boldsymbol{u}^\dagger \frac{d\boldsymbol{u}}{dr_*} = - i \omega
\left|\boldsymbol{u}(-\infty) \right|^2 $. With these two
considerations, the energy balance like equation,
Eq.~\eqref{eq:stability1},
becomes
\begin{align}
i\omega A + \omega^2 B =\int^{r_*^{(\infty)}}_{-\infty} 
\left(\left| \boldsymbol{D}\boldsymbol{u}\right|^2+
\boldsymbol{u}^\dagger
\bar{\boldsymbol{V}} 
\boldsymbol{u}\right) dr_* + C \,,\label{eq:s_deform}
\end{align}
where
\begin{align}
& \bar{\boldsymbol{V}} = \boldsymbol{V}+
\frac{d\boldsymbol{S}}{dr_*}-\boldsymbol{S}^2 \,
\nonumber\\
& A=\left|\boldsymbol{u}(-\infty) \right|^2 \,,\,
B=\int^{r_*^{(\infty)}}_{-\infty}\left|
\boldsymbol{u} \right|^2 dr_*\,,
\label{eq:Vbar}\\
&C = - \eval{\left(\boldsymbol{u}^\dagger 
\frac{d \boldsymbol{u}}{dr_*}\right)}_{r_*=r_*^{(\infty)}} 
-\left(\boldsymbol{u}^\dagger \boldsymbol{S}\boldsymbol{u}\right)
\biggr\rvert_{r_*=-\infty}^{r_*=r_*^{(\infty)}} \nonumber\,\,.
%\label{eq:C}
\end{align}
In order to proceed, we must require that the effective potential
$\bar{\boldsymbol{V}}$ must be positive definite and that $C$ must be
a real positive number. The main reason for the introduction of
$\boldsymbol{S}$ is related to the cases where $\boldsymbol{V}$ is not
positive definite, as one can choose an $\boldsymbol{S}$ so that
a modified potential $\bar{\boldsymbol{V}}$ is positive definite. 
This procedure is called
the $S$-deformation technique.  The requirement of $C$ being a real
positive number can be achieved by the combination of the boundary
condition at spatial infinity and the choice of $\boldsymbol{S}$. One
now can obtain the imaginary part of Eq.~\eqref{eq:s_deform} as
\begin{align}
\mathfrak{Re}(\omega) A + 2 \mathfrak{Re}(\omega) 
\mathfrak{Im}(\omega) B = 0\,.\label{eq:imaginarys_deform}
\end{align}
Since $A$ and $B$ are positive and nonzero, one has that 
$\mathfrak{Im}(\omega)<0$ if $\mathfrak{Re}(\omega) \neq 0$. 
In the case $\mathfrak{Re}(\omega) = 0$, one can
consider the real part of Eq.~\eqref{eq:s_deform}, written as
\begin{align}
&- \mathfrak{Im}(\omega) A - \mathfrak{Im}(\omega)^2 B 
\notag \\
&= \int^{r_*^{(\infty)}}_{-\infty} 
\left(\left| \boldsymbol{D}\boldsymbol{u}\right|^2+
\boldsymbol{u}^\dagger
\bar{\boldsymbol{V}} 
\boldsymbol{u}\right) dr_* + C \,.\label{eq:reals_deform}
\end{align}
Since the right-hand side is positive with the considerations above,
with $A$ and $B$ also positive, then $\mathfrak{Im}(\omega)< 0$.

Therefore, the method used here guarantees that if $\boldsymbol{V}$ is
Hermitian, if one can pick an $\boldsymbol{S}$ such that
$\boldsymbol{V}$ is positive definite, and if the choice of
$\boldsymbol{S}$ and the boundary condition at spatial infinity lead
to a real positive $C$, then the QNM frequencies have a
negative imaginary part, yielding thus linear stability.
Let us apply this technique to 
Proca and Maxwell field perturbations.

\subsubsection{Proca field perturbations}

For scalar-type Proca field perturbations, this technique can only be
addressed for the monopole mode, $\ell = 0$, with the potential given
by $V_{s11}$ in Eq.~\eqref{eq:matrixpotential}.  In this case,
although the monopole potential is negative in a region near the
horizon, one can consider the effective potential $\bar{{V}}
={V}+\frac{dS}{dr_*}-S^2 = f\mu^2$, see Eq.~\eqref{eq:Vbar},
with $S = \frac{(d-2)f}{2r}$.  One
has that $\bar{{V}}\geq0$, $S(r=r_{\rm h})=0$, $S
|u|^2\rvert_{r=+\infty} \sim \lim_{r\to +\infty}
r^{-\sqrt{(d-3)^2+4\mu^2l^2}} = 0$, and also $C=0$ due to the
Dirichlet boundary condition, so that the results from
Eqs.~\eqref{eq:imaginarys_deform} and~\eqref{eq:reals_deform} can be
used.  Schwarzschild-AdS is then linearly stable against monopole
Proca field perturbations.
Higher multipoles of scalar-type Proca field perturbations are
governed by the coupled potential Eqs.~\eqref{eq:scalarpotential_sads}
and~\eqref{eq:matrixpotential}, which is non-Hermitian. This leads to
additional terms in Eqs.~\eqref{eq:imaginarys_deform}
and~\eqref{eq:reals_deform}, making it impossible to draw any conclusions 
about stability through this technique. One might argue that there could be an
$\boldsymbol{S}$ function transforming the non-Hermitian potential
into a Hermitian one.  However, this is not the case, as for the
S-deformation technique to be useful, one also needs to ensure that
$\boldsymbol{S}$ itself is Hermitian.  From the relation between the
old potential and the transformed one, see Eq.~\eqref{eq:Vbar}, if
$\boldsymbol{\bar{V}} \equiv
\boldsymbol{V}+\frac{d\boldsymbol{S}}{dr_*}-\boldsymbol{S}^2$ is
Hermitian, $\boldsymbol{V}$ needs to be Hermitian.  Thus, in this
case, one can only test stability by computing numerically the
QNM spectrum.

For vector-type Proca field perturbations,
the potential $V_{\rm v}$ ruling them given in
Eq.~\eqref{eq:vectorpotential_sads}, is positive in the region of
interest. One can therefore choose $S = 0$ and due to the Dirichlet
boundary condition, $C$ also vanishes.  It follows then from
Eqs.~\eqref{eq:imaginarys_deform} and \eqref{eq:reals_deform} that
Schwarzschild-AdS is linearly stable against such perturbations.

\subsubsection{Maxwell field perturbations}

For the scalar-type Maxwell perturbations, the potential is given by 
Eq.~\eqref{eq:scalarpotentialmaxwell_sads}. One must split the analysis 
into the different  dimensions. 
For $d=4$, the potential is positive and Hermitian,
the Dirichlet condition leads to 
a vanishing $C$, and so one has $S = 0$. So there is 
linear stability. Note that in $d=4$, the scalar-type Maxwell
potential  is equal 
to the vector-type Maxwell potential,
For $d=5$, 
the potential is unbounded from below, such that 
$V\rightarrow-\infty$ as $r\rightarrow\infty$. One can choose an $S$ as
$S = \frac{f}{2r}$ so that $\bar{V}$ becomes positive. 
Moreover, by using the Dirichlet-Neumann condition, which we
are using in $d=5$,
and using the 
chosen $S$, surprisingly, the term $C$ vanishes.
Therefore, we have the necessary requisites to apply
Eqs.~\eqref{eq:imaginarys_deform} and~\eqref{eq:reals_deform}.
So
there is linear stability.  For $d\geq6$, the potential dips to
negative values near the horizon.  One must then choose $S =
\frac{(d-4)f}{2r}$ to have $\bar{V} \geq 0$.  By using the chosen $S$
and the Dirichlet boundary condition, the term $C$ vanishes, and so
from Eqs.~\eqref{eq:imaginarys_deform} and~\eqref{eq:reals_deform}
there is linear stability.

For vector-type Maxwell perturbations, the potential is given by 
Eq.~\eqref{eq:vectorpotential_sadsm}. It 
is positive and Hermitian and 
also the Dirichlet boundary condition ensures a vanishing $C$, and so from 
Eq.~\eqref{eq:imaginarys_deform} and~\eqref{eq:reals_deform} with $S = 0$, 
there is linear stability.

\vfill

\subsection{Isospectrality between scalar-type and vector-type 
perturbations in large black holes}\label{sec:isospectral}

\subsubsection{Initial considerations}

Proca fields in $d=4$ do not show isospectrality between scalar-type
and vector-type perturbations.  On the other hand, it is well-known
that scalar-type and vector-type Maxwell field perturbations are
exactly isospectral in $d=4$ \cite{cardoso_lemos2001}.
Here we investigate 
isospectrality in $d$ dimensions and show in what conditions
it might occur.

\subsubsection{Proca field perturbations and isospectrality}

For Proca, 
looking at the scalar potential matrix $\mathbb{V_{\rm s}}$ with components 
in Eqs.~\eqref{eq:matrixpotential}
and the vector potential $V_{\rm v}$ in Eq.~\eqref{eq:vectorpotential_sads}, 
respectively,
it is clear they are
completely different, even in $d=4$. This may indicate that
isospectrality between the two types of
perturbation for a Proca field can never be realized. Indeed, 
this is supported by numerical calculations in the Sec.~\ref{sec:num_results} and 
in~\cite{tiago_sads2022} for $d=4$.

\subsubsection{Maxwell field perturbations and isospectrality}

For Maxwell, we look
at the scalar potential $V_{\rm sm}$
and the vector potential $V_{\rm vm}$ in
Eqs.~\eqref{eq:scalarpotentialmaxwell_sads}
and \eqref{eq:vectorpotential_sadsm}, respectively.
If we restrict their expressions to $d=4$ it becomes clear
that the potentials are equal and so
the scalar and vector perturbations
are isospectral.
On the other hand, from Eqs.~\eqref{eq:scalarpotentialmaxwell_sads}
and~\eqref{eq:vectorpotential_sadsm}
it is immediate
that $V_{\rm sm}$
and  $V_{\rm vm}$
are
different and it may implicate that
isospectrality breaks for higher dimensions, $d\geq5$. This is 
in fact supported by the numerical results in the Sec.~\ref{sec:num_results}.

It turns out that for the Maxwell field,
as one
analyzes large black holes with $\frac{r_{\rm h}}{l}\gg \ell$, the
scalar-type and vector-type Maxwell perturbations tend to be
isospectral in higher dimensions, although nothing in the potentials
suggests such behavior.
Indeed,
Maxwell scalar-type perturbations are governed
by the potential given in 
Eq.~\eqref{eq:scalarpotentialmaxwell_sads},
which for
large black holes in $d\geq5$
reduces to
\begin{align}\label{eq:scalar_potential_large}
V_{\rm sm}^{\frac{r_{\rm h}}{l}\gg \ell}=
f\biggl(&\frac{(d-4)(d-6)}{4l^2}
\notag\\ &-\frac{(3d-8)(d-4)}{4l^2}
\biggl(\frac{r_{\rm h}}{r}\biggr)^{d-1}\biggr)\,.
\end{align}
Maxwell vector-type perturbations are governed
by the potential of Eq.~\eqref{eq:vectorpotential_sadsm}, which for
large black holes in $d\geq5$ can be written as
\begin{equation}\label{eq:vector_potential_large}
V_{\rm vm}^{\frac{r_{\rm h}}{l}\gg \ell}=
f\biggl(\frac{(d-2)(d-4)}{4l^2}+\frac{d(d-4)}{4l^2}
\biggl(\frac{r_{\rm h}}{r}\biggr)^{d-1}\biggr)\,.
\end{equation} 
The potentials of Eqs.~\eqref{eq:scalar_potential_large}
and 
\eqref{eq:vector_potential_large} 
can be written as
\begin{align}\label{eq:isospectral0}
&V_{\rm sm}^{\frac{r_{\rm h}}{l}\gg \ell} =  W^2+\frac{dW}{dr_*}
\,,
\notag\\ &
V_{\rm vm}^{\frac{r_{\rm h}}{l}\gg \ell} = W^2-\frac{dW}{dr_*}\,,
\\ &
W(r) \equiv -\frac{(d-4)}{2l^2}
\left(1-\biggl(\frac{r_{\rm h}}{r}\biggr)^{d-1}\right)r\,.\notag
\end{align}
Following a procedure developed by Chandrasekhar
\cite{chandra_mathematical_BH1983}, 
see also \cite{cardoso_lemos2001}, one can check that
Eqs.~\eqref{eq:sch_u12}
and~\eqref{eq:sch_u3m}  can be written as
\begin{align}
\label{eq:isospectral1}
&u_{12}=\frac{1}{\sqrt{\beta-\omega^2}}
\left(W u_{3}+\frac{d u_{3}}{dr_*}\right)
\,,
\notag\\
&u_3=\frac{1}{\sqrt{\beta-\omega^2}}
\left(-W u_{1 2}+\frac{d u_{1 2}}{dr_*}\right)
\,,
\end{align}
respectively.
Supposing that $\omega$ is a QNM of $u_3$, one must have
$u_3 = \alpha_3^{r=r_{\rm h}} {\rm e}^{-i\omega r_*}$
when $r\rightarrow r_{\rm h}$ and
$u_3 = 0$ when $r \rightarrow \infty$.
The behavior of $u_{1 2}$ near the horizon and
near spatial infinity then yields, 
from Eq.~\eqref{eq:isospectral1},
$u_{1 2} =  -\alpha_3^{r=r_{\rm h}}
{\rm e}^{-i\omega r_*}$ when $r\rightarrow r_{\rm h}$ and
$u_{1 2} = \frac{1}{\sqrt{-\omega^2}}
\left(Wu_3+f \frac{du_3}{dr}
\right)\bigr\rvert_{r=\infty}$ when
$r \rightarrow \infty$.
The behavior of $u_3$ near spatial infinity is given by 
Eq.~\eqref{eq:infbehaviourMAXWELL2}, so that 
$\left(Wu_3+f \frac{du_3}{dr}\right)\bigr\rvert_{r=\infty} 
\sim \frac{d-3}{l^2} r^{2-\frac{d}2}\bigr\rvert_{r=\infty}$,
which vanishes for $d\geq5$, i.e., $u_{12}\to0$. 
Thus, if $\omega$ is a QNM frequency of $u_3$, 
it is also a QNM frequency of $u_{12}$
when $r \rightarrow \infty$,
and so the scalar-type
and vector-type modes tend to isospectrality
when $\frac{r_{\rm h}}{l}\gg \ell$. 
Note that we do not consider modes that have
$\omega^2=0$, as it is not expected 
that they exist.

\subsection{Low purely damped  modes for large
black holes}\label{sec:siopsis}

\subsubsection{Initial considerations}

Scalar-type and vector-type gravitational perturbations in large
Schwarzschild-AdS black holes possess low frequency modes, which play
a role in the AdS/CFT correspondence, as they describe the behavior of
the dual field theory in the hydrodynamic regime.  These gravitational
modes were
found numerically  \cite{cardoso_lemos2001,cardoso_lemos_konoplya2003}
and analytically \cite{zanchin2006}
in $d=4$, and 
were
found numerically
\cite{gubser2007} in $d=5$.  Using a
perturbative approach, in \cite{siopsis2007}
it was found the analytical expression
for these modes in $d$ dimensions, whose results agree not only with
the previous numerical results, but also with previous analytical
results from the dual field theory side.

In this section, we employ
the perturbative method of \cite{siopsis2007} to study if such
low-frequency modes exist in Proca and Maxwell fields.
We mention in passing Proca field modes
and argue that for scalar-type Proca perturbations
the approach does not seem to be
feasible, as this would require solving a coupled system, while 
the results may follow for the
vector-type Proca perturbations from the vector-type Maxwell case,
indeed, in $d\geq4$, vector-type Proca perturbations
do not possess such low-frequency modes. 
We perform an exhaustive analysis for Maxwell field perturbations.
For scalar-type Maxwell perturbations
we show that in $d=4$ there are no such modes.
In $d\geq5$ we show that
scalar-type Maxwell perturbations in large Schwarzschild-AdS
black holes have purely damped modes that scale with the inverse of
the radius of the black hole.
In $d\geq4$, vector-type Maxwell perturbations
do not possess low-frequency modes.

We must note that in
\cite{cardoso_lemos2001,tiago_sads2022}, for $d=4$, both
the Proca and Maxwell fields have a finite number of almost purely
damped high modes for large black holes, both in the scalar and the
vector sector. However, these modes scale with the event horizon
radius and so they must not be confused with the low purely damped
modes that we are analyzing here.

\subsubsection{Proca field perturbations}

For Proca scalar field perturbations and modes
we will not pursue an analytical
study as this would require solving perturbatively the coupled system of
Eq.~\eqref{eq:sch_coupled}. The existence of such modes will be
considered numerically.

For vector field perturbations and modes,
we will analyze the corresponding modes
in the Maxwell theory and show that
the analysis 
can be extended to the vector-type Proca perturbations by
taking into account the mass $\mu$ of the field.

\subsubsection{Maxwell field perturbations}

We now do first scalar-type perturbations
for the Maxwell field.
To check for the existence of the low pure damped modes, we must
analyze the solution in the limit of very large black holes, i.e.,
$r_{\rm h} \to \infty$, while keeping
$\hat{l}^2$, $\hat{\ell}$, and $\hat{\omega}$ small but fixed,
these being defined
through
\begin{align}
\label{eq:newdefinitions}
\hat{l}^2=\frac{l^2}{r_{\rm h}^2}\,,\quad
\hat{\ell}^2=\hat{l}^2\ell(\ell+d-3)\,,\quad
\hat{\omega} = \hat{l}^2\omega r_{\rm h}
\,.
\end{align}
Following \cite{siopsis2007}, we use a transformation
from the variable $r$ to $\zeta$ and from the function
$u_{1 2}$ to $U_{12}$ by
\begin{align}
\label{eq:newvariables}
\hskip -0.2cm
\zeta=\left(\frac{r_{\rm h}}{r}\right)^{d-3}\hskip -0.1cm, \quad
u_{1 2}(\zeta)=(1-\zeta)^{
\frac{i\hat{\omega}}{(d-1)+(d-3)\hat{l}^2}}
U_{12}(\zeta)
\,.
\end{align}
Then, with these new definitions and variables 
Eq.~\eqref{eq:sch_u12} turns into
\begin{equation}\label{eq:siopsis2}
A_{\hat{l}^2} \partial^2_\zeta U_{12} 
+ B_{\hat{l}^2,\hat{\omega}}\partial_\zeta U_{12} 
+ C_{\hat{l}^2, \hat{\omega},\hat{\ell}^2}U_{12} = 0\,,
\end{equation}
with coefficients
\begin{align}
\hskip -0.2cm
A_{\hat{l}^2}\hskip 0.15cm =\hskip 0.15cm& (d-3)^2 \left(\zeta^3-
\zeta^{\frac{2d-8}{d-3}}\right)
+{(d-3)^2 (\zeta-1) \zeta^2}\hat{l}^2,\nonumber\\
%\end{align}
%\begin{align}
B_{\hat{l}^2,\hat{\omega}}=&\hspace{1mm}(d-3)
\left((2 d-5) \zeta^2-(d-4)\zeta^{\frac{d-5}{d-3}}\right)
\notag\\
&+(d-3) \left((2 d-5) \zeta^2 + (2-d) \zeta\right)\hat{l}^2
\notag\\
&+\frac{2 i (d-3)^2 \left(\zeta^{\frac{2d-8}{d-3}}-
\zeta^3\right)}{(d-1) (\zeta-1)}\hat{\omega}
+\mathcal{O}(\varepsilon^2)\,,\\
%\end{align}
%\begin{align}
C_{\hskip -0.03cm\hat{l}^2\hskip -0.05cm,
\hat{\omega},\hat{\ell}^2}\hskip -0.05cm
=&\frac{(d-4)}{4} 
\left((d-6) \zeta^{-\frac{2}{d-3}}+(8-3 d) \zeta\right)\nonumber\\
&+\frac{1}{4}(d-4) \left((d-2) +(8-3 d) \zeta \right)
\hat{l}^2\notag\\
&-\frac{i (d-3) \zeta}{(d-1) (\zeta-1)^2} 
\biggl((d+u-4)\zeta^{-\frac{2}{d-3}}\notag\\&+\zeta
(d (u-2)-2 \zeta+5)\biggr)\hat{\omega}
+\hat{\ell}^2+\mathcal{O}(\varepsilon^2)\,,\nonumber
\end{align}
where an expansion up to first order was done in
$\hat{l}^2$,
$\hat{\ell}^2$,
and $\hat{\omega}$, and the parameter $\varepsilon$ can be
understood as the smallness of these three parameters, meaning
$\hat{l}^2 \sim \varepsilon$,
$\hat{\ell}^2 \sim \varepsilon$, and
$\hat{\omega} \sim \varepsilon$.
The idea now is to consider a
perturbative expansion of the solution for $U_{12}$ as
\begin{align}
\label{perturbationU12}
U_{12} = U^{(0)}_{12}+
U^{(1)}_{12}+\mathcal{O}(\varepsilon^2)\,,
\end{align}
with $\varepsilon \ll 1$,
where $\mathcal{O}(U^{(0)}_{12}) = \mathcal{O}(1)$ and
$\mathcal{O}(U^{(1)}_{12}) = \mathcal{O}(\varepsilon)$.  At zeroth
order, we obtain
\begin{equation}\label{eq:siopsis_zerothorder}
\mathcal{H}_0 U^{(0)}_{12} = 0, \quad\quad
\mathcal{H}_0 \equiv A_0 \partial^2_\zeta + B_{0,0}\partial_\zeta +
C_{0,0,0}\,,
\end{equation}
with
$A_0=A_{\hat{l}^2=0}$,
$B_{0,0}=B_{\hat{l}^2=0,\hat{\omega}=0}$,
$C_{0,0,0}=C_{\hat{l}^2=0,\hat{\omega}=0,\hat{\ell}^2=0}$.
The solutions to Eq.~\eqref{eq:siopsis_zerothorder} are
written as
\begin{equation}\label{eq:solution_u0}
U^{(0)}_{12} (\zeta) = \tilde{\gamma}_{12}^{(0)}
\tilde{U}^{(0)}_{12}(\zeta)  +
\hat{\gamma}_{12}^{(0)} \hat{U}^{(0)}_{12}(\zeta)\,,
\end{equation}
with $\tilde{\gamma}_{12}^{(0)}$ and
$\hat{\gamma}_{12}^{(0)}$ constants, and
\begin{align}
&\tilde{U}^{(0)}_{12} (\zeta) =
\zeta^{\frac{d-4}{2(d-3)}}\,,\quad
\hat{U}^{(0)}_{12}(\zeta)=\zeta^{\frac{d-4}{2(d-3)}}
\int^\zeta d\zeta'
\frac{\mathcal{W}(\zeta')}{\zeta'^{\frac{d-4}{(d-3)}}},
\label{eq:solUhat}
\end{align}
where $\mathcal{W}(\zeta)$ is the Wronskian with the following
dependence $\mathcal{W}(\zeta) \sim
\zeta^{-\frac{d-4}{d-3}}\left(1-\zeta^\frac{d-1}{d-3}\right)^{-1}$.
The boundary condition at the horizon, described in these coordinates
as $\zeta=1$, imposes $\hat{\gamma}_{12}^{(0)}  = 0$ in
Eq.~\eqref{eq:solution_u0} for all spacetime dimensions, since
$\hat{U}^{(0)}_{12} \sim \ln(1-\zeta)$.
For $d=4$, the Dirichlet boundary condition is only
satisfied if
$\tilde{\gamma}_{12}^{(0)} = 0$, which yields the trivial solution
$U^{(0)}_{12}=0$. This implies the absence of the low pure damped
modes for the case of $d=4$.
The
Dirichlet-Neumann boundary condition for
$d=5$
and the
Dirichlet boundary
condition for $d > 5$
at spatial infinity, described by $\zeta=0$, are automatically
satisfied.
Carrying on for the case of $d\geq 5$,
at first order Eqs.~\eqref{eq:siopsis2}-\eqref{perturbationU12}
yield
\begin{equation}\label{eq:siopsis_firstorder}
\mathcal{H}_0 U^{(1)}_{12} = - \mathcal{H}_1 U^{(0)}_{12}\,,
\end{equation}
where $\mathcal{H}_1 =  A_{\hat{l}^2}\partial^2_\zeta+
B_{\hat{l}^2,\hat{\omega}}\partial_\zeta 
+ C_{\hat{l}^2,\hat{\omega},\hat{\ell}^2} - \mathcal{H}_0$. 
It can be shown that $\left(A_{\hat{l}^2} \partial^2_\zeta 
+ B_{\hat{l}^2,0}\partial_\zeta+C_{\hat{l}^2,0,0}\right)
\Tilde{U}^{(0)}_{12} = 0$, 
so that if one neglects the $\hat{l}^2$ terms in the forthcoming
calculations, the results still remain valid at
$\mathcal{O}(\hat{l}^2)$.
The particular solutions to
Eq.~\eqref{eq:siopsis_firstorder} can be written as
\begin{align}\label{eq:solution_u1}
U^{(1)}_{12}(\zeta) = \,& \hat{U}^{(0)}_{12}(\zeta)
\int^{\zeta}_1 d\zeta ' \frac{\tilde{U}^{(0)}_{12} 
\mathcal{H}_1 U^{(0)}_{12}}{A_0 \mathcal{W}} \notag\\
+&\tilde{U}^{(0)}_{12}(\zeta)\int^{\zeta}_1
d\zeta ' \frac{\hat{U}^{(0)}_{12} 
\mathcal{H}_1 U^{(0)}_{12}}{A_0 \mathcal{W}} \,,
\end{align} 
where the lower limit in the integrals was chosen to be $\zeta'=1$ so
that the boundary condition at the horizon is automatically
satisfied. The possible divergent term at spatial infinity is the
first line in Eq.~\eqref{eq:solution_u1}, i.e., the term proportional
to $\hat{U}^{(0)}_{12}$. Imposing the boundary condition at spatial
infinity, one must require that $\int_1^0 d\zeta '
\frac{\tilde{U}^{(0)}_{12} \mathcal{H}_1 U^{(0)}_{12}}{A_0
\mathcal{W}} = 0$.  This last condition, together with $U^{(0)}_{12}
\sim \zeta^{\frac{d-4}{2(d-3)}}$, see Eq.~\eqref{eq:solUhat}, gives
precisely the equation for the low pure damped QNM
frequencies
\begin{equation}\label{eq:siopsis_final}
\omega = -i \frac{\ell (\ell+d-3)}{(d-3)r_{\rm h}}\,.
\end{equation}
Note again that these modes do not exist 
for $d=4$, since Eq.~\eqref{eq:siopsis2} has no
nontrivial solutions satisfying the appropriate boundary conditions.
The modes of Eq.~\eqref{eq:siopsis_final} only exist
for $d\geq5$.

We continue with the analysis now for
vector-type perturbations
of the Maxwell field.
For vector-type Maxwell field perturbations, we can perform the same
transformations as in the scalar-type perturbations,
which now are
\begin{align}
\label{eq:newvariablesu3}
\hskip -0.2cm
\zeta=\left(\frac{r_{\rm h}}{r}\right)^{d-3}\hskip -0.2cm, \quad
u_{3}(\zeta)=(1-\zeta)^{
\frac{i\hat{\omega}}{(d-1)+(d-3)\hat{l}^2}}
U_{3}(\zeta)
\,,
\end{align}
see Eq.~\eqref{eq:newvariables}.
Doing the same procedure that was done for
scalar-type perturbations, the vector-type solutions at zeroth order
of $\hat{\ell}^2 \sim \varepsilon$, $\hat{\omega} \sim \varepsilon$
and $\hat{l}^2 \sim \varepsilon$ are
\begin{equation}\label{eq:solution_u0_vt}
U^{(0)}_{3} (\zeta) = \tilde\gamma_{3}^{(0)}
\tilde{U}^{(0)}_{3}(\zeta) + \hat\gamma_{3}^{(0)} 
\hat{U}^{(0)}_{3}(\zeta)\,,
\end{equation}
where $\tilde\gamma_{3}^{(0)}$ and
$\hat\gamma_{3}^{(0)} $
are constants, with 
\begin{align}
\tilde{U}^{(0)}_{3} (\zeta)
\hskip -0.1cm = \hskip -0.1cm
\zeta^{-\frac{d-4}{2(d-3)}},
\,
\hat{U}^{(0)}_{3}(\zeta)
\hskip -0.1cm = \hskip -0.1cm
\zeta^{-\frac{d-4}{2(d-3)}} 
\int^\zeta \hskip -0.15cm  d\zeta'
\zeta'^{\frac{d-4}{(d-3)}}\mathcal{W}(\zeta')\,,   
\end{align}
and the Wronskian being $\mathcal{W} \sim
\zeta^{-\frac{d-4}{d-3}}\left(1-\zeta^\frac{d-1}{d-3}\right)^{-1}$.
Since $\tilde{U}^{(0)}_{3}$ does not satisfy the Dirichlet boundary
condition at $\zeta = 0$ for any spacetime dimension, one needs to
impose $\tilde\gamma_{3}^{(0)} = 0$ in Eq.~\eqref{eq:solution_u0_vt}.
As before, 
the boundary condition at the horizon, described in these coordinates
as $\zeta=1$, imposes $\hat{\gamma}_2^{(0)}  = 0$ in
Eq.~\eqref{eq:solution_u0_vt} for all spacetime dimensions, since
$\hat{U}^{(0)}_3 \sim \ln(1-\zeta)$.
Thus,
there are no nontrivial perturbative solutions, and vector-type
Maxwell perturbations do not exhibit low pure damped modes.  Such
conclusions may be extended to the vector-type Proca perturbations by
considering $\hat{\mu}^2 = \hat{l}^2\mu^2 r_{\rm h}^2$ as a
perturbation, with
$\mathcal{O}(\hat{\mu}^2)=\mathcal{O}(\varepsilon)$.

\subsubsection{Comments}

A comment is in order regarding the results from the two
last sections.  In
Sec.~\ref{sec:isospectral}, it was proved that, for large black
holes, Maxwell scalar-type and vector-type modes tend to
isospectrality.  On the other hand, the study from Sec.~\ref{sec:siopsis}
reveals that there are low frequency modes for scalar-type
perturbations that do not exist for vector-type perturbations.  There
is an apparent contradiction between the two results, but this is only
due to the degree of the approximation in both results.
Note, that Eq.~\eqref{eq:isospectral0} only holds when one
neglects the angular momentum number $\ell$ and one
considers the large black hole approximation, i.e., 
$f(r) \sim \frac{r^2}{l^2}-\frac{r_{\rm h}^2}{l^2}
(\frac{r_{\rm h}}{r})^{d-3}$. If $\ell$ is restored
and the full expression for $f(r) $ is considered,
then there is no $W$ that satisfies
Eq.~\eqref{eq:isospectral0}, and isospectrality cannot be
established. Furthermore, the low pure damped modes scale with the
inverse of the event horizon radius and so it is expected that for
infinite large black holes these modes tend to a zero frequency. 
In turn, note from
Eq.~\eqref{eq:siopsis_final}, that if one neglects the angular
momentum number $\ell$, there are no low frequency scalar-type
modes.
Thus,
the results
from Sec.~\ref{sec:isospectral}
and Sec.~\ref{sec:siopsis}
are consistent. In fact, we checked numerically that such low
frequency modes do not exist for the Schr\"{o}dinger-like equation
with potential Eq.~\eqref{eq:scalar_potential_large} and
$f(r) \sim \frac{r^2}{l^2}-\frac{r_{\rm h}^2}{l^2}
(\frac{r_{\rm h}}{r})^{d-3}$, see below
Section~\ref{sec:num_results}.

%%%%%%%%%%%%%%%%%%%%%%%%%%%%%%%%%%%%%%%%%%%%%%%%%%%%%%%%%%%%%%%%%%%%%%
\section{Numerical integration methods for the computation of the 
quasinormal mode frequencies
of Proca and Maxwell fields in Schwarzschild-AdS}
\label{sec:numerics_integration}
%%%%%%%%%%%%%%%%%%%%%%%%%%%%%%%%%%%%%%%%%%%%%%%%%%%%%%%%%%%%%%%%%%%%%%

\subsection{Numerical integration methods for the Proca field:
Decoupled and coupled Schr\"odinger-like equations}

\subsubsection{The shooting method}

\noindent {\it (a) Scalar-type $\ell=0$ monopole Proca
field and vector-type Proca field: Decoupled Schr\"odinger-like equation}

\vskip 0.2cm

\noindent
The numerical integration method to obtain
the QNMs for the Proca field
in $d$-dimensions
can be
based on integrating the Schr\"{o}dinger-like equation using
the shooting method
\cite{pani2013,chandra_detweiler1975,konoplya_stability2008,
phd_zhidenko2009,dolan2012}.
One starts from a point near the horizon,
with radius $r_i=r_{\rm h}
(1+\epsilon)$, $\epsilon \ll 1$, where it is imposed the boundary
condition at the horizon, Eq.~\eqref{eq:bc_horizon}.
The method works 
for Proca monopole modes, Proca vector-type perturbation modes,
and Maxwell modes. We explain now in detail
the method applied for Proca vector-type perturbation modes
which is the more involved case, the other
cases follow with ease.

The expansion of the Proca function $u_3$
evaluated at $r_i$ is
given by
\begin{align}\label{eq:numerical_bc_horizon}
&u_3 (r_i) = (r_i-r_{\rm h})^{
-\frac{i\omega}{f'(r_{\rm h})}}\sum_{j=0}^{N_i}
a_{(3)j} (r_i-r_{\rm h})^j \notag\,,\\
&\partial_r u_3 (r_i)=
(r_i-r_{\rm h})^{-\frac{i\omega}{f'(r_{\rm h})}-1} \notag\\
&\times\sum_{j=0}^{N_i} \left( j a_{(3) j} - \frac{i\omega}
{f'(r_{\rm h})}a_{(3)j}\right)(r_i-r_{\rm h})^{j}\,,
\end{align}
where $N_i$ is the order at which the expansion is truncated, and
$a_{(3)j}$ are coefficients of the expansion in $r$ near the
horizon. The coefficients ${a}_{(3)j}$ are found by substituting
$u_3(r)=(r-r_{\rm h})^{-\frac{i\omega}{f'(r_{\rm h})}}
\sum_{j=0}^{N_i} a_{(3)j} (r-r_{\rm h})^j$ in Eq.~\eqref{eq:sch_u3}
and equating the coefficients corresponding to each power term
$(r-r_{\rm h})$, in the first $N_i$ expansion terms.
This expansion of the Proca function $u_3$
evaluated at $r_i$ as given in Eq.~\eqref{eq:numerical_bc_horizon}
is taken as the initial condition for the initial
value problem.
At a large
radius $r_f = \kappa r_{\rm h}$, with $\kappa \gg 1$, one also expands
the field using the asymptotic behavior that obeys the boundary
conditions that
we impose, in this case the Dirichlet boundary
condition. The expansion of the function $u_3$, now evaluated at
$r_f$, is given by
\begin{align}\label{eq:numerical_bc_far}
&u_{3} (r_f) = u_3^{r\rightarrow \infty}(r_f)
\sum_{j=0}^{N_f} b_{(3) j} r_f^{-j} \notag\,,\\
&\partial_r u_{3} (r_f)= \frac{\partial_r u_3^{r\rightarrow \infty}(r_f)}
{u_3^{r \rightarrow \infty}(r_f)}u_{3} (r_f) \notag\\
& - u_3^{r\rightarrow \infty}(r_f)\sum_{j=0}^{N_f}
j  b_{(3) j} {r_f}^{-j-1}\,, 
\end{align}
where $u_3^{r\rightarrow \infty}(r) =
r^{-\frac{1}{2}-\frac{1}{2}\sqrt{(d-3)^2 + 4\mu^2 l^2}}$ and $b_{(3)
j}$ are the expansion coefficients at very large radius.  Again, the
coefficients $b_{(3)j}$ can be determined in terms of $b_{(3)0}$ by
inserting the series $u_3 (r) = u_3^{r\rightarrow \infty}(r)
\sum_{j=0}^{N_f} b_{(3) j} r^{-j}$ into Eq.~\eqref{eq:sch_u3} and
equating the coefficients corresponding to each power term $\frac1r$, in
the first $N_f$ expansion terms.
This expansion of the Proca function $u_3$
evaluated at $r_f$ as given in Eq.~\eqref{eq:numerical_bc_far}
is taken as  the initial condition for a distinct initial value
problem.

One now chooses, without loss of generality, $a_{(3)0} = 1$ and
$b_{(3)0} = 1$.  The method now consists in performing two
integrations, through numerical methods, of Eq.~\eqref{eq:sch_u3}. One
integration is carried from $r=r_i$ up to the middle point $r=r_m$,
i.e., $r_i < r_m < r_f$, using the initial values of
Eq.~\eqref{eq:numerical_bc_horizon}. In turn, the other integration is
carried from $r=r_f$ down to $r=r_m$, using the initial values of
Eq.~\eqref{eq:numerical_bc_far}.  Since the two solutions, say $u_{3
r_i} (r)$ and $u_{3 r_f} (r)$, must be linearly dependent at
$r=r_m$, the problem amounts to finding the root of the Wronskian
\begin{align}\label{eq:wronskianu3}
\mathcal{W}(u_{3 r_i},u_{3 r_f};r_m) =
(u_{3 r_i} u'_{3 r_f} 
- u_{3 r_f} u'_{3 r_i})\big\vert_{r=r_m} = 0 \,,
\end{align}
which must be satisfied for the QNM frequency  $\omega$.
We note that the Wronskian is independent of $r$, so that the choice
of $r=r_m$ does not affect the results.

We can apply this method for
the monopole scalar-type Proca field $u_1$ with $u_2=0$ and
with potential $V_{s11}$ in Eq.~\eqref{eq:matrixpotential},
for the vector-type Proca field $u_3$ as just explained,
for 
the scalar-type Maxwell field
$u_{12}$ with potential given
in Eq.~\eqref{eq:scalarpotentialmaxwell_sads},
and for the vector-type Maxwell field $u_3$
in
Eq.~\eqref{eq:vectorpotential_sadsm}.
The
expansions near the horizon are similar, while for the expansions in the
far radius $r_f$, one must use the asymptotic behavior obeying the
boundary conditions for each field.

\vskip 0.4cm

\noindent {\it (b)
Scalar-type $\ell\geq 1$ Proca field: Coupled equations}

\vskip 0.2cm

\noindent
For the coupled modes, i.e., $\ell \geq 1$ scalar-type Proca field
modes in $d$-dimensions, one must extend the shooting
method procedure dimensions according to
\cite{pani2013,dolan2012}.
We perform the expansion of the functions $u_1$ and
$u_2$ near the horizon at $r_i = r_{\rm h}(1 + \epsilon)$, with
$\epsilon \ll 1$, as
\begin{align}\label{eq:numerical_bc_horizonu1u2}
&u_k (r_i) = (r_i-r_{\rm h})^{-\frac{i\omega}{f'(r_{\rm h})}}
\sum_{j=0}^{N_i} 
a_{(k)j} (r_i-r_{\rm h})^j \,,\notag\\
&\partial_r u_k (r_i)=(r_i-r_{\rm h})^{
-\frac{i\omega}{f'(r_{\rm h})}-1} \notag\\
&\times\sum_{j=0}^{N_i} \left( j a_{(k) j} -
\frac{i\omega}{f'(r_{\rm h})}
a_{(k)j}\right)(r_i-r_{\rm h})^{j}\,,
\end{align}
for $k=\{1,2\}$ in this case, with $N_i$ being the truncation order,
where $a_{(k)j}$ are the expansion coefficients near the horizon.  By
putting the expansion in the system of coupled equations
Eq.~\eqref{eq:sch_coupled}, one determines the coefficients of the
expansion $a_{(k')j}$ in function of $a_{(k)0}$. One also performs the
expansion of the functions $u_1$ and $u_2$ at a large radius $r_f =
\kappa r_{\rm h}$, where $\kappa \gg 1$, as
\begin{align}
\label{eq:numerical_bc_faru1u2}
&u_{k} (r_f) = u_k^{r\rightarrow \infty}(r_f)
\sum_{j=0}^{N_f} b_{(k) j} {r_f}^{-j} \notag\,,\\
&\partial_r u_{k} (r_f)= \frac{\partial_r u_k^{r\rightarrow \infty}(r_f)}
{u_k^{r\rightarrow \infty}(r_f)}u_{k} (r_f) \notag\\
& - u_k^{r\rightarrow \infty}(r_f)\sum_{j=0}^{N_f}
j  b_{(k) j} {r_f}^{-j-1}\,, 
\end{align}
where $u_k^{r\rightarrow \infty}(r) =
r^{-\frac{1}{2}-\frac{1}{2}\sqrt{(d-3)^2 + 4\mu^2 l^2}}$ and $b_{(k)
j}$ are the expansion coefficients at very large radius. One then uses
Eq.~\eqref{eq:sch_coupled} to determine the $b_{(k') j}$ in function
of $b_{(k) 0}$, by considering the first $N_f$ terms of the expansion
in $\frac1r$.

The expansions in Eqs.~\eqref{eq:numerical_bc_horizonu1u2}
and~\eqref{eq:numerical_bc_faru1u2} are treated as initial values for
the integration of the coupled system in Eq.~\eqref{eq:sch_coupled},
as in the decoupled case. Since the coupled system in
Eq.~\eqref{eq:sch_coupled} is linear, the integration with initial
values given in Eq.~\eqref{eq:numerical_bc_horizonu1u2} yields a
linear combination of independent functions with coefficients
$a_{(k)0}$. We thus perform two integrations, from $r_i$ to $r_m$, for
the pairs $(a_{(1)0},a_{(2)0}) = (1,0)$ and $(a_{(1)0},a_{(2)0}) =
(0,1)$, yielding the functions $(u_{1 r_i})_{a_{(1)0}}$ and $(u_{2
r_i})_{a_{(1)0}}$ for the first pair, and $(u_{1 r_i})_{a_{(2)0}}$ and
$(u_{2 r_i})_{a_{(2)0}}$ for the second pair.  Similarly, the
integration of Eq.~\eqref{eq:sch_coupled} with initial conditions
Eq.~\eqref{eq:numerical_bc_faru1u2} gives a linear combination of
independent functions with coefficients $b_{(k)0}$, and we perform two
integrations, from $r_f$ to $r_m$, for the pairs $(b_{(1)0},b_{(2)0})
= (1,0)$ and $(b_{(1)0},b_{(2)0}) = (0,1)$, with solutions $(u_{1
r_f})_{b_{(1)0}}$ and $(u_{2 r_f})_{b_{(1)0}}$ for the first pair, and
$(u_{1 r_f})_{b_{(2)0}}$ and $(u_{2 r_f})_{b_{(2)0}}$ for the second
pair.  Linear dependence of $u_1$ and $u_2$ can be encapsulated in the
vanishing of the Wronskian of the system at $r_m$
\begin{align}\label{eq:wronskianu1u2}
&\mathcal{W}(\bm{u}_{r_i}, \bm{u}_{r_f};r_m) = \notag \\
&\det\begin{pmatrix}
(u_{1 r_i})_{a_{(1)0}} &
(u_{1 r_i})_{a_{(2)0}} & (u_{1 r_f})_{b_{(1)0}} 
& (u_{1 r_f})_{b_{(2)0}} \\
(u_{2 r_i})_{a_{(1)0}} & (u_{2
r_i})_{a_{(2)0}} & (u_{2 r_f})_{b_{(1)0}} 
& (u_{2 r_f})_{b_{(2)0}} \\
(u'_{1 r_i})_{a_{(1)0}} & (u'_{1
r_i})_{a_{(2)0}} & (u'_{1 r_f})_{b_{(1)0}} 
& (u'_{1 r_f})_{b_{(2)0}} \\
(u'_{2 r_i})_{a_{(1)0}} & (u'_{2
r_i})_{a_{(2)0}} & (u'_{2 r_f})_{b_{(1)0}} 
& (u'_{2 r_f})_{b_{(2)0}} 
\end{pmatrix}\notag\\
&= 0\,,
\end{align}
where $\bm{u}_{r_i}$ and $\bm{u}_{r_f}$ are the vectors of
functions resulting from the integration of the coupled system with
initial values at $r_i$ and $r_f$, respectively, and the determinant
is evaluated at $r=r_m$. The vanishing of the Wronskian in
Eq.~\eqref{eq:wronskianu1u2} yields the QNM frequencies
of the scalar-type Proca field. We note that the method of finding the
zero of the Wronskian in the case of AdS has great accuracy due to the
asymptotic decay of the regular solutions.

\subsubsection{The Horowitz-Hubeny method}

One can also apply the Horowitz–Hubeny numerical method to the Proca
equations in $d$-dimensions. This method is well suited for determining
QNMs in asymptotically AdS spacetimes and has become one
of the standard approaches for studying perturbations in such
backgrounds. Although the method has been clearly described in several
works in the literature, for completeness we provide a brief review of
its implementation in Appendix \ref{sec:app_horowitz_hubeny}.
We have compared the results of
the two methods we use.

\subsection{Numerical integration methods for the Maxwell
field: Decoupled equations}

The numerical integration shooting
method to obtain the QNMs for
the Maxwell field in $d$ dimensions
is based on integrating the Schr\"{o}dinger-like
equation also starting from a point near the horizon, with radius
$r_i=r_{\rm h} (1+\epsilon)$, $\epsilon \ll 1$.  Since the boundary
condition at the horizon for Maxwell field perturbations is the same
as for Proca field perturbations, i.e., it is given by
Eq.~\eqref{eq:bc_horizonMaxwell}, the method for decoupled equations
apply straightforwardly here.  One expands $u_{12}$ and $u_3$ of
Eqs.~\eqref{eq:sch_u12} and \eqref{eq:sch_u3m}, respectively in the
same way as we did for the Proca field in
Eqs.~\eqref{eq:numerical_bc_horizon} and \eqref{eq:numerical_bc_far}
and numerically find the functions.

For the Maxwell field in $d$ dimensions one can
also apply the
Horowitz-Hubeny method
and 
compare it as we did with the shooting method specified above.

%%%%%%%%%%%%%%%%%%%%%%%%%%%%%%%%%%%%%%%%%%%%%%%%%%%%%%%%%%%%%%%%%%%%%%%
\section{Numerical analysis of the Proca and Maxwell quasinormal modes}
\label{sec:num_results}
%%%%%%%%%%%%%%%%%%%%%%%%%%%%%%%%%%%%%%%%%%%%%%%%%%%%%%%%%%%%%%%%%%%%%%%

\subsection{Numerical parameters}

We now proceed with the presentation and the analysis of the numerical
results for the QNMs.
The numerical results of this section were obtained using the
integration methods outlined in
Sec.~\ref{sec:numerics_integration}. The shooting method has proven to be
highly reliable for computing the QNMs, as the boundary
conditions imposed require one to minimize the dominating piece of the
solution at spatial infinity. When feasible, the results were
cross-checked with the Horowitz-Hubeny method, see
also Sec.~\ref{sec:numerics_integration} and
Appendix~\ref{sec:app_horowitz_hubeny}. However, the Horowitz-Hubeny
method suffers from some subtleties.  Namely, for sufficiently small
black holes, typically $\frac{r_{\rm h}}{l} \leq 0.5$, the method's
convergence properties worsen, leading to increased computing time and
numerical error contamination.  Moreover, the Frobenius series,
see Eq.~\eqref{eq:exp_U}, is only applicable if the region of interest, $x
\in (0,x_h]$, lies within the series' radius of convergence. While it
can be shown that this condition is satisfied for large black holes
and $d \leq 7$, such cannot be assured for smaller black holes or higher 
dimensions.

For the numerical integration, we have used $N_i = N_f=6$ for the
order of the expansions near the horizon and in the far region. We
have fixed the near-horizon radius to be $r_i = 1.01 r_{\rm h}$, i.e.,
$\epsilon = 0.01$, and the far region radius $r_f = 10^{11} r_{\rm
h}$, i.e., $\kappa=10^{11}$.  The intermediate radius at which we
minimize the Wronskian was set at $r_m = 0.67 r_f$.  A reasonable
variation in these values gives the same results within our precision.
The numerical tool used to integrate the differential equations was
\textit{NDSolve} and the numerical tool used to find the root of the
Wronskian was \textit{FindRoot} from \textit{Mathematica}.

\subsection{General characterization of the quasinormal
mode frequency spectrum}

We establish the parameter space for the QNMs and conduct
a detailed analysis of their properties across various regions within
this space.  In Schwarzschild-AdS spacetime, there are two different
length scales: the radius of the black hole, $r_{\rm h}$, and the
radius of curvature of AdS, $l$.  The Schr\"{o}dinger-like equations
are invariant under a rescaling on $r$, $\mu$, and $\omega$, which
corresponds to choosing the units of the physical quantities. In
particular, one can perform the rescaling  $r
\rightarrow \frac{r}{l}$, $\omega \rightarrow \omega l$, $\mu
\rightarrow \mu l$, so that the equations are independent of
$l$. Thus, without loss of generality, in the numerical results we fix
$l=1$.

In asymptotically AdS spacetimes, the qualitative behavior of the
QNM spectrum highly depends on the radius of the black
hole  \cite{cardoso_lemos2001,cardoso_lemos2001cylindrical,
cardoso_lemos_konoplya2003}, so it is useful to
do a separate study for three different types of black holes,
according to their radii: large black holes, with $\frac{r_{\rm h}}{l}
\gg 1$, intermediate black holes, with $\frac{r_{\rm h}}{l} \simeq 1$,
and small black holes, with $\frac{r_{\rm h}}{l} \ll 1$. In addition,
the QNM spectrum depends on the dimension $d$ of the
spacetime as well as on the perturbation-related parameters: the type
of perturbation, either scalar-type and vector-type, the mass of the
field, $\mu$, the angular momentum number, $\ell$, and the overtone
number, $k$.  Our analysis focuses on $4,5,6,7$-dimensional
Schwarzschild-AdS spacetimes, generalization to other higher
dimensions can be made.

\begin{figure*}[t]
\centering
\includegraphics[width=.65\textwidth]{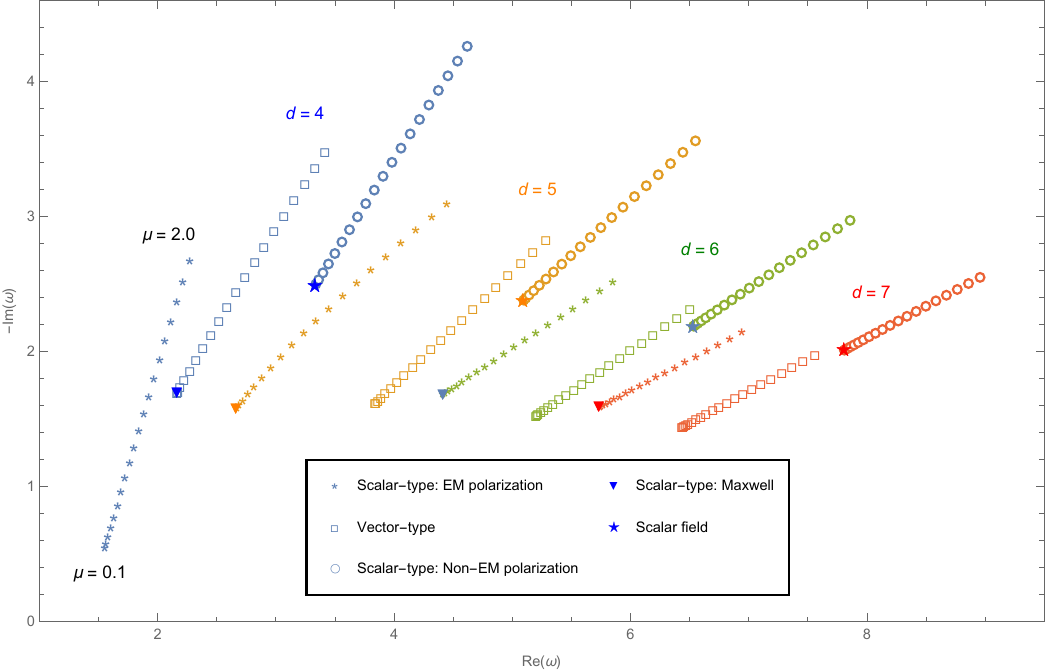}
\caption{Fundamental QNM frequencies of $\ell=1$
scalar-type with electromagnetic polarization
(asterisks),
vector-type (squares), and scalar-type with nonelectromagnetic polarization
(circles), Proca perturbations with field masses $\mu =0.1, 0.2,
...,2.0$, in a $r_{\rm h}=1$ Schwarzschild-AdS black hole with $d=4$
(blue), $d=5$ (orange), $d=6$ (green), and $d=7$ (red), spacetime
dimensions.  Scalar-type Maxwell (triangles) and scalar field (stars)
QNM frequencies are also plotted.}
\label{fig:graph_mass}
\end{figure*}

In order to interpret correctly the spectrum for scalar-type Proca
perturbations, we make the distinction between the two modes described
by the coupled system Eq.~\eqref{eq:sch_coupled}.
For that, we  define 
two polarizations.
One
is the nonelectromagnetic polarization, which contains the
physical degree of freedom of the monopole mode of the Proca field and
reduces to the pure-gauge degree of freedom in the electromagnetic
limit. Indeed, Eq.~\eqref{eq:rw2} allows us to interpret the
nonelectromagnetic polarization as a scalar field polarization.
The other is the electromagnetic polarization, which
corresponds to the physical degree of freedom in the electromagnetic
limit.
Assuming that the QNMs change smoothly when the field
acquires mass, it is expected that the
nonelectromagnetic 
and 
electromagnetic polarization modes approach those of
scalar field perturbations
and scalar-type Maxwell perturbations, respectively, in
the small-mass limit with a fixed $r_{\rm h}$. However, such behavior
does not happen in the case of $d=4$, as we shall see below.
We note a
difference in the nomenclature
``scalar-type'' and ``vector-type'' adopted
in \cite{dolan2012,tiago_sads2022} with the one used here.  In
\cite{dolan2012,tiago_sads2022}, the ``scalar-type'' and ``vector-type''
polarizations were motivated by their mode behavior, i.e.,
``scalar-type'' behaves as a scalar field and ``vector-type'' behaves as a
vector field, and correspond to our scalar-type nonelectromagnetic
polarization and vector-type/scalar-type electromagnetic
polarizations, respectively. However, the nomenclature adopted in
\cite{dolan2012,tiago_sads2022} to distinguish between the two polarizations
only makes sense in $d=4$, where the scalar-type and vector-type
Maxwell modes are exactly isospectral.  In higher-dimensional
spacetimes, isospectrality no longer happens, making the aforementioned
nomenclature unsuitable.

We present in Fig.~\ref{fig:graph_mass}
some of 
the numerical results we have worked out,
specifically,
the fundamental QNM frequencies of $\ell=1$
 scalar-type with electromagnetic polarization
(asterisks),
vector-type (squares), and scalar-type with nonelectromagnetic polarization
(circles), Proca perturbations with field masses $\mu =0.1, 0.2,
...,2.0$, in a $r_{\rm h}=1$ Schwarzschild-AdS black hole with $d=4$
(blue), $d=5$ (orange), $d=6$ (green), and $d=7$ (red), spacetime
dimensions.  Scalar-type Maxwell (triangles)
QNM frequencies are also shown.
To be complete
scalar field (stars)
QNM frequencies are also plotted.
The scalar fields $\phi$
obey the scalar equation $\mathcal{D}^{s=0}_\ell\phi=0$,
where $\mathcal{D}^{s=0}_\ell$ is the operator
defined in Eq.~\eqref{eq:operator_scalarfield}.

To complete these results we
present the numerical results of the fundamental QNM
frequencies in four tables in the Appendix~\ref{sec:app_tables}
for the typical range of
the parameters in the literature, namely for $4,5,6,7$-dimensional
Schwarzschild-AdS black holes with radius $r_{\rm h}\in [0.05,1,100]$
and Proca mass $\mu \in [0,0.1,0.2,0.3,0.4,0.5]$.  Generically, the
QNM frequencies increase with the dimension and with the
Proca mass in these ranges.

%\newpage
%\centerline{}
%\newpage

\subsection{Analysis of the dependence of the quasinormal modes on
the Proca field mass}

We now give a specific analysis of the dependence of the QNM
frequencies on the Proca field mass.  In Fig.~\ref{fig:graph_mass} it
is shown the effect of the mass of the field on the $\ell=1$ QNMs for
$4,5,6,7$-dimensional Schwarzschild-AdS black holes with size $r_{\rm
h}=1$.  For all spacetime dimensions, both real and imaginary parts of
the frequencies increase in magnitude with increasing mass, in
agreement with what was found for $d=4$ in
\cite{konoplya_massive2006,tiago_sads2022}.  The increase of the real
part is related to the decrease of the Compton wavelength of the Proca
particle, $\lambda_{\rm c} = \frac1\mu$. On the other hand, the
increase of the imaginary part is a consequence of the barrier
becoming less pronounced with increasing mass. Note that such effect
on the QNMs is suppressed for higher-dimensional spacetimes, as these
already provide a mass term to the potential, i.e.,
$\mu^2_{\mathrm{eff}} = \mu^2+\frac{(d-2)(d-4)}{4l^2}$.

For $d \geq 5$, the scalar-type Proca modes with electromagnetic
polarization reach, in the massless limit, the scalar-type Maxwell
modes.  Such does not happen, however, for $d=4$: in this case, the
scalar-type and vector-type sectors are isospectral, and the
transition from massless to massive regimes is not smooth. Indeed, for
$\mu = 0$, we have $\mu_{\mathrm{eff}}=0$, and the potential does not
diverge at infinity, whereas for $\mu > 0$, we have
$\mu_{\mathrm{eff}}>0$, and the potential diverges. This discontinuity
is visible in Fig.~\ref{fig:graph_mass} already for $\mu=0.1$. The
analysis is not done for smaller mass, because the method with the
coupled system gives results dependent on the initialization of
\textit{FindRoot}, which indicates a failure of accuracy of the
method.

Lastly, note from the figure
that scalar-type Proca modes with nonelectromagnetic
polarization reach the scalar field modes, in the massless limit
and for all spacetime dimensions, in agreement with
Eq.~\eqref{eq:rw2}.

\subsection{Analysis of the dependence of the quasinormal
modes on the black hole radius}

\subsubsection{Large black holes}

\centerline{\it (a) Ordinary modes of large black holes }
\vskip 0.2cm

We now
analyze the dependence of the
QNMs  on large black hole horizon radius
for ordinary modes.
As shown in \cite{hubeny2000}, the
ordinary modes are QNMs in which their frequencies for
large Schwarzschild-AdS black holes should scale linearly with the
radius of the black hole, i.e., $\omega \sim r_{\rm h}$.
The results for scalar-type Proca perturbations and 
for vector-type Proca
perturbations are similar.

\begin{figure}[h]
\centering
\includegraphics[width=.48\textwidth]{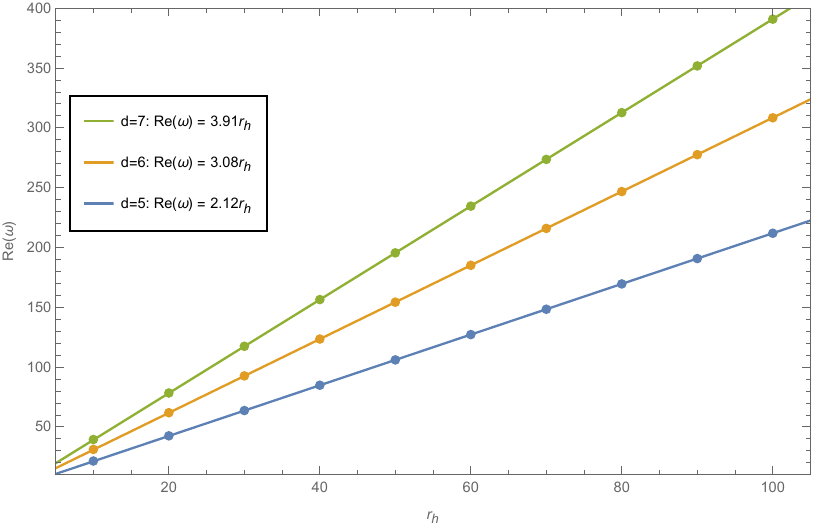}
\caption{Real part of the QNM frequencies of $\ell=1$,
$\mu=0.5$, vector-type Proca perturbations in $5,6,7$-dimensional
Schwarzschild-AdS black holes, as a function of the black hole radius,
in the large black hole regime.  Ordinary modes scale linearly with
the radius of the black hole.}
\label{fig:large_BH}
\end{figure}

In Fig.~\ref{fig:large_BH} the results for the real part of the
QNM frequencies of $\ell=1$ vector-type Proca
perturbations in $5,6,7$-dimensional Schwarzschild-AdS black holes, as
a function of the black hole radius, in the large black hole regime,
are displayed.  The scaling of the frequency $\omega$ with the horizon
radius $r_{\rm h}$ appears in a clear way.

In Tables~\ref{tab:large_BH_Proca} and~\ref{tab:large_BH_Maxwell} we
show the fundamental QNM frequencies of $\ell=1$
scalar-type and vector-type Proca perturbations and $\ell=1$
scalar-type and vector-type Maxwell perturbation in $d=7$ large
Schwarzschild-AdS black holes, respectively.
For Proca perturbations, Table~\ref{tab:large_BH_Proca} displays
isospectrality between the electromagnetic polarized scalar-type modes
and the vector-type modes in the limit of very large $r_{\rm h}$,
although we have not found an analytical proof for this statement.
For Maxwell perturbations, the isospectrality between the two sectors
is clearly displayed in Table \ref{tab:large_BH_Maxwell} in the limit of
very large $r_{\rm h}$, in agreement with the analytical proof of
Section~\ref{sec:isospectral}.
\begin{table}[h]
\centering
\resizebox{0.48\textwidth}{!}{\begin{tblr}{|c|c|c|c|c|}
$r_{\rm h}$ & $\omega$ (Scalar-type, NonEM) & $\omega$ (Scalar-type, EM) & $\omega$ (Vector-type)   \\\hline
$500$ &   $2638.34 - 1390.90i$              &       $2054.37 - 1083.43i$             &      $ 2054.37 - 1083.43i$         \\
$250$ &    $1319.18 - 695.448i$                   &  $1027.19 - 541.716i$               &    $1027.20 - 541.715i$          \\
$100$ &        $527.702 - 278.173i$                  & $410.897 - 216.682i $               &   $410.906 - 216.680i$        \\
$50$ &        $263.902 - 139.076i$               &   $ 205.484 - 108.334 i $        &            $205.502 - 108.330i$   \\
$25$ &         $132.054 - 69.5159i$                &   $102.813 - 54.1533i$           &       $102.849 - 54.1441i$     \\
$10$ &        $53.1090 - 27.7452i$               &  $ 41.3223 - 21.6225i$        &      $ 41.4135 - 21.5997i$          \\
\end{tblr}}
\caption{Fundamental QNMs of $\mu=1$ Proca $\ell=1$
perturbations in $d=7$ large Schwarzschild-AdS black holes.  For large
black holes, scalar-type nonelectromagnetically polarized stand apart,
whereas scalar-type electromagnetically polarized perturbations  
and 
vector-type Proca perturbations tend to isospectrality.}
\label{tab:large_BH_Proca}
\end{table}
\begin{table}[h]
\centering
 \resizebox{0.34\textwidth}{!}{\begin{tblr}{|c|c|c|c|}
$r_{\rm h}$ & $\omega$ (Scalar-type) &  $\omega$ (Vector-type)   \\\hline
$500$ &    $1918.30 - 999.501i$                         &             $1918.30 - 999.501i$        \\
$250$ &    $ 959.155 - 499.749i$                  &              $959.159 - 499.748i$             \\
$100$ &       $383.681 - 199.896i $                     &       $383.691 - 199.894i$            \\
$50$ &        $191.875 - 99.9413i $                &              $191.894 - 99.9367i$    \\
$25$ &          $96.0056 - 49.9572i$              &              $96.0447 - 49.9481i$    \\
$10$ &       $38.5930 - 19.9454i$                 &          $38.6901 - 19.9228i$      \\
\end{tblr}}
\caption{Fundamental QNMs of Maxwell $\ell=1$
perturbations in large $d=7$ Schwarzschild-AdS black holes. For
sufficiently large black holes, the scalar-type and vector-type
sectors of Maxwell perturbations are isospectral, in agreement with
the analytical study of Section~\ref{sec:isospectral}. }
\label{tab:large_BH_Maxwell}
\end{table}

%\newpage
%\centerline{}
%\newpage

%\centerline{}

\newpage

\vskip 0.55cm
\centerline{\it (b) Purely damped modes of large black holes }
\vskip 0.2cm

\begin{figure}[h]
\centering
\includegraphics[width=.48\textwidth]{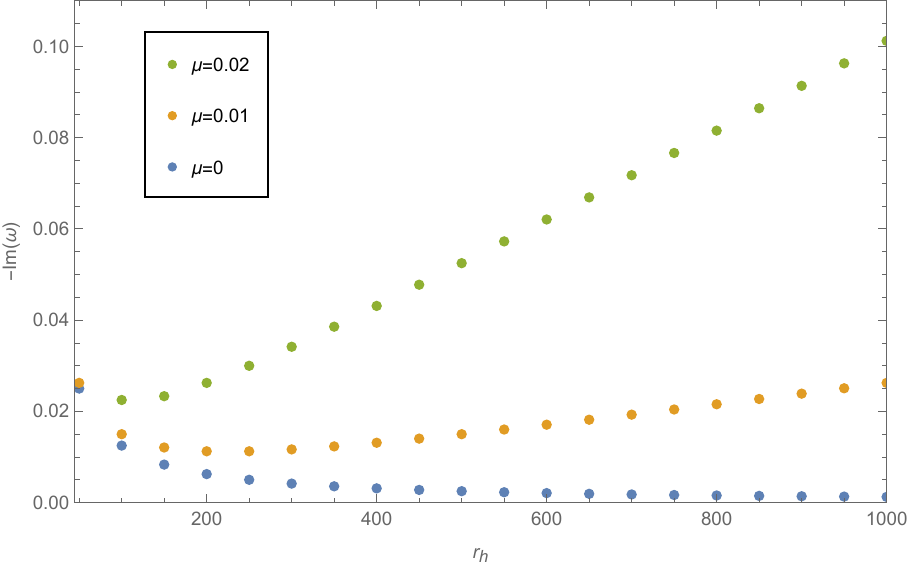}
\caption{Purely damped modes of
$\mu = 0.01$ Proca
(orange), and $\mu=0.02$ Proca (green), $\ell=1$ scalar-type
perturbations in $7$-dimensional large Schwarzschild-AdS black holes
are plotted 
as a function of the black hole radius. Purely damped modes of Proca
perturbations scale linearly with the black hole radius.
Purely damped modes of
Maxwell (blue) scalar-type
perturbations in $7$-dimensional large Schwarzschild-AdS black holes
are plotted
as a function of the black hole radius. Purely damped modes of
Maxwell perturbations scale linearly with the inverse of the
radius, in contrast
with Proca  modes.}
\label{fig:siopsis1}
\end{figure}

In addition to the ordinary modes, we have analyzed
numerically the purely damped
modes of Proca perturbations.  In Fig.~\ref{fig:siopsis1}, we present
the results for Proca scalar perturbations in $7$-dimensional large
Schwarzschild-AdS black holes, as a function of the black hole
radius. Purely damped modes of Proca perturbations scale linearly with
the black hole radius as shown.  We have not pursued an analytic study
of Proca perturbations, as this would require solving
the coupled system given in Eq.~\eqref{eq:sch_coupled}.  Thus, further
investigation is needed, but our numerical results indicate that, when
the field acquires mass, the low-frequency modes start to scale
linearly with the black hole radius as indicated in the figure.

\begin{figure}[h]
\centering
\includegraphics[width=.48\textwidth,height=.33\textwidth]
{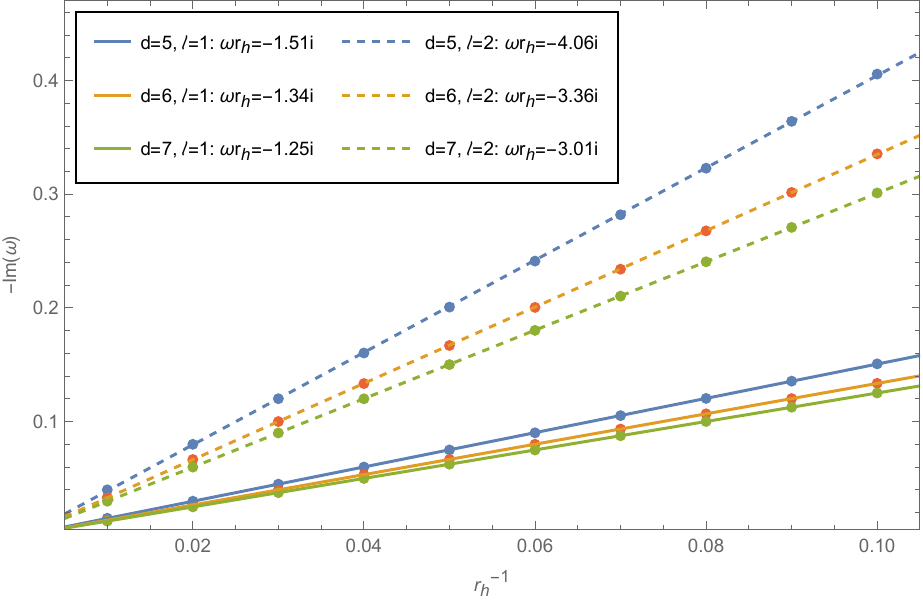}
\caption{Purely damped modes of
scalar-type Maxwell $\ell=1$ (solid
lines) and $\ell=2$ (dashed lines) perturbations in $5$-dimensional
(blue), $6$-dimensional (orange), and $7$-dimensional (green), large
Schwarzschild-AdS black holes, as a function of the inverse of the
black hole radius. The numerical results show support for
Eq.~\eqref{eq:siopsis_final}. Note the linear scaling
with $r_{\rm h}^{-1}$.}
\label{fig:siopsis}
\end{figure}

In addition to the ordinary modes, we have as well analyzed
numerically the purely damped
modes of Maxwell  perturbations.
 In Fig.~\ref{fig:siopsis}, we present
the results for  $\ell=1,2$ Maxwell perturbations in
 $d=5,6,7$ dimensions.
The numerical results confirm our analytic procedure 
of Sec.~\ref{sec:siopsis}
that
scalar-type Maxwell perturbations in  $d \geq 5$
large Schwarzschild-AdS black holes exhibit purely damped
QNM frequencies scaling linearly with the inverse of the
radius.
Furthermore, no purely damped
modes were found for scalar-type perturbations in $d=4$
Schwarzschild-AdS black holes, neither for vector-type perturbations
in all spacetime dimensions, once again confirming the results of
Sec.~\ref{sec:siopsis}.

\subsubsection{Intermediate black holes}

\begin{figure}[h]
\centering
\includegraphics[width=.48\textwidth]{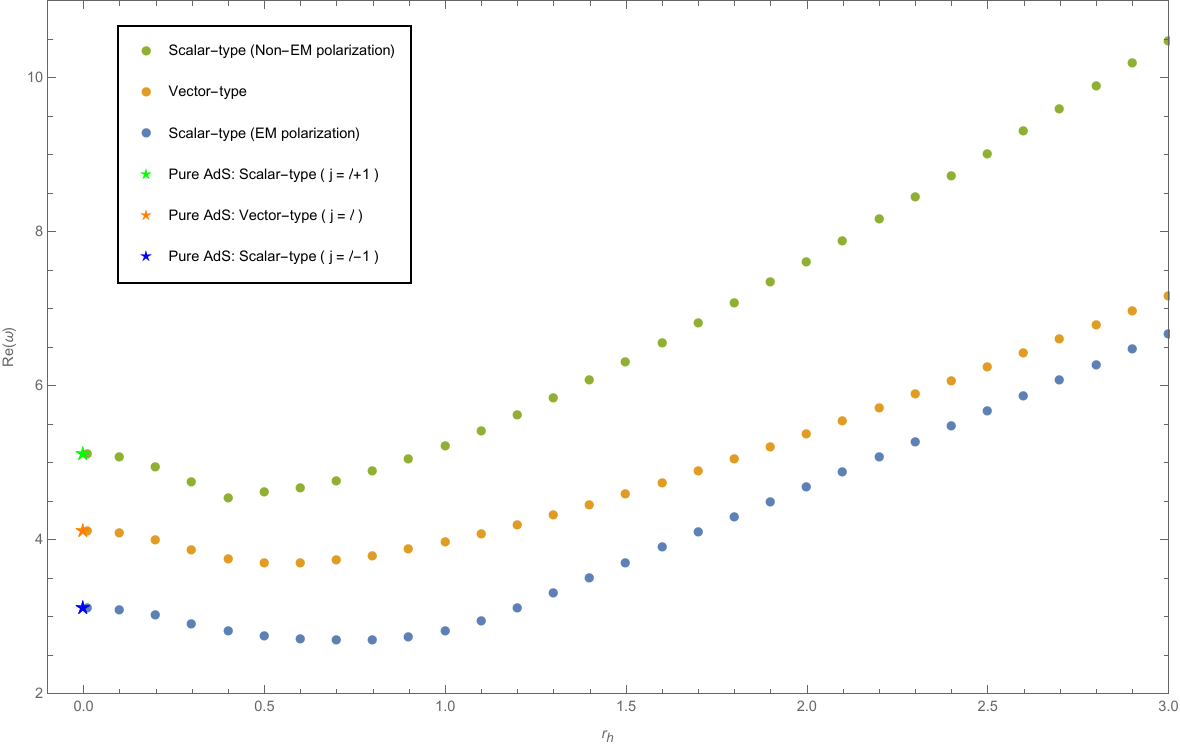}
\caption{Real part of the fundamental QNM 
frequencies of $\ell=1$, $\mu = 0.5$, scalar-type with 
electromagnetic polarization (blue), vector-type (orange), 
and scalar-type with nonelectromagnetic polarization (green) 
Proca perturbations in $5$-dimensional intermediate and 
small Schwarzschild-AdS black holes
plotted as a function of the 
black hole radius. The pure AdS normal mode frequencies 
(stars) were also plotted \cite{lopes2024}.}
\label{fig:small_BH}
\end{figure}

We now analyze the case of intermediate black holes.  As we
decrease the horizon radius from the large black hole case, the
QNMs start to deviate from the linear scaling as
one approaches $r_{\rm h} \simeq 1$. This is shown in
Fig.~\ref{fig:small_BH}, where it can be seen that the real part of
the frequencies approaches a minimum for $r_{\rm h}<1$, meaning that
there are differently-sized black holes having QNM
frequencies with the same real part.
This may be due to the fact that
the black hole starts to have a comparable size with the Compton
wavelength of the field, $\lambda_{\rm c} = \frac{1}{\omega}$.

%\newpage

\subsubsection{Small black holes}

We now analyze the case of small black holes.
As one approaches the $r_{\rm h} \rightarrow 0$ limit, the
QNM frequencies approach those of pure AdS spacetime, see
Fig.~\ref{fig:small_BH}, studied in \cite{lopes2024}. Scalar-type Proca
modes with
nonelectromagnetic
and 
electromagnetic  polarizations
approach
the scalar-type $j=\ell+1$
and
the scalar-type $j=\ell-1$
modes of pure AdS, respectively,
where $j$ stands for the total angular momentum
number of the modes. Likewise, vector-type Proca modes approach the
vector-type $j=\ell$ modes of pure AdS.
On the other hand, as one approaches the $r_{\rm h} \rightarrow 0$ limit,
the imaginary part of the frequencies starts decreasing as an
$\ell,d$-dependent power of $r_{\rm h}$, which is difficult to
determine numerically, due to the poor convergence properties of the
methods for very small black holes. An analytical approach for small
black holes is done further below using matched asymptotic expansions.
The vanishing of the imaginary part for very small black holes is
intuitively clear, since the modes only decay due to the presence of the
black hole and there is no energy dissipated through spatial
infinity. However, the fact that the frequencies approach those of
pure AdS was not necessarily expected, because the purely ingoing
boundary condition at $r=r_{\rm h}$ in the $r_{\rm h} \rightarrow 0$
limit does not reduce to the regularity boundary condition imposed at
the origin of pure AdS in \cite{lopes2024}. Such is in accordance,
however, with results for other types of perturbations
\cite{cardoso_lemos_konoplya2003}.

%%%%%%%%%%%%%%%%%%%%%%%%%%%%%%%%%%%%%%%%%%%%
\section{Matched asymptotic expansions for small
Schwarzschild-AdS black holes}\label{sec:matching}
%%%%%%%%%%%%%%%%%%%%%%%%%%%%%%%%%%%%%%%%%%%%

%%%%%%%%%%%%%%%%%%%%%%%%%%%%%%%%%%%%%%%%%%%%%%%%
\subsection{Preamble}
%%%%%%%%%%%%%%%%%%%%%%%%%%%%%%%%%%%%%%%%%%%%%%%%

Both the numerical shooting integration method and the Horowitz-Hubeny
method begin to fail for sufficiently small black holes.  In this
section, we investigate if an analytical procedure for such cases can
be developed, following
\cite{dolan2012,cdly2004,cardoso_dias_yoshida2006,wang_herdeiro_matching2014}.
We do not fix the AdS length $l$ here, as it is important for the
approximations that we are taking.

The spacetime is separated initially into two different
regions: the near region, the region near the black hole
horizon, verifying
$\frac{r-r_{\rm h}}{l} \ll \frac1{\omega l}$,
where the AdS curvature may be neglected, and the far region,
the region far from the black hole
horizon,
verifying 
$\frac{r-r_{\rm h}}{l} \gg \frac{r_{\rm h}}{l} $, where the effect of
the black hole may be neglected.
The regime we work is the small black hole case, i.e.,
$\frac{r_{\rm h}}{l} \ll 1$, and the large wavelength
perturbation mode approximation, i.e.,
small wave frequency $\omega$ approximation,
$\omega r_{\rm h} \ll 1$.
After imposing the usual boundary conditions at the horizon
and at spatial infinity, the QNM frequencies are
determined from the matching condition of the two solutions in the
overlap region, defined through
$\frac{r_{\rm h}}{l}  \ll \frac{r-r_{\rm h}}{l}  \ll
\frac1{\omega l}$.
Thus in brief there are three regions, namely,
\begin{align}\label{eq:nearr}
\frac{r-r_{\rm h}}{l} \ll \frac1{\omega l}\,, \quad\quad\quad
{\rm near\; region},
\end{align}
\begin{align}\label{eq:overlapr}
\hskip 1.1cm
\frac{1}{\omega l}\gg \frac{r-r_{\rm h}}{l}  \gg
 \frac{r_{\rm h}}{l},\;\;
{\rm overlap\; region},\quad
%
%\frac{r_{\rm h}}{l}  \ll \frac{r-r_{\rm h}}{l}  \ll
%\frac1{\omega l}\,, \quad\quad\quad
%{\rm overlap\; region},
\end{align}
\begin{align}\label{eq:farr}
\frac{r_{\rm h}}{l}\
\ll
\frac{r-r_{\rm h}}{l}\,, \quad\quad\quad
{\rm far\; region}.
%\frac{r-r_{\rm h}}{l} \gg \frac{r_{\rm h}}{l}\,, \quad\quad\quad
%{\rm far\; region}.
\end{align}
In the regime given by Eqs.~\eqref{eq:nearr}-\eqref{eq:farr},
scalar-type Proca
perturbations with $\ell=0$, i.e., monopole perturbations,
have decoupled equations
and thus the monopole case
can be handled through a matching asymptotic expansion
technique. 
The same happens for the 
vector-type
Proca perturbations.
Indeed, for monopole and  
vector-type
Proca perturbations,
the associated decoupled
Schr\"{o}dinger-like equations can be solved analytically in each
region and then overlapped.
Maxwell perturbations can also be handled in this regime.
On the other hand for this regime, i.e., the
regime given by Eqs.~\eqref{eq:nearr}-\eqref{eq:farr},
the coupled modes associated to $\ell \geq 1$ scalar-type Proca
perturbations remain nontrivially coupled in the near region,
and for such
perturbations it is not possible to perform a matching of the
solutions.

In what follows, we study the
scalar monopole and vector-type
Proca perturbations in $d$ dimensions.
We also do the matching of the Maxwell field in $d$ dimensions
which has not been previously done.

\subsection{Proca perturbations: Matched asymptotic expansions
in the
scalar monopole
and vector-type cases}

\subsubsection{Scalar-type Proca perturbations: Monopole case}

\vskip 0.15cm
\centerline{\it Near region}
\vskip 0.15cm

To deal with the near region we define the unitless
radial coordinate $z$ by
\begin{align}\label{eq:z}
z \equiv z(r)=1-\left(\frac{r_{\rm h}}{r}\right)^{d-3}\,,
\end{align}
meaning $z=0$ at the horizon 
and $z=1$ at spatial infinity.
From  Eq.~\eqref{eq:u1}
together with Eq.~\eqref{eq:matrixpotential},
the $\ell=0$ 
Schr\"{o}dinger-like equation in the near region
is then
\begin{align}\label{eq:monopole_near}
&z(1-z)\frac{d^2 u^{\mathrm{near}}_1}{d z^2}+ 
\left(1-z\left(2+\frac{1}{d-3}\right)\right)
\frac{d u^{\mathrm{near}}_1 }{d z}+\notag\\
&+\frac{1}{(d-3)^2}\biggl(\frac{\omega^2
r_{\rm h}^2}{z(1-z)}-\frac{\mu^2 r_{\rm h}^2}{1-z}\notag\\
&-\frac{d(d-2)z}{4(1-z)}+\frac{(d-2)(d-3)}{2}\biggr)
u^{\mathrm{near}}_1 = 0 \,,
\end{align}
where we assume that
$\mu (r-r_{\rm h}) \ll 1$ for some Proca mass $\mu$ and so this term can 
be neglected, as we only consider leading order terms in $r_{\rm h}$, but keep 
the term $\omega r_{\rm h}$. The solution to Eq.~\eqref{eq:monopole_near} is
given by the Gaussian hypergeometric function $_2 F_1$ \cite{abramowitz} 
as $u_1^{\mathrm{near}}(z) = c z^{-i\overline{\omega}}
\left(1-z\right)^{\frac{d-2}{2(d-3)}}
{_2F_1}\left[2+\frac{2}{d-3}-i\overline{\omega},-i\overline{\omega};1-
2i\overline{\omega};z\right]$,
for some constant $c$ and
$\overline{\omega} \equiv \frac{\omega r_{\rm h}}{d-3}$.

The near region solution in the overlap region
for $d\geq6$ is
\begin{align}\label{eq:monopole_near_final}
&u_1^{\rm{near}} = c a_1^{\rm{near}}
\biggl(\frac{r}{r_{\rm h}}\biggr)^{-\frac{d-2}{2}} 
+ c b_1^{\rm{near}}
\biggl(\frac{r}{r_{\rm h}}\biggr)^{\frac{d}{2}},\\
&a_1^{\rm{near}} = \frac{\Gamma[1-2i\overline{\omega}]
\Gamma[-1-\frac{2}{d-3}]}{\Gamma[1-i\overline{\omega}]
\Gamma[-1-\frac{2}{d-3} -i\overline{\omega}]} 
\,,
\nonumber\\
&b_1^{\rm{near}} =
\frac{\Gamma[1-2i\overline{\omega}]\Gamma[1+\frac{2}{d-3}]}
{\Gamma[2+\frac{2}{d-3}-
i\overline{\omega}]
\Gamma[-i\overline{\omega}]}\,,
\label{eq:betanearmono}
\end{align}
with $c={\rm constant}$.
The ratio between the coefficients from the near region solution is 
\begin{align}
& \frac{a_1^{\rm{near}}
r_{\rm h}^{d-1}}{b_1^{\rm{near}}} =
-\frac{1 + \frac{2}{d-3}}{i\overline{\omega}}r_{\rm h}^{d-1}
\,,\label{eq:ratiomonopolenear}
\end{align}
where we have used that $\Gamma[1 - i\bar{\omega}] = 
-i \bar{\omega}\Gamma[-i \bar{\omega}]$ and we have neglected 
the contribution of $\bar{\omega}$ in the remaining $\Gamma$'s.
While the expression in Eq.~\eqref{eq:ratiomonopolenear}
seems to be formally valid only for 
$d\geq 6$, it is also valid for $d=4$ and $d=5$.
For these specific cases, one must use the 
correct transformation to obtain the behavior at the overlap
region. This yields a leading order term containing a
logarithm. However, the dominant term is $\psi[-i\overline{\omega}]$,
where $\psi[x]$ is the digamma function, i.e.,
$\psi[x]=\frac{d(\ln\Gamma[x])}{dx}$.
It diverges for $\omega r_{\rm h}\ll
1$ but can be regularized by $\frac{1}{\Gamma[-1-\frac{2}{d-3} - i
\overline{\omega}]}$. The resulting asymptotic behavior yields the
same as Eq.~\eqref{eq:monopole_near_final} with the coefficient ratio
given by Eq.~\eqref{eq:ratiomonopolenear}.

\vskip 0.25cm
\centerline{\it Far region}
\vskip 0.15cm

To deal with the far region we define the unitless
radial coordinate $y$ as 
\begin{align}\label{eq:ycoordinate}
y\equiv y(r)= \left(1+\frac{r^2}{l^2}\right)^{-1},
\end{align}
such that $y=0$ at spatial infinity 
and $y=1$ at the event horizon.

Using the approximation $\frac{r-r_{\rm h}}{l} \gg \frac{r_{\rm h}}{l}$
and 
Eq.~\eqref{eq:u1} with $\ell = 0$ and $u_2=0$, we can
obtain the Proca perturbation equation for $u_1$
in the far region
regime. It is
\begin{align}\label{eq:vt_farscalar}
&y (1-y) \frac{\partial^2 u^{\mathrm{far}}_1}{\partial y^2}
+\left(\frac{1}{2}-y\right)\frac{\partial
u^{\mathrm{far}}_1}{\partial y}\notag\\
&+\biggl(\frac{\omega^2l^2}{4}-
\frac{(d-2)d}{16(1-y)}\notag\\
&-\frac{1}{4y}\biggl(\mu^2l^2+\frac{(d-4)(d-2)}{4}\biggr)
\biggr)u^{\mathrm{far}}_1 = 0 \,.
\end{align}
The solution of Eq.~\eqref{eq:vt_farscalar} obeying the Dirichlet boundary
condition at $y=0$ is $u_1^{\mathrm{far}}(y)= c
y^{\frac{1}{4}\left(2\gamma - 1\right)}
(1-y)^{\frac{d}{4}}{_2F_1}[\eta_1,\sigma_1;\gamma;y]$, for
some constant $c$, with $\eta_1 = \frac{1}{4} \left(2\gamma
+d-1 + 2\omega l\right)$, $\sigma_1 = \frac{1}{4}
\left(2\gamma+d-1 - 2\omega l\right)$, $\gamma =
1+\frac{1}{2}\sqrt{(d-3)^2+4\mu^2l^2}$.

Using the linear
transformation law of the hypergeometric function $y\rightarrow 1-y$,
the behavior in the overlap region of the solution
of Eq.~\eqref{eq:vt_farscalar} for noninteger
$\gamma-\eta_1 - \sigma_1 = - \frac{d-1}{2}$, i.e., for even
$d$ is
\begin{align}\label{eq:vt_far_2scalar}
&u_1^{\mathrm{far}}(r)\hspace{-0.5mm} = \hspace{-0.5mm}c
a_1^{\rm{far}} \left(
\frac{r}{l}\right)^{\hspace{-0.5mm}-\frac{d-2}{2}}
\hspace{-0.5mm}+ cb_1^{\rm{far}}
\left(\frac{r}{l}\right)^{\hspace{-0.5mm} \frac{d}{2}}
\hspace{-0.5mm},\\
& a_1^{\rm{far}}
= \frac{\Gamma[\eta_1+
\sigma_1-\gamma]}{\Gamma[\eta_1]\Gamma[\sigma_1]}\,,\,
b_1^{\rm{far}} = \frac{\Gamma[\gamma-\eta_1-\sigma_1]}
{\Gamma[\gamma-\eta_1]\Gamma[\gamma-\sigma_1]}
\,.\label{eq:alphabetafarscalar}
\end{align}
The ratio between the coefficients for the far region
solution in the overlap region is
\begin{align}
&\frac{a_1^{\rm{far}} l^{d-1}}{b_1^{\rm{far}}} = 
 -\frac{k!\Gamma[\frac{d-1}{2}]\Gamma[k+\gamma]\,
 i\delta l^{d}}
{2\Gamma[\gamma+k+\frac{d-1}{2}](\frac{d+1}{2})_k},
\label{eq:ratiomonopolefar}
\end{align}
where the QNM frequency was expanded as $\omega =
\omega_{\rm{AdS}} + i \delta$ and only leading order in $|\delta| \ll
\omega_\mathrm{AdS}$ was considered, with $\omega_{\rm{AdS}}l =
\gamma + 2 k + \frac{d-1}{2}$, $k$ being the overtone number and a
non-negative integer, and
where $(x)_n \equiv \frac{\Gamma[x+n]}{\Gamma[x]}$
is the Pochhammer symbol
for a real number $x$. In order to obtain the expression in 
Eq.~\eqref{eq:ratiomonopolefar}, we have used that, 
for an integer $k$ and a small $\delta$, $\lim_{z\rightarrow -k} 
\frac{1}{\Gamma[z + i \delta]} = (-1)^k k! i \delta$ and 
also $\Gamma[z + i \delta] = \Gamma[z]$ for noninteger $z$.

\vskip 0.15cm
\vskip 2.15cm
\centerline{\it Overlap region}
\vskip 0.15cm

The matching of the two solutions in the overlap region is performed
by equating the ratio in Eq.~\eqref{eq:ratiomonopolenear} with the
ratio in Eq.~\eqref{eq:ratiomonopolefar}, i.e.,
$\frac{a_1^{\rm{near}}r_{\rm h}^{d-1}}{b_1^{\rm{near}}} =
\frac{a_1^{\rm{far}} l^{d-1}}{b_1^{\rm{far}}}$. The real part
of the correction to the QNMs, i.e.,
$\mathfrak{\delta}l$, for the monopole case is
\begin{align}\label{eq:delta_monopole}
&\mathfrak{Re}{\delta} = -\frac{2(d-1)
\Gamma[\frac{d-1}{2}+k+\gamma]}
{k!\Gamma[\frac{d-1}{2}]\Gamma[k+ \gamma]}
\left(\frac{d+1}{2}\right)_k\notag \\
&\times\biggl(\frac{r_{\rm h}}{l}\biggr)^{d-2}\frac{1}{
\omega_{\mathrm{AdS}}}\quad.
\end{align}
Note that $\delta$ scales linearly with the inverse of the pure
AdS frequency, whereas for the vector-type perturbation,
as we will see, it scales
linearly with the pure AdS frequency.  Also, $\mathfrak{Re} \delta
\sim (\frac{r_{\rm h}}{l})^{d-2}$, which is useful for comparison with the
numerical results, as can be
seen in Tables~\ref{tab:num_vs_an_monopole1}
and~\ref{tab:num_vs_an_monopole2}.
Note that the agreement between numerical and
analytical results for
monopole perturbations is highly satisfactory.
Indeed, since the monopole frequencies decrease slowly
with the radius of the black hole, we were able to obtain numerical
results for small black holes, for which the approximations of the
analytical method are good.

\begin{table}[h]
\centering
\begin{tabular}{c|c|c|c|c}
&$k$ & $\mu$ & $\mathfrak{Re}\delta_{n} l$
& $\mathfrak{Re}\delta_{a}
l$ \\\hline
 & $0$ & $0.5$ &$-5.66689\e{-8}$ &$-5.66321\e{-8}$\\
 $d=4$ & $0$ & $2.0$ & $-8.88933\e{-8}$ &$-8.88073\e{-8}$\\
& $1$ & $0.5$ & $-1.64019\e{-7} $ &$-1.63822\e{-7}$\\\hline
& $0$ & $0.5$ &$-1.28297\e{-11}$ &$-1.28296\e{-11}$\\
 $d=5$& $0$ & $2.0$ & $-2.09443\e{-11}$ &$-2.09443\e{-11}$\\
& $1$ & $0.5$ & $-5.03700\e{-11}$ &$-5.03698\e{-11}$
\end{tabular}
\caption{Comparison between numerical results for the real part of
$\delta l$, $\mathfrak{Re}\delta_{n}l$, and analytical results
obtained from Eq.~\eqref{eq:delta_monopole}, $\mathfrak{Re}\delta_a
l$, for different parameters of the monopole Proca perturbation in
$d=4,5\,$ Schwarzschild-AdS black holes,
with radius $\frac{r_{\rm h}}{l}=
10^{-4}$.}
\label{tab:num_vs_an_monopole1}
\end{table}

\begin{table}[h]
\centering
\begin{tabular}{c|c|c|c|c}
& $k$ & $\mu$ & $\mathfrak{Re}\delta_{n} l$ &
$\mathfrak{Re}\delta_{a} l$ \\\hline
 & $0$ & $0.5$ &$-2.30944\e{-9}$ &$-2.30877\e{-9}$\\
 $d=6$ & $0$ & $2.0$ & $-3.66837\e{-9}$ &$-3.66693\e{-9}$\\
& $1$ & $0.5$ & $-1.14208\e{-8} $ &$-1.14145\e{-8}$\\\hline
& $0$ & $0.5$ &$-1.51407\e{-11}$ &$-1.51388\e{-11}$\\
 $d=7$& $0$ & $2.0$ & $-2.30083\e{-11}$ &$-2.30046\e{-11}$\\
& $1$ & $0.5$ & $-9.01694\e{-11}$ &$-9.01485\e{-11}$
\end{tabular}
\caption{Comparison between numerical results for the real part of
$\delta l$, $\mathfrak{Re}\delta_n l$, and analytical results obtained
from Eq.~\eqref{eq:delta_monopole}, $\mathfrak{Re}\delta_a l$, for
different parameters of the monopole Proca perturbation in $d=6,7\,$
Schwarzschild-AdS black holes, with radius $\frac{r_{\rm h}}{l}=3\e{-3}$.}
\label{tab:num_vs_an_monopole2}
\end{table}

\vskip 3cm

\subsubsection{Vector-type Proca perturbations}

\centerline{\it Near region}
\vskip 0.15cm
In the near region, the background is essentially Schwarzschild,
and one can use the approximation $\frac{r-r_{\rm h}}{l} \ll
\frac{1}{\omega l}$ to rewrite Eq.~\eqref{eq:uv} as
\begin{align}\label{eq:vt_near}
&z(1-z)\frac{d^2 u^{\mathrm{near}}_3}{d z^2} 
+ \left(1-z\left(2+\frac{1}{d-3}\right)\right)
\frac{d u^{\mathrm{near}}_3 }{d z}+\notag\\
&+\frac{1}{(d-3)^2}\biggl(\frac{\omega^2 r_{\rm h}^2}{z(1-z)}
-\frac{(\ell+1)(\ell+d-4)+\mu^2 r_{\rm h}^2}{1-z}\notag\\
&-\frac{(d-4)(d-6)z}{4(1-z)}-\frac{(d-4)(d-3)}{2}\biggr)
u^{\mathrm{near}}_3 = 0 \,,
\end{align}
with $z \equiv z(r)=1-(\frac{r_{\rm h}}{r})^{d-3}$,
see Eq.~\eqref{eq:z}, meaning $z=0$ at the
horizon and $z=1$ at spatial infinity, and where it was assumed that,
in the near region, $\mu (r-r_{\rm h}) \ll 1$.  Moreover, the term
$\mu r_{\rm h} \ll 1$ may be neglected, since it provides a minor
correction to $(\ell+1)(\ell+d-4) \gtrsim 1$.  The solution to
Eq.~\eqref{eq:vt_near} obeying the ingoing boundary condition,
Eq.~\eqref{eq:bc_horizon}, at $z=0$, is given in terms of the Gaussian
hypergeometric function  $_2 F_1$ \cite{abramowitz} as 
$u_3^{\rm{near}}= c z^{-i\overline{\omega}}
(1-z)^{\frac{2\ell+d-4}{2(d-3)}}
{_2F_1}\left[1+\frac{\ell-1}{d-3}-i\overline{\omega},
1+\frac{\ell+1}{d-3}-i\overline{\omega};
1-2i\overline{\omega};z\right]$, for some constant
$c$, with
$\overline{\omega} \equiv \frac{\omega r_{\rm h}}{d-3}$.

One now must obtain the functional behavior of the
near region solution in the
overlap region, by performing the transformation $z\rightarrow 1-z$ of
the hypergeometric function and consider the leading
order terms of $\frac{r_{\rm h}}{r} \ll 1$.  However, one must be
careful to do such transformation as its validity depends on whether
$-1 - \frac{2\ell}{d-3}$ is an integer
or not, i.e., whether $\frac{2\ell}{d-3}$ is an integer or not.  For a
noninteger $\frac{2\ell}{d-3}$, we have at the overlap region
\begin{align}\label{eq:vt_near_final}
&u_3^{\mathrm{near}} (r)\hspace{-.5mm} =\hspace{-.5mm} 
ca_3^{\rm{near}}
\left(\frac{r}{r_{\rm h}}\right)^{\hspace{-1mm}-
\frac{2\ell + d-4}{2}}
\hspace{-1mm}+ c b_3^{\rm{near}} 
\left(\frac{r}{r_{\rm h}}\right)^{\hspace{-1mm}
\frac{2\ell + d -2}{2}}\hspace{-1.5mm},\\
& a_3^{\rm{near}} = \frac{\Gamma[-1-\frac{2\ell}{d-3}]}
{\Gamma[-\frac{\ell-1}{d-3}-i\overline{\omega}]
\Gamma[-\frac{\ell+1}{d-3}-i\overline{\omega}]}\,,\\
&b_3^{\rm{near}} = \frac{\Gamma[1+\frac{2\ell}{d-3}]}
{\Gamma[1+\frac{\ell-1}{d-3}-i\overline{\omega}]
\Gamma[1+\frac{\ell+1}{d-3}-i\overline{\omega}]}\,.\label{eq:alphabetanear}
\end{align}
From the definition of the near region,
Eq.~\eqref{eq:vt_near_final} holds approximately in the region $0 \ll
r-r_{\rm h} \ll \frac{1}{\omega}$. To perform the matching, the relevant
quantity is in fact the ratio between the coefficients of the two
leading terms. This is because one can always absorb a common factor
into the constant $c$. We have for small $\overline{\omega}$ and
noninteger $\frac{2\ell}{d-3}$
\begin{align}\label{eq:vt_near_final_2}
&\frac{a^{\rm{near}}_3 r_{\rm h}^{2\ell +
d-3}}{b^{\rm{near}}_3} =
\frac{\Gamma^2[1+\frac{\ell-1}{d-3}]\Gamma^2[1+\frac{\ell+1}{d-3}]r_{\rm
h}^{2\ell + d-3}} {(1+\frac{2\ell}{d-3})
\Gamma^2[1+\frac{2\ell}{d-3}]}\notag
\\
&\times\Biggl(\frac{\sin{\left(\frac{\pi(\ell-1)}{d-3}\right)}
\sin{\left(\frac{\pi(\ell+1)}{d-3}\right)}}
{\pi\sin{\left(\frac{2\ell\pi}{d-3}\right)}}
+i\overline{\omega}\Biggr)\,,
\end{align}
where the equality is approximate.
For integer $\frac{2\ell}{d-3}$, i.e.,  for $
-1-\frac{2\ell}{d-3} \equiv -m$, where $m$ is a positive integer,
then $a^{\rm{near}}_3$ diverges, Eq.~\eqref{eq:vt_near_final}
is not valid, and the upshot is that it is not possible
to make the matching.
Indeed, in this case, one must take the correct transformation
\cite{abramowitz} giving the leading terms $\left(\frac{r}{r_{\rm
h}}\right)^{\frac{2\ell + d-2}{2}}$ and
$\frac{1}{\Gamma[-\frac{\ell-1}{d-3}-i\overline{\omega}]
\Gamma[-\frac{\ell+1}{d-3}-i\overline{\omega}]}
\left(\frac{r}{r_{\rm h}}\right)^{-\frac{2\ell + d-4}{2}}
((d-3)\ln\left(\frac{r_{\rm h}}{r}\right)-\psi[1] - \psi[m +1] +
\psi[[1+\frac{\ell-1}{d-3}-i\overline{\omega}] 
+ \psi[1+\frac{\ell+1}{d-3}-i\overline{\omega}])$, where $\psi[x]$ 
is the digamma
function, i.e., $\psi[x]=\frac{d\ln\Gamma[x]}{dx}$.  As we are interested
in $\frac{r_{\rm h}}{r}\ll1$, the logarithmic term dominates in the
second leading order term and so the behavior of the solution for
integer $\frac{2\ell}{d-3}$ is described by $\left(\frac{r}{r_{\rm
h}}\right)^{\frac{2\ell + d-2}{2}}$ and $\left(\frac{r}{r_{\rm
h}}\right)^{-\frac{2\ell + d-4}{2}}\ln\left(\frac{r_{\rm
h}}{r}\right)$. Thus, for this case, the functional matching is not
possible as the far region does not have such asymptotic behavior.

\vskip 0.25cm
\centerline{\it Far region}
\vskip 0.15cm
%here

In the far region, the background is essentially pure AdS
and the effect of the 
black hole can be neglected.
To deal with the far region we define the unitless
radial coordinate $y$ as 
$y\equiv y(r)= \left(1+\frac{r^2}{l^2}\right)^{-1}$,
see Eq.~\eqref{eq:ycoordinate},
such that $y=0$ at spatial infinity 
and $y=1$ at the event horizon. 
Using the approximation $\frac{r-r_{\rm h}}{l} \gg \frac{r_{\rm h}}{l}$, 
Eq.~\eqref{eq:uv} can be written as
\begin{align}\label{eq:vt_far}
&y (1-y) \frac{\partial^2 u^{\mathrm{far}}_3}{\partial y^2}
+\left(\frac{1}{2}-y\right)\frac{\partial
u^{\mathrm{far}}_3}{\partial y}\notag\\
&+\biggl(\frac{\omega^2l^2}{4}-
\frac{(2\ell+d-4)(2\ell+d-2)}{16(1-y)}\notag\\
&-\frac{1}{4y}\biggl(\mu^2l^2+\frac{(d-4)(d-2)}{4}\biggr)
\biggr)u^{\mathrm{far}}_3 = 0 \,.
\end{align}
The solution of Eq.~\eqref{eq:vt_far} obeying the Dirichlet boundary
condition at $y=0$ is 
$
u_3^{\mathrm{far}}(y)= c
y^{\frac{1}{4}\left(2\gamma - 1\right)}
(1-y)^{\frac{2\ell+d-2}{4}}{_2F_1}[\eta_\ell,\sigma_\ell;\gamma;y]$,
for
some constant $c$, with $\eta_\ell = \frac{1}{4} (2\gamma
+2\ell+d-3 + 2\omega )$, $\sigma_\ell = \frac{1}{4}
\left(2\gamma+2\ell+d-3 - 2\omega l\right)$, $\gamma =
1+\frac{1}{2}\sqrt{(d-3)^2+4\mu^2l^2}$.

One now must obtain the functional behavior of the
far region solution in the
overlap region. Using  the linear
transformation law of the hypergeometric function $y\rightarrow 1-y$,
the behavior of the solution in the overlap region for noninteger
$\gamma-\eta_\ell - \sigma_\ell = -\ell - \frac{d-3}{2}$, i.e., for even
$d$ is
\begin{align}\label{eq:vt_far_2}
&u_3^{\mathrm{far}}(r)\hspace{-0.5mm} = \hspace{-0.5mm}c
a_3^{\rm{far}} \left(
\frac{r}{l}\right)^{\hspace{-0.5mm}-\frac{2\ell + d-4}{2}}
\hspace{-0.5mm}+ cb_3^{\rm{far}}
\left(\frac{r}{l}\right)^{\hspace{-0.5mm} \frac{2\ell + d-
2}{2}}\hspace{-0.5mm},\\
& a_3^{\rm{far}}
= \frac{\Gamma[\eta_\ell+
\sigma_\ell-\gamma]}{\Gamma[\eta_\ell]\Gamma[\sigma_\ell]}\,,\,
b_3^{\rm{far}} = \frac{\Gamma[\gamma-\eta_\ell-\sigma_\ell]}
{\Gamma[\gamma-\eta_\ell]\Gamma[\gamma-\sigma_\ell]}
\,,\label{eq:alphabetafar}
\end{align}
Since one is in the AdS regime, Eq.~\eqref{eq:vt_far_2} may be
simplified by considering that the QNM frequencies are
given by the pure AdS normal mode frequencies,
$\omega_{\mathrm{AdS}}l=2k+\ell+\frac{d-3}{2}+\gamma_3$, where $k$ is a
non-negative integer, see also \cite{lopes2024}, plus a small correction,
$\delta$, which is complex in general. The frequencies can then be
written as $\omega = \omega_{\mathrm{AdS}}+i\delta$, with
$|\delta| \ll |\omega_{\mathrm{AdS}}|$, 
and the problem of finding $\omega$ reduces to the problem of finding
the correction $\delta$.  We note that the stability study of
Sec. \ref{sec:stability} implies that $\mathfrak{Re} (\delta)<0$.  The
hypergeometric coefficients become
$\eta_\ell = \gamma+k+\ell+\frac{d-3}{2}+i\frac{\delta l}{2}$,
$\sigma_\ell = -k-i\frac{\delta l}{2}$, $\gamma-\eta_\ell =
-k-\ell-\frac{d-3}{2}-i\frac{\delta l}{2}$, $\gamma-\sigma_\ell =
\gamma+k+i\frac{\delta l}{2}$.  With these considerations, we can
now evaluate the ratio between the two leading order terms for small
$\delta$, valid for even $d$, yielding
\begin{align}\label{eq:vt_far_final}
& \frac{a^{\rm{far}}_3 l^{2\ell + d-3}}{b^{\rm{far}}_3} = 
-\frac{k!\Gamma[\ell+\frac{d-3}{2}]\Gamma[k+\gamma]\,
i\delta l^{2\ell + d-2}}
{2\Gamma[\gamma+k+\ell+\frac{d-3}{2}](1+\ell+\frac{d-3}{2})_k},
\end{align}
where $(x)_n \equiv \frac{\Gamma[x+n]}{\Gamma[x]}$
is the Pochhammer symbol
for a real number $x$.  When $\ell+\frac{d-3}{2}=m$, with $m$
being a positive integer, i.e., for odd $d$, $b^{\rm{far}}_3$ diverges
with $\Gamma[\gamma-\eta_\ell-\sigma_\ell]=
\Gamma[-m]$, and the expression
in Eq.~\eqref{eq:vt_far_final} becomes formally invalid.  In this
case, one must pick the correct transformation $y\rightarrow 1-y$,
yielding
the leading terms $a^{\rm{far}}_3
\left(\frac{r}{l}\right)^{-\frac{2\ell+d-4}{2}}$ 
and $\frac{(-1)^{m+1}}{m!\Gamma[\gamma-
\eta_\ell]\Gamma[\gamma-\sigma_\ell]}
\left(\frac{r}{l}\right)^{\frac{2\ell+d-2}{2}}
(-2\ln(\frac{r}{l})-\psi[1]-\psi[m+1]+\psi[\eta_\ell]+\psi[\sigma_\ell])$.
The second leading term needs to be evaluated carefully, since
$\sigma_\ell$ and $\gamma-\eta_\ell$ are negative integers at zeroth order
in $\delta$. This means the logarithmic term is dropped and only the
term $\frac{\psi[\sigma_\ell]}{\gamma[\gamma - \eta_\ell]}$ survives.
Using the small $\delta$ approximation, the behavior becomes the same
as Eq.~\eqref{eq:vt_far_final} with integer $\ell+\frac{d-3}{2}=m$.
Thus, the far region solution behaves as Eq.~\eqref{eq:vt_far_final}
for both even and odd $d$.

Note that the scalar monopole
case follows from here, by putting $\ell = 1$ in the
vector-type perturbations.  Indeed, the
far region solution for the monopole perturbation in the overlap
region is simply Eq.~\eqref{eq:vt_far_2scalar} with $\ell=1$.
Note also that in
 $u_{1}^{\rm{far}} = c a_1^{\rm{far}}
\left(\frac{r}{l}\right)^{- \frac{d-2}{2}} 
+ b_1^{\rm{far}}\left(\frac{r}{l}\right)^{\frac{d}{2}}$
of Eq.~\eqref{eq:vt_far_2scalar} 
the $a_1^{\rm{far}}$ and
$b_1^{\rm{far}}$ 
are simply 
$a_1^{\rm{far}} = a_3^{\rm{far}}|_{\ell = 1}$
and $b_1^{\rm{far}} = b_3^{\rm{far}}|_{\ell = 1}$,
see also Eq.~\eqref{eq:alphabetafar}.

%alpha2 is eta2, beta2 is sigma2.

\vskip 0.15cm
\centerline{\it Overlap  region}
\vskip 0.15cm

With the near and far region behavior in the overlap region described
by $r_{\rm h} \ll r \ll \frac{1}{\omega}$,
we now proceed with the functional
matching between the two regions.  The previous analysis reveals that
such matching is only possible for noninteger $\frac{2 \ell}{d-3}$.
Otherwise, the near region solution presents a logarithmic term in the
overlap region that does not match with the expected behavior from the
far-region solution. The matching condition is obtained by equating
the two ratios, i.e., $\frac{\alpha_3^{\rm{near}}r_{\rm h}^{2\ell +
d-3}} {b_3^{\rm{near}}} = \frac{\alpha_3^{\rm{far}}l^{2\ell +
d-3}}{b_3^{\rm{far}}}$, given in Eq.~\eqref{eq:vt_near_final_2}
and in Eq.~\eqref{eq:vt_far_final}.  This gives an expression for both
real and imaginary part of $\delta$. There is more interest in the
real part of $\delta$ to give the estimation of the imaginary part of
the QNM frequencies.  Only considering the leading order of
$\frac{r_{\rm h}}{l}$, we have that the real part of $\delta$ defined
as $\mathfrak{Re}\delta$ is
\begin{align}\label{eq:delta_final_3}
&\mathfrak{Re}\delta =-\frac{2\Gamma\left[k+\ell+\frac{d-3}{2}+
\gamma\right]
\Gamma^2\left[1+\frac{\ell+1}{d-3}\right]}{k!\left(2\ell+d-3\right)
\Gamma\left[\gamma+k\right]}\notag\\
&\times\frac{\Gamma^2\left[1+\frac{\ell-1}{d-3}\right]
\biggl(\ell+\frac{d-1}{2}\biggr)_k}{\Gamma^2\left[1+
\frac{2\ell}{d-3}\right]
\Gamma\left[\ell+\frac{d-3}{2}\right]}
\left(\frac{r_{\rm h}}{l}\right)^{2\ell+d-2} \omega_{\mathrm{AdS}}\,,
\end{align}
where $\gamma = 1 + \frac{1}{2}\sqrt{(d-3)^2 + 4 \mu^2 l^2}$.  We
note that $\mathfrak{Re} \delta \sim
(\frac{r_{\rm h}}{l})^{2\ell+d-2}$, which
is the same result for scalar field perturbations in $d$-dimensional
Reissner-Nordstr\"{o}m-AdS black holes \cite{wang_herdeiro_matching2014}.
Also, $\mathfrak{Re} \delta<0$, as expected from the linear stability
analysis.  We recall that Eq.~\eqref{eq:delta_final_3} only holds for
noninteger $\frac{2\ell}{d-3}$.  This means that it does not formally
hold for $d=4,5$, and neither for the specific values of $\ell =
\frac{m (d-3)}{2}$, $m = 1,2,...$ when $d\geq 6$.

The validity of Eq.~\eqref{eq:delta_final_3} could be in principle
tested by fitting the numerical results to $\mathfrak{Re}\delta = -P
(\frac{r_{\rm h}}{l})^Q$, with $P,Q$ parameters, and comparing $P,Q$
with the ones predicted by the analytical expression.  As stated in
the beginning of this section, the numerical results are not accurate
enough for sufficiently small black holes. Typically, for $\omega$
with $|\mathfrak{Im}(\omega l)| \lesssim 10^{-11}$, the
QNM frequencies obtained from numerically integrating the
Schr\"{o}dinger-like equations oscillate substantially when one
changes the initial guess or the domain of integration.  Since the
magnitude of the imaginary part of the frequencies decreases as
$(\frac{r_{\rm h}}{l})^{2\ell+d-2}$,
this means that the numerical results are
only accurate for $\frac{r_{\rm h}}{l} \gtrsim 0.01$.  Thus, numerical
verification of Eq.~\eqref{eq:delta_final_3} is still needed, possibly
using the Breit-Wigner resonance method approached in
\cite{cardoso_breit_wigner2009,pani2013}.  Comparison between analytical and
numerical results is done in Table~\ref{tab:num_vs_an}.  The
analytical and numerical results of Table~\ref{tab:num_vs_an} agree up
to the second significant figure. A better agreement could in
principle be achieved for smaller black holes, although for that to
happen our numerical methods need to be improved.
\begin{table}[h]
\centering
\begin{tabular}{c|c|c|c|c}
& $k$ & $\mu$ & $\mathfrak{Re}\delta_{n} l$ &
$\mathfrak{Re}\delta_{a} l$ \\\hline
& $0$ & $0$ &$-1.758\e{-9}$ &$-1.738\e{-9}$\\
$d=6$ & $0$ & $0.5$ & $ -1.906\e{-9}$ &$-1.883\e{-9}$\\
& $1$ & $0.5$ & $-1.849\e{-8}$ &$-1.809\e{-8}$\\\hline
& $0$ & $0$  &$-7.725\e{-11}$ &$-7.680\e{-11}$\\
$d=7$& $0$ & $0.5$ & $-8.188\e{-11}$ &$-8.139\e{-11}$\\
& $1$ & $0.5$ & $-8.665\e{-10}$ &$-8.572\e{-10}$
\end{tabular}
\caption{Comparison between numerical results for the
real part of $\delta l$, 
$\mathfrak{Re}\delta_n l$, and 
analytical results obtained from Eq.~\eqref{eq:delta_final_3},
$\mathfrak{Re}\delta_a l$, 
for different parameters of the vector-type $\ell=1$ 
Proca perturbation in $d=6,7$ Schwarzschild-AdS black holes,
with radius $\frac{r_{\rm h}}{l}=0.02$.}
\label{tab:num_vs_an}
\end{table}

Interestingly, we have found that the numerical results obey
Eq.~\eqref{eq:delta_final_3} even for integer $\frac{2\ell}{d-3}$, which
seems to indicate that the solutions can still be matched as if there
was no logarithmic term. Such might be due to the approximations done
in $\omega r$ and $\frac{r}{l}$, in order to solve analytically the
Sch\"{o}dinger-like equation in the near region, although further
investigation is still needed in this regard.

Note that the agreement between analytical and numerical results for
monopole perturbations is much more satisfactory than for vector-type
perturbations.  Indeed, since the monopole frequencies decrease slowly
with the radius of the black hole, we were able to obtain numerical
results for smaller black holes, for which the approximations of the
analytical method are better.

\subsection{Maxwell perturbations: Matched asymptotic expansions}

\subsubsection{Scalar-type Maxwell perturbations}

\vskip 0.15cm
\centerline{\it Near  region}
\vskip 0.15cm

We now perform the matching asymptotic expansions for the Maxwell field
in a Schwarzschild-AdS black hole background, a
procedure that, to our knowledge, has not yet been carried out
explicitly in $d$ spacetime dimensions. While similar techniques have
been extensively applied to scalar and gravitational fields, the
analogous analysis for the Maxwell field remains unexplored in
higher-dimensional settings. This analysis is crucial for
understanding the asymptotic structure of electrodynamics in arbitrary
dimensions and may provide insight into the behavior of
electromagnetic radiation, conserved charges, and symmetry structures
at null infinity.

We start by treating the scalar-type case of the Maxwell field. The 
scalar-type modes are described by Eqs.~\eqref{eq:sch_u12} and
~\eqref{eq:scalarpotentialmaxwell_sads}. By performing the transformation
in Eq.~\eqref{eq:z},
i.e., $z
\equiv z(r)=1-\left(\frac{r_{\rm h}}{r}\right)^{d-3}$,
the equation for these Maxwell modes in the near region 
becomes 
\begin{align}\label{eq:near_u12}
    &z(1-z)\frac{d^2 u_{12}^{\rm near}}{dz^2}
    + \left( 1 - z\left(2 + \frac{1}{d-3}\right)\right)
    \frac{du_{12}^{\rm near}}{dz} \notag\\
    & + \frac{1}{(d-3)^2}\left( 
    \frac{\omega^2 r_h^2}{z(1-z)} - \frac{l(l+d-3)}{1-z}\right.\notag\\
    &\left. -\frac{(d-2)(d-4)z}{4(1-z)} + 
    \frac{(d-4)(d-3)}{2}
    \right)u_{12}^{\rm near} = 0\,,
\end{align}
where $u_{12}^{\rm near}$ is the solution $u_{12}$ of the scalar-type 
modes in the near region. Equation~\eqref{eq:near_u12} can be solved 
by a hypergeometric function, having thus 
$u_{12}^{\rm near} = c z^{-\frac{i \omega r_h}{d-3}} 
(1-z)^{\frac{2\ell +d -4}{2(d-3)}} \prescript{}{2}{F}_1\left[
    \frac{\ell}{d-3}-i\bar{\omega}, 2 + \frac{\ell}{d-3}-i\bar{\omega}, 
    1 - 2i\bar{\omega};z\right]$, where $c$ is a constant and
$\bar{\omega} = \frac{\omega r_h}{d-3}$. Now, we must evaluate the solution 
asymptotically in the overlap region by studying the behaviour of the 
solution for $z\rightarrow 1$. One can do this by applying the transformation
$z\rightarrow 1-z$ of the hypergeometric function. If $\frac{2\ell}{d-3}$ is 
a non-integer, then the solution in the overlap region is described by
\begin{align}
    &u_{12}^{\rm near} = c a_{12}^{\rm near} 
    \left(\frac{r_h}{r}\right)^{\frac{2\ell + d -4}{2}}
    + c b_{12}^{\rm near} \left(\frac{r}{r_h}\right)^{\frac{2\ell + 
    d -2}{2}}\,,\nonumber\\
    &a_{12}^{\rm near} = \frac{\Gamma\left[1 - 2i\bar{\omega}\right]
    \Gamma\left[ -1-\frac{2\ell}{d-3}\right]}{
        \Gamma\left[1 - \frac{\ell}{d-3}- i\bar{\omega}\right] 
        \Gamma\left[ -1-\frac{\ell}{d-3}- i\bar{\omega}\right]}\,,\,
	\nonumber\\
	&
    b_{12}^{\rm near} = \frac{\Gamma\left[1 - 2i\bar{\omega} \right]
     \Gamma\left[1 
    + \frac{2\ell}{d-3} \right]}{
        \Gamma\left[ \frac{\ell}{d-3} - i\bar{\omega}\right]
        \Gamma\left[2 + \frac{\ell}{d-3} - i\bar{\omega}\right]}\,.
\end{align}
To perform the matching, we must calculate the ratio between 
$a_{12}^{\rm near}$ and $b_{12}^{\rm near}$ for small $\bar{\omega}$. 
For a noninteger $\frac{2\ell}{d-3}$, we have 
\begin{align}
    \frac{a_{12}^{\rm near} r_h^{2\ell + d - 3}}{b_{12}^{\rm near}}
    = &\frac{(d-3)\Gamma^2[\frac{\ell}{d-3}]\Gamma^2[2+\frac{\ell}{d-3}]
    r_h^{2\ell + d - 3}}{(2\ell + d-3)\Gamma^2[1 + \frac{2\ell}{d-3}]}\notag \\
    &\times \left( \frac{\tan(\pi\frac{\ell}{d-3})}{2\pi} 
    + i \bar{\omega}\right)\,,
\end{align} 
where the equality is approximate. For the case of $\frac{2\ell}{d-3}$ being 
an integer, the solution evaluated in the overlap region has a logarithmic 
term which forbids the matching.

\vskip 0.15cm
\centerline{\it Far  region}
\vskip 0.15cm

To analyze the scalar-type modes of the Maxwell field 
in the far region, we need to perform the transformation 
in Eq.~\eqref{eq:ycoordinate},
i.e., $y\equiv y(r)=
\left(1+\frac{r^2}{l^2}\right)^{-1}$,  to Eqs.~\eqref{eq:sch_u12} and
~\eqref{eq:scalarpotentialmaxwell_sads}, and only consider 
leading order terms in $r_{\rm h}$, as we are in the approximation 
$\frac{r-r_{\rm h}}{l}\gg \frac{r_{\rm h}}{l}$. The equation for the field in 
this approximation becomes 
\begin{align}\label{eq:u12far}
    &y(1-y)\frac{\partial^2 u_{12}^{\rm far}}{\partial y^2} 
    + \left(\frac{1}{2} - y \right)
    \frac{\partial u_{12}^{\rm far}}{\partial y}\notag \\
    &+\left( \frac{\omega^2 l^2}{4} - \frac{\ell(\ell + d-3)}{4(1-y)}
    \right.\notag \\
    &\left. - \frac{(d-2)(d-4)}{16 y (1-y)} + \frac{d-4}{4y}
    \right)u_{12}^{\rm far} = 0\,. 
\end{align}
The solution to Eq.~\eqref{eq:u12far}, in the far region, 
that obeys the Dirichlet boundary conditions at infinity 
is described by the 
hypergeometric function as 
$u_{12}^{\rm far} = y^{\frac{d-4}{4}}(1-y)^{\frac{2\ell + d -2}{4}}
\prescript{}{2}{F}_{1}[\frac{\ell +d-3 +\omega l}{2}, 
\frac{\ell + d -3 - \omega l}{2}; \frac{d-3}{2};y]$ for $d\geq 5$
and for $d=4$, it is 
$u_{12}^{\rm far} = y^{\frac{1}{2}}(1 - y)^{\frac{\ell + 1}{2}}
\prescript{}{2}{F}_1[1 + \frac{\ell + \omega l}{2}, 
1 + \frac{\ell - \omega l}{2};\frac{3}{2};y]$. However, since for 
$d=4$, the near region is not valid, we shall also omit the
$d=4$ case from the analysis.

The solution in the far region can be evaluated at the overlap region, 
at $y = 1$, 
by performing the transformation $y\rightarrow 1-y$ in the hypergeometric 
function. For noninteger $-\ell - \frac{d-3}{2}$ and for $d\neq 4$, 
the solution can be expanded as 
\begin{align}
    &u_{12}^{\rm far}(r) = c a_{12}^{\rm far} 
    \left(\frac{r}{l}\right)^{-\frac{2\ell + d -4}{2}}
    + c b_{12}^{\rm far} 
    \left( \frac{r}{l}\right)^{\frac{2\ell + d -2}{2}} \\
    & a_{12}^{\rm far} = \frac{\Gamma[\ell + \frac{d-3}{2}]}
    {\Gamma[\frac{\ell +d-3 +\omega l}{2}]
    \Gamma[\frac{\ell +d-3 -\omega l}{2}]}\,,\\
    & b_{12}^{\rm far} = \frac{\Gamma[-\ell - \frac{d-3}{2}]}
    {\Gamma[- \frac{\ell + \omega l}{2}]
    \Gamma[- \frac{\ell - \omega l}{2}]}\,,
\end{align}
where $\omega l = \omega_{\rm AdS}l + i \delta l$, with 
$\omega_{\rm AdS}l = 2k + \ell + d - 3$. Recall that this expression 
only describes the case $d\geq 5$. The ratio between the two coefficients 
for small $\delta$ becomes
\begin{align}
    \frac{a_{12}^{\rm far} l^{2\ell + d -3}}{b_{12}^{\rm far}}
    = -\frac{k!\Gamma[\ell + \frac{d-3}{2}]
    \Gamma[k + \frac{d-3}{2}] l^{2\ell + d - 2}}{2\Gamma[k + \ell +d -3]
    \left( \ell + \frac{d-1}{2}\right)_k}i\delta\,.
\end{align}

\vskip 0.15cm
\centerline{\it Overlap  region}
\vskip 0.15cm

We can then connect the near region and the far region expressions 
in the overlap region. The matching condition is given by ensuring that 
the field and its first derivative are continuous, meaning that 
$\frac{a_{12}^{\mathrm{far}} l^{2\ell + d -3}}{b_{12}^{\mathrm{far}}} = 
\frac{a_{12}^{\mathrm{near}} r^{2\ell + d -3}_\mathrm{h}}{b_{12}^{\mathrm{near}}}$. 
This gives an expression for the real and imaginary part of $\delta$. We shall 
only consider $\mathfrak{Re}\delta$ since it describes the decay of the modes, 
and its expression is
\begin{align}\label{eq:scalarsmallrh}
    &\mathfrak{Re} \delta = - \frac{2
    \Gamma^2[\frac{\ell}{d-3}] \Gamma^2[2 + \frac{\ell}{d-3}]}{
    (2\ell + d -3)\Gamma^2[1 + \frac{2\ell}{d-3}]\Gamma[\ell + \frac{d-3}{2}]}
    \notag\\
    &\times \frac{\Gamma[k + \ell + d -3]}{k!\Gamma[k + \frac{d-3}{2}]}
    \left(\ell + \frac{d-1}{2}\right)_k\left(\frac{r_{\rm h}}{l}\right)^{2\ell + d-2} 
    \omega_{\mathrm{AdS}}\,,
\end{align}
which is only valid for non-integer 
$\frac{2\ell}{d-3}$ and non-integer $-\ell - \frac{d-3}{2}$. Again, one can 
observe that the decay goes as $\mathfrak{Re}\delta \sim \left(\frac{r_{\rm h}}{l}
\right)^{2\ell + d -2}$. We have compared the analytic results in 
Eq.~\eqref{eq:scalarsmallrh} with the numerical results, which are summarized 
in Table~\ref{tab:num_vs_anscalar}. The values of the analytical expression 
match the numerical ones up to the second significant digit.

\begin{table}[h]
\centering
\begin{tabular}{c|c|c|c|c}
& $\ell$ & $\mathfrak{Re}\delta_{n} l$ &
$\mathfrak{Re}\delta_{a} l$ \\\hline
\multirow{2}{*}{$d=6$} & $1$ &$-1.156\e{-4}$ &$-1.017\e{-4}$\\
& $4$ & $ -1.153\e{-11}$ &$-7.552\e{-12}$\\\hline
$d=8$& $1$ & $-1.342\e{-5}$ &$-1.513\e{-5}$
\end{tabular}
\caption{Comparison between numerical results for the
real part of $\delta l$, 
$\mathfrak{Re}\delta_n l$, and 
analytical results obtained from Eq.~\eqref{eq:scalarsmallrh},
$\mathfrak{Re}\delta_a l$, 
for different parameters of the scalar-type $k=0$ 
electromagnetic Maxwell perturbation in $d=6,8$ Schwarzschild-AdS black holes,
with radius $\frac{r_{\rm h}}{l}=0.01$.}
\label{tab:num_vs_anscalar}
\end{table}

\subsubsection{Vector-type Maxwell perturbations}

Regarding the vector-type Maxwell perturbations, the QNM
frequencies
$\omega = \omega_{\mathrm{AdS}} + i \delta$ follow 
from the vector-type Proca perturbations 
by setting $\mu = 0$. In particular, the real part of 
$\delta$ for the vector-type eletromagnetic 
perturbations is given by Eq.~\eqref{eq:delta_final_3} 
with $\mu=0$, and so  $\gamma = 1 + \frac{1}{2}\sqrt{(d-3)^2}$,
yielding
\begin{align}\label{eq:delta_final_3max}
&\mathfrak{Re}\delta =-\frac{2\Gamma\left[k+\ell+ d-2\right]
\Gamma^2\left[1+\frac{\ell+1}{d-3}\right]}{k!\left(2\ell+d-3\right)
\Gamma\left[\frac{d-1}{2}+k\right]}\notag\\
&\times\frac{\Gamma^2\left[1+\frac{\ell-1}{d-3}\right]
\biggl(\ell+\frac{d-1}{2}\biggr)_k}{\Gamma^2\left[1+
\frac{2\ell}{d-3}\right]
\Gamma\left[\ell+\frac{d-3}{2}\right]}
\left(\frac{r_{\rm h}}{l}\right)^{2\ell+d-2} \omega_{\mathrm{AdS}}\,,
\end{align}
where 
$\omega_{\mathrm{AdS}} = 2k + \ell + d - 2$. Note that, 
like the Proca case, this expression is only valid for non-integer 
$\frac{2\ell}{d-3}$. The comparison between the analytical expression 
and the numerical values are displayed in Table~\ref{tab:num_vs_an} when 
$\mu = 0$.

%%%%%%%%%%%%%%%%%%%%%%%%%%%%%%%%%%%%%%%%%%%%%%%%%
\section{Conclusions}\label{sec:conclusions}
%%%%%%%%%%%%%%%%%%%%%%%%%%%%%%%%%%%%%%%%%%%%%%%%%%

We studied the QNMs of Proca and Maxwell field perturbations in 
$d$-dimensional Schwarzschild-AdS spacetimes
with Dirichlet boundary conditions at infinity.

We reviewed how the Proca field equations in a static and 
spherically symmetric $d$-dimensional background reduce to a set of 
three radial wave-like equations: one equation describing the $d-3$
vector-type modes of the Proca field, which are completely decoupled,
and two coupled equations describing the two scalar-type modes of the
Proca field.  Unlike in pure AdS, in Schwarzschild-AdS the scalar-type
modes are nontrivially coupled, which makes their analysis
substantially more involved. We also showed how the 
Maxwell field equations, as a zero mass limit
of the Proca equations, can be obtained.
The two coupled Proca degrees
of freedom were distinguished in the massless limit, where there is a
pure gauge degree of freedom that obeys the Klein-Gordon equation, and
a Maxwell physical degree of freedom that obeys the scalar-type
Maxwell equation.
With such distinction made, we inferred by back
propagation which of the two coupled Proca modes gives the gauge
and physical modes
in the massless limit.
The Proca nonelectromagnetic polarization mode was shown to
give the Maxwell gauge mode,
whereas the 
Proca electromagnetic polarization mode was shown to
give the Maxwell physical mode.
In $d=4$, however, the scalar-type
Maxwell mode does not follow smoothly from the massless limit of the
Proca mode with electromagnetic polarization. Such happens due to the
vanishing of the effective mass,
$\mu^2_{\mathrm{eff}}=\mu^2+\frac{(d-2)(d-4)}{4l^2}$, which changes
the behavior of the potential at infinity. In this case, the
Maxwell scalar-type modes obey exactly the same equation as the vector-type
modes, and the Maxwell modes can be obtained smoothly from the
massless limit of the vector-type Proca modes.

The QNMs of these fields were obtained numerically for
$4,5,6,7$-dimensional Schwarzschild-AdS black holes. Generalization to
other
higher dimensions should be straightforward.
For all the dimensions $d$ and all the perturbation types, both real and
imaginary parts of the frequencies increased with increasing mass of
the Proca field, in agreement with previous works for $d=4$.
Proca field perturbations were found to show no isospectrality.
On the other hand
 Maxwell
perturbations of
scalar-type and vector-type
were found to be isospectral in the large black hole
regime, for all the spacetime dimensions. Such isospectrality was
proved analytically, using the mathematical machinery developed by
Chandrasekhar.  Taking into consideration Proca perturbations, our
numerical results suggest that the isospectrality is preserved between
Maxwell scalar-type and vector-type modes with electromagnetic
polarization.

 The
effect of the black hole radius on the Proca QNM frequencies
was also investigated. For large black holes, with $r_{\rm h} \gg l$,
most Proca mode frequencies scale linearly with the black hole radius, in
accordance with previous works.
Scalar-type Maxwell perturbations in $d\geq5$ large Schwarzschild-AdS
black holes also exhibit purely imaginary low frequency modes, whose
expression was found analytically to scale with the inverse of the
black hole radius and with the square of the angular momentum number
of the perturbation. The numerical results showed good support for the
analytical expression. Such low frequency modes were found in previous
studies for vector-type gravitational perturbations, and have a clear
meaning within the AdS/CFT correspondence, being associated to the
hydrodynamic modes of the dual field theory. It would be interesting
to check if the results from the CFT side match ours.
When the radius of the black hole is of the same order as the radius
of AdS, $r_{\rm h} \simeq l$, the linear behavior breaks down, with
the real part of the frequencies approaching a minimum, which may be
due to the black hole starting to have a comparable size with the
Compton wavelength of the field. In the limit of vanishing radius, the
frequencies approach those of pure AdS. Namely, scalar-type Proca
frequencies with electromagnetic and nonelectromagnetic polarizations
approach, respectively, the scalar-type Proca frequencies of pure AdS
with total angular momentum number $j = \ell-1$ and
$j=\ell+1$. Similarly, vector-type Proca frequencies approach the
vector-type $j=\ell$ frequencies of pure AdS.
When the radius of the black hole is much less than the radius
of AdS, $r_{\rm h} \ll l$, one is in the
small black hole regime. Numerical methods do not yield
reliable results in this regime. 
However,
we were able to obtain analytical expressions for the frequencies' imaginary
part, $\delta$, of vector-type and monopole Proca perturbations, 
and also Maxwell perturbations,
in the
small black hole regime.  This was done by
matching asymptotic
expansions of the solution in an overlap region, which forms if the
Compton wavelength of the Proca particle is much larger than the
radius of the black hole. For vector-type perturbations the imaginary
part scales as $\delta \sim (\frac{r_{\rm h}}{l})^{2\ell+d-2}
\omega_{\mathrm{AdS}}$, whereas for monopole perturbations it scales
as $\delta \sim (\frac{r_{\rm h}}{l})^{d-2}
\omega_{\mathrm{AdS}}^{-1}$, where
$\omega_{\mathrm{AdS}}$ is the corresponding pure AdS normal mode
frequency.  The numerical results show support for the analytical
expressions, although it would be interesting to implement more
suitable numerical methods to verify our analytical predictions.

\acknowledgments{
The authors thank the financial support from Funda\c{c}\~ao
para a Ci\^encia e Tecnologia - FCT through the
project No. UIDB/00099/2025 and project No. UIDP/00099/2025, 
namely T.~F.~thanks for the financial support through the grant FCT 
no. RD 1415.
}

\appendix

\setcounter{table}{0}

%%%%%%%%%%%%%%%%%%%%%%%%%%%%%%%%%%%%%%%%%%%%%%%%%%%%%%%%%%%%%%%%%%%%%%%%
\section{Horowitz-Hubeny method for computation of quasinormal modes
in Schwarzschild-AdS}
\label{sec:app_horowitz_hubeny}
%%%%%%%%%%%%%%%%%%%%%%%%%%%%%%%%%%%%%%%%%%%%%%%%%%%%%%%%%%%%%%%%%%%%%%%%

\subsection{Preamble}

In Sec.~\ref{sec:numerics_integration} we introduced our own numerical
method
based on the shooting method, and mentioned that
we have also used, as way of
comparing the results,
the  Horowitz-Hubeny method \cite{hubeny2000}.
We now present 
the Horowitz-Hubeny method. This method is widely used 
to compute QNMs in asymptotically AdS spacetimes, using the 
fact that the eigenvalue problem involves an equation with regular 
singular points together with the Dirichlet boundary condition.

\subsection{Horowitz-Hubeny numerical integration method for the Proca
field: Decoupled and coupled Schr\"odinger-like equations}

\subsubsection{Scalar-type $\ell=0$ monopole Proca
field and vector-type Proca field: Decoupled Schr\"odinger-like equation}

The Horowitz-Hubeny method in the Proca field case
can be applied to the monopole case of
the scalar-type Proca field described by $u_1$ with $u_2=0$, since in
this case, the coupled system of equations reduces to a single
equation.
It can also be applied to the vector-type Proca field.
We explain the method applied to the vector-type Proca field,
the other cases follow with ease.

The vector-type sector of the Proca field is described by a single
Schr\"odinger-like equation in Eq.~\eqref{eq:sch_u3}. It is useful to
rewrite the vector-type radial function $u_3(r)$ as $u_3(r) = {\rm
e}^{-i\omega r_*} \psi_3(r)$, where $\psi_3(r)$ obeys now 
\begin{equation}\label{eq:intermediate}
\partial_r^2 \psi_3 + \frac{\left(f'-2i\omega\right)}{f}
\partial_r \psi_3 - \frac{{V_{\rm v}}}{f^2} \psi_3 = 0 \,.
\end{equation}
The equation has formally $d+1$ regular singular points, with the
zeroes of $f$ corresponding to $d-1$ regular singular 
points, with $r=r_{\rm h}$ being one of them, together with the 
points $r=0$ and $r=+\infty$. The $r=+\infty$ singularity is studied 
by writing $x=\frac{1}{r}$, with range $x\in (0, \frac{1}{r_{\rm h}}]$, 
which is always finite in the range of interest. 
By doing such transformation, Eq.~\eqref{eq:intermediate} becomes
\begin{equation}\label{eq:sch_x}
    s(x)\left(x-x_{\rm h}\right) \partial_x^2 \psi_3
    + t(x) \partial_x \psi_3
+ \frac{v(x)}{(x-x_{\rm h})}\psi_3= 0 \,,
\end{equation}
where 
\begin{align}
    &s(x) = \frac{x^4 f}{x-x_{\rm h}} \notag\,,\\
    &t(x) = x^2 \partial_x \left(x^2 f\right)
    + 2i\omega x^2\notag\,,\\
&v(x) = -\frac{\left(x-x_{\rm h}\right) V_{\rm v}(x)}{f} \,,
\end{align}
with $f$ and $V_{\rm v}$ written in terms of $x$. The Fuchs' theorem
guarantees that Eq.~\eqref{eq:sch_x} admits a Frobenius solution near
each of its regular singular points, namely, near $x_{\rm h}$.  The
radius of convergence of the solution will be given by the minimum
distance between $x=x_{\rm h}$ and the other $d$ singular points in
the complex plane. The expansion near the horizon $x=x_{\rm h}$ is
described by
\begin{equation}\label{eq:exp_U}
    \psi_3 (x) = (x-x_{\rm h})^{\alpha}
    \sum_{m=0}^\infty a_{(3) m} (x-x_{\rm h})^m \,,
\end{equation}
where the exponent
$\alpha$ is a constant determined from the indicial 
equation after imposing 
the boundary condition at the horizon,
and $a_{(3) m}$ are coefficients that usually depend on the frequency 
i.e., $a_{(3) m} \equiv a_{(3) m}(\omega)$, with $a_{(3) m} \neq 0$.
The polynomials $s(x)$, $t(x)$,
and $u(x)$ can also be expanded near the horizon, 
i.e., $s(x) = \sum^\infty_{p=0} s_p \left(x-x_{\rm h}\right)^p$, $t(x) = 
\sum^\infty_{p=0} t_p \left(x-x_{\rm h}\right)^p$, and 
$v(x) = \sum^\infty_{p=0} v_p \left(x-x_{\rm h}\right)^p$, 
where $s_p, t_p$, and $v_p$ are coefficients independent of $x$. 
Substituting Eq.~\eqref{eq:exp_U} and the expanded polynomials 
$s(x)$, $t(x)$, and $v(x)$ in Eq.~\eqref{eq:sch_x}, one gets
\begin{align}\label{eq:recursive1}
\sum_{m,p=0}^{\infty} &\left[ s_p
(m+\alpha)(m+\alpha-1)+t_p(m+\alpha)+v_p \right]\notag\\
&\times a_{(3) m} \left(x-x_{\rm h}\right)^{m+\alpha+p} = 0\,.
\end{align}
The term with $m=p=0$ gives rise to the indicial equation for $\alpha$
\begin{equation}\label{eq:indicial}
s_0 \alpha(\alpha-1)+t_0\alpha+v_0 = 0\quad.
\end{equation}
Taking into account that $s_0 = - x_{\rm h}^2 f'(r_{\rm h}) $,
$t_0 = 2x_{\rm h}^2\left(i\omega-\frac{f'(r_{\rm h})}{2}\right)$ and 
$v_0 = 0$, the solutions to Eq.~\eqref{eq:indicial} are $\alpha = 0$
or $\alpha = \frac{2i\omega}{f'(r_{\rm h})}$, 
corresponding, respectively, to ingoing and outgoing modes at the
horizon.  The boundary condition Eq.~\eqref{eq:bc_horizon} or
Eq.~\eqref{eq:bc_horizonMaxwell} sets $\alpha = 0$, so that only
ingoing modes are allowed near the horizon.  Setting this in
Eq.~\eqref{eq:recursive1}, relabelling indexes and equating the terms
for each power of $(x-x_{\rm h})$, one arrives at the recursion
relation for the coefficients $a_{(3)j}$,
\begin{align}\label{eq:recursive}
&a_{(3)j} =-\frac{1}{P_j}\notag\\
&\times \sum_{m=0}^{j-1}  \left[ m(m-1)s_{j-m} + m t_{j-m} +
v_{j-m}\right]a_{(3)m} \,,
\end{align}
where $P_j = j(j-1)s_0+jt_0$. Without loss of generality, one can set
$a_{(3) 0} = 1$ and solve recursively Eq.~\eqref{eq:recursive} up to a
given order, $\mathrm{N}$.  The Dirichlet boundary condition at
spatial infinity then yields
\begin{equation}\label{eq:dirichlet_decoupled}
\psi_3 (x=0) = \sum^\infty_{m=0} a_{(3)m}(\omega)
\left(-x_{\rm h}\right)^m = 0 \,,
\end{equation}
which can be solved numerically for $\omega$ by truncating the sum at
a sufficiently large $\mathrm{N}$. To see whether the $\mathrm{N}$
chosen is sufficiently large, one simply checks if the solution of the
$\mathrm{N}$-term sum is sufficiently close to the solution of the
$(\mathrm{N}+1)$-term sum, within the desired accuracy.

This method is also used in 
the monopole case of
the scalar-type Proca field described by $u_1$ with $u_2=0$, since in
this case, the coupled system of equations reduces to a single
equation. Indeed, one can follow the method
 explained with the
replacement of the potential 
$V_{s11}$ for the monopole scalar-type Proca field.

\subsubsection{Scalar-type $\ell\geq 1$ Proca field: Coupled equations}

Scalar-type Proca field perturbations are described by the coupled
system Eq.~\eqref{eq:sch_coupled}, so that the Horowitz-Hubeny
technique from the previous section needs to be slightly modified, as
in \cite{pani2013,tiago_sads2022}.  We define the vector
$\boldsymbol{u}(r) = (u_1, u_2)^T$ and make the transformation
$\boldsymbol{u}(r) = {\rm e}^{-i\omega r_*} \boldsymbol{\psi}(r)$ in
the coupled system, so that Eq.~\eqref{eq:sch_coupled} can be written
as
\begin{equation}
     s(x)\left(x-x_{\rm h}\right) \partial_x^2
     \boldsymbol\psi 
 + t(x) \partial_x \boldsymbol{\psi} 
 + \frac{\boldsymbol{v}(x)}{(x-x_{\rm h})}
 \boldsymbol{\psi}= 0 \,,
\end{equation}
with $\boldsymbol{v}(x)=-\frac{\left(x-x_{\rm h}\right)
\boldsymbol{V_{\rm s}}}{f}$. 
After expanding $\boldsymbol{\psi}(x)$ near the horizon as
\begin{equation}\label{eq:coupled_expansion}
    \boldsymbol{\psi}(x) = \sum^\infty_{m=0}
    \boldsymbol{a}_j (\omega) (x-x_{\rm h})^j \,,
\end{equation}
where $\boldsymbol{a}_j$ are coefficients of the series and they are
vectors, as well as expanding $s(x)$ with coefficients $s_j$, $t(x)$
with coefficients $t_j$ and $\boldsymbol{v}(x)$ with coefficients
$\boldsymbol{v}_j$, one obtains a recurrence relation for the
coefficients $\boldsymbol{a}_j$, which can be related to the
coefficient $\boldsymbol{a}_0$ as $\boldsymbol{a}_j (\omega)=
\boldsymbol{{B}}_j (\omega) \boldsymbol{a_0}$ with matrices
$\boldsymbol{{B}}_j$ determined by
\begin{align}\label{eq:coupled_recursion2}
 &\boldsymbol{{B}}_j = -\frac{1}{P_j} \notag\\
 &\times\sum_{m=0}^{j-1}  
 \left[\left( m(m-1)s_{j-m} + m t_{j-m}\right)\boldsymbol{\mathbf{I}} 
 + \boldsymbol{v}_{j-m}\right]\boldsymbol{{B}}_m \,,
\end{align}
The Dirichlet boundary condition at spatial infinity then yields
$\boldsymbol\psi (x=0) =
\left(\sum^\infty_{j=0}
\boldsymbol{B}_j(\omega)(-x_{\rm h})^j\right)\boldsymbol{a}_0
= 0$, whose nontrivial solution fixes the values of $\omega$ through
\begin{equation}\label{eq:coupled_findroot}
\det\left(\sum^\infty_{j=0}\boldsymbol{B_j}(\omega)(-x_{\rm h})^j
\right) = 0 \,.
\end{equation}
As in the decoupled case, 
Eq.~\eqref{eq:coupled_findroot} can be solved numerically 
for $\omega$ by truncating the sum at a sufficiently large $\mathrm{N}$.

\subsection{Horowitz-Hubeny numerical integration method for the Maxwell
field: Decoupled equations}

The Maxwell field also yields decoupled equations. So under Dirichlet
boundary condition, the method for decoupled Proca equations given
above can in principle be applied.

The Horowitz-Hubeny numerical integration method can be applied to the
scalar-type Maxwell field $u_{12}$.  Indeed, one can follow the same
method as explained above with the replacement of the potential
$V_{\rm v}$ by the potential $V_{s}$ for the scalar-type Maxwell
field.  However, it cannot be applied to the specific case of the
scalar-type Maxwell field in $d=5$ since Dirichlet-Neumann boundary
conditions are imposed and not Dirichlet.

The Horowitz-Hubeny numerical integration method can also be applied
to the vector-type Maxwell field $u_3$ with $\mu=0$.  One can also
follow the same method with the replacement of the potential $V_{\rm
v}$ by the potential $V_{\rm v}$ with $\mu = 0$ for the vector-type
Maxwell field.

%%%%%%%%%%%%%%%%%%%%%%%%%%%%%%%%%%%%%%%%%%%%%%%%%%%%%%%%%%%
\section{Numerical values for the Proca quasinormal modes}
\label{sec:app_tables}
%%%%%%%%%%%%%%%%%%%%%%%%%%%%%%%%%%%%%%%%%%%%%%%%%%%%%%%%%%%

We present the numerical values in
Tabs.~\ref{tab:monopole_mass}-\ref{tab:vt_mass}
of this Appendix, of the Proca
QNMs in four, five, six and seven dimensions, for values
of the black hole horizon $r_{\rm h} \in [0.05,1,100]$ and for values
of the Proca mass $\mu\in [0.1,0.2,0.3,0.4,0.5]$, for the scalar-type
and vector-type modes, and their polarizations,
see also Sec.~\ref{sec:num_results}. We also include the
electromagnetic limit for the electromagnetic polarized scalar-type
and vector-type modes.
\vskip -1cm

\onecolumngrid

\begin{table}[H]
\centering
\resizebox{0.95\textwidth}{!}{
\begin{tblr}{c|c|c|c|c|c}
$r_{\rm h}$ & $\mu$ & $\omega$ $(d=4)$ & $\omega$ $(d=5)$ & $\omega$ $(d=6)$ & $\omega$ $(d=7)$  \\ \cline{1-6}
\SetCell[r=5]{}$100$ &$0.1$&$ 185.569 - 267.526 i$&$ 312.426 - 275.165 i $&$ 413.971 - 269.616 i$&$ 501.060 - 261.424 i$\\
&$0.2$&$ 187.346 - 270.819 i $&$313.801 - 276.650 i $&$415.038 - 270.457 i$&$ 501.916 - 261.962 i$\\
&$0.3$&$ 190.106 - 275.942 i$&$  316.047 - 279.075 i$&$ 416.800 - 271.846 i$&$  503.334 - 262.853 i$\\
&$0.4$&$ 193.636 - 282.504 i$&$ 319.103 - 282.376 i$&$  419.234 - 273.765 i$&$ 505.304 - 264.092 i$\\
&$0.5$&$ 197.742 - 290.148 i $&$ 322.892 - 286.471 i$&$ 422.310 - 276.190 i$&$ 507.811 - 265.669 i$\\\cline{1-6}
\SetCell[r=5]{}$1$&$0.1 $&$2.80724 - 2.68313 i$&$  4.58507 - 2.55966 i$&$ 6.01308 - 2.35363 i$&$ 7.27221 - 2.16280 i$\\
&$0.2 $&$2.83328 - 2.71753 i$&$ 4.60375 - 2.57442 i$&$ 6.02701 - 2.36168 i$&$ 7.28309 - 2.16779 i$\\
&$0.3 $&$2.87379 - 2.77101 i$&$  4.63430 - 2.59855 i$&$  6.05003 - 2.37497 i$&$ 7.30115 - 2.17606 i$\\
&$0.4 $&$2.92572 - 2.83946 i$&$  4.67589 - 2.63137 i$&$  6.08184 - 2.39332 i$&$ 7.32625 - 2.18756 i$\\
&$0.5 $&$2.98623 - 2.91912 i$&$  4.72755 - 2.67209 i$&$ 6.12207 - 2.41652 i$&$ 7.35822 - 2.20219 i$\\\cline{1-6}
\SetCell[r=5]{}$0.05$&$0.1$ &$2.86292 - 1.94677\e{-2} i$ &$3.98937 - 1.75825\e{-3}i$ &$5.00162 - 1.81866\e{-4} i$ &$6.00231 - 1.94173\e{-5} i$\\
&$0.2$ &$2.88910 - 1.98707\e{-2} i$  &$4.00405 - 1.77507\e{-3} i$& $5.01155 - 1.83002\e{-4} i$& $6.00979 - 1.95086\e{-5} i$\\
&$0.3$ &$2.92988 - 2.05020\e{-2} i$ &$4.02805 - 1.80268\e{-3} i$ &$5.02797 - 1.84886\e{-4} i$ & $6.02219 - 1.96605\e{-5} i$\\
&$0.4$& $2.98222 - 2.13182\e{-2} i$ &$4.06075 - 1.84049\e{-3} i$&$5.05066 - 1.87503\e{-4} i$ & $6.03942 - 1.98727\e{-5} i$\\
&$0.5$& $3.04330 - 2.22800\e{-2} i$  & $4.10136 - 1.88781\e{-3} i$  &$5.07935 - 1.90836\e{-4} i$ & $6.06136 - 2.01447\e{-5} i$
\end{tblr}
}
\caption{Fundamental modes of $\ell=0$ Proca monopole scalar
perturbations in $4,5,6,7$-dimensional Schwarzschild-AdS spacetime
with $r_{\rm h}=100$, $r_{\rm h}=1$ and $r_{\rm h}=0.05$, for
different values of the mass of the field.}
\label{tab:monopole_mass}
\end{table}

\vskip -4.0cm
\centerline{}
\vskip 7cm
\centerline{}

\begin{table}[H]
\centering
\resizebox{0.95\textwidth}{!}{
\begin{tblr}{c|c|c|c|c|c}
$r_{\rm h}$ & $\mu$ & $\omega$ $(d=4)$ & $\omega$ $(d=5)$ & $\omega$ $(d=6)$ & $\omega$ $(d=7)$  \\ \cline{1-6}
\SetCell[r=5]{}$100$ 
&$0.1$&$ 185.577 - 267.524 i$&$ 312.433 - 275.163 i $&$ 413.978 - 269.614 i$&$ 501.067 - 261.422 i$\\
&$0.2$&$ 187.354 - 270.817 i $&$313.808 - 276.648 i $&$415.045 - 270.455 i$&$ 501.922 - 261.960 i$\\
&$0.3$&$ 190.114 - 275.939 i$&$  316.054 - 279.073 i$&$ 416.807 -  271.844 i$&$  503.340 - 262.852 i$\\
&$0.4$&$193.644 - 282.501 i$&$ 319.110 - 282.374 i$&$   419.241 - 273.763 i$&$ 505.310 - 264.090 i$\\
&$0.5$&$ 197.750 - 290.146 i $&$ 322.899 - 286.469 i$&$ 422.316 - 276.188 i$&$507.817 - 265.667i$\\\cline{1-6}
\SetCell[r=5]{}$1$
&$0.1 $&$3.33865 - 2.50108 i$&$  5.09680 - 2.38187 i$&$  6.53418 - 2.18970 i$&$ 7.80878 - 2.01206 i$\\
&$0.2 $&$3.36178 - 2.53440 i$&$5.11406 - 2.39630   i$&$  6.54725 - 2.19759 i $&$ 7.81908 - 2.01696 i$\\
&$0.3 $&$3.39782 - 2.58616 i$&$ 5.14229 - 2.41988 i$&$  6.56884 - 2.21062i$&$ 7.83617 -  2.02509 i$\\
&$0.4 $&$3.44410 - 2.65233 i$&$  5.18076 - 2.45195 i$&$   6.59870 - 2.22862 i $&$  7.85993 - 2.03638 i$\\
&$0.5 $&$3.49817 - 2.72926 i$&$  5.22860 - 2.49174 i $&$ 6.63649 - 2.25137 i$&$ 7.89019 - 2.05076 i$\\\cline{1-6}
\SetCell[r=5]{}$0.05$
&$0.1$ &$3.91824 - 4.75561\e{-5} i$ &$4.99732- 7.18144\e{-6} i$ &$6.00264 -8.72484\e{-7} i$ &$7.00244 - 9.88835\e{-8} i$\\
&$0.2$ &$3.94538 - 4.89806\e{-5} i$  &$5.01207 - 7.27222\e{-6} i$& $6.01258 - 8.79410\e{-7} i$& $7.00991 - 9.94624\e{-8} i$\\
&$0.3$ &$3.98765 - 5.12361\e{-5} i$ &$5.03619 - 7.42189\e{-6} i$ &$6.02901 - 8.90921\e{-7}  i$ & $ 7.02231 - 1.00415\e{-7} i$\\
&$0.4$& $4.04191 - 5.41980\e{-5} i$ &$5.06904 - 7.62824\e{-6} i$&$ 6.05171 - 9.06995\e{-7}  i$ & $7.03954 - 1.01750\e{-7} i$\\
&$0.5$& $4.10524 - 5.77538\e{-5} i$  & $5.10986 - 7.88857\e{-6} i$  &$6.08042 - 9.27537\e{-7} i$ & $7.06149 - 1.03470\e{-7} i$
\end{tblr}
}
\caption{Fundamental modes of $\ell=1$ nonelectromagnetic polarized
scalar-type Proca perturbations in $4,5,6,7$-dimensional
Schwarzschild-AdS spacetime with $r_{\rm h}=100$, $r_{\rm h}=1$ and
$r_{\rm h}=0.05$, for different values of the mass of the field.}
\label{tab:proca_plus_pol}
\end{table}

\begin{table}[H]
\centering
\resizebox{0.95\textwidth}{!}{
\begin{tblr}{c|c|c|c|c|c}
$r_{\rm h}$ & $\mu$ & $\omega$ $(d=4)$ & $\omega$ $(d=5)$ & $\omega$ $(d=6)$ & $\omega$ $(d=7)$  \\ \cline{1-6}
\SetCell[r=5]{}$100$& $0$ & $0. - 150.048 i$ & $200.013  - 200.003 i$ &$299.447 - 200.490 i$&$383.681 - 199.896 i$\\
&$0.1$&$ 0.-152.097i$&$200.509 - 200.499 i$&$ 299.812 - 200.768 i$& $383.969 - 200.073  i$\\
&$0.2$&$0.-158.412i$&$201.990 - 201.980 i $&$300.902 - 201.599 i$&$384.830 - 200.604   i$\\
&$0.3$&$0.-168.740i $&$204.413 - 204.403 i$&$302.704 - 202.972  i$&$386.257 - 201.485    i$\\
&$0.4$&$0.-183.124 i$&$ 207.714 - 207.703 i$&$ 305.194 - 204.871  i$&$388.242 - 202.709   i$\\
&$0.5$&$0.-202.409  i $&$ 211.814 - 211.803   i$&$308.345 - 207.272   i$&$ 390.770 - 204.269  i$\\\cline{1-6}
\SetCell[r=5]{}$1$& $0$ &$2.16302 - 1.69909 i$ & $2.66237 - 1.58299i$ & $4.41406 - 1.68956 i$ & $5.73731 - 1.59713 i$\\
&$0.1 $&$ 1.55730 - 0.552855 i$&$2.66878 - 1.59009 i$  &$4.41877 - 1.69234  i$&$5.74090 - 1.59879  i$\\
&$0.2$&$ 1.56806 - 0.584611  i$  &$2.68791 - 1.61115  i$&$4.43286 - 1.70062  i$&$5.75162 - 1.60373  i$\\
&$0.3$&$1.58497  -  0.634989 i$  &$2.71949 - 1.64544  i$&$4.45614 - 1.71432  i$&$5.76942 - 1.61194  i$\\
&$0.4$&$1.60693 - 0.700879 i$  &$2.76308 - 1.69187  i$&$ 4.48836 - 1.73324  i$&$5.79417 - 1.62335   i$\\
&$0.5$&$1.63296 - 0.779477 i$  &$2.81810 - 1.74904  i$&$4.52915 - 1.75717   i$&$5.82571 - 1.63789  i$\\\cline{1-6}
\SetCell[r=5]{}$0.05$ & $0$ & $2.93223 - 5.39171\e{-5} i$ & $2.99617 - 5.74443\e{-6} i$ & $3.99948 - 1.65071\e{-6} i$ & $4.99994 - 3.35996\e{-7} i$\\
&$0.1$& $1.98737 - 7.62153\e{-6}i$&$ 3.00114 - 5.79865\e{-6}  i$ &$4.00281 - 1.65889\e{-6} i$ &$5.00243 - 3.37073\e{-7} i$\\
&$0.2$ &$2.01476 - 8.30304\e{-6}i $  &$ 3.01589 - 5.96200\e{-6} i      $&    $4.01275 - 1.68350\e{-6}i$& $ 5.00991 - 3.40313\e{-7}i$\\
&$0.3$ &$2.05739 - 9.44648\e{-6} i$ &$3.04000 - 6.23648\e{-6} i$ &$4.02917 - 1.72478\e{-6} i$ & $5.02231 - 3.45740\e{-7} i$\\
&$0.4$& $2.11204 - 1.10687\e{-5}i$ &$ 3.07285 - 6.62545\e{-6}i $&$ 4.05187 - 1.78312\e{-6}i$ & $ 5.03954 - 3.53396\e{-7} i$\\
&$0.5$& $2.17574 - 1.31994\e{-5} i$  & $ 3.11366 - 7.13369\e{-6}  i$  &$4.08058 - 1.85907\e{-6} i$ & $5.06149 - 3.63339\e{-7} i$
\end{tblr}
}
\caption{Fundamental modes of $\ell=1$ electromagnetically polarized
scalar-type Proca perturbations and the scalar-type Maxwell modes in
$4,5,6,7$-dimensional Schwarzschild-AdS spacetime with $r_{\rm
h}=100$, $r_{\rm h}=1$ and $r_{\rm h}=0.05$, for different values of
the mass for the case of the Proca field. The case $\mu=0$ denotes the
Maxwell mode and not the massless limit for $d=4$.}
\label{tab:proca_minus_pol}
\end{table}

\begin{table}[H]
\centering
\resizebox{0.95\textwidth}{!}{\begin{tblr}{c|c|c|c|c|c}
$r_{\rm h}$ & $\mu$ & $\omega$ $(d=4)$ & $\omega$ $(d=5)$ & $\omega$ $(d=6)$ & $\omega$ $(d=7)$  \\ \cline{1-7}
\SetCell[r=6]{}$100$ & $0$&$0.-150.048 i$   & $200.026 - 199.995 i$ &     $ 299.458 - 200.486 i$ &   $383.691 - 199.894 i$ \\
&$0.1$  &$0.- 152.185i$ &$200.525 - 200.493 i$    &  $299.823 -  200.765i$ &    $383.979 - 200.071 i$     \\
&  $0.2$ &$0.-158.505i$  &$202.007 - 201.975 i$     & $300.914 - 201.596 i$ &    $384.839 - 200.602 i$     \\   &  $0.3$& $0.-168.844i$ &$204.429 - 204.398 i$     & $302.715 - 202.969 i$ &   $386.267 - 201.483 i$   \\
&  $0.4$& $0.-183.251i$ &$207.729 - 207.698 i$     & $305.205 - 204.868 i$ &   $388.252 - 202.707 i$   \\
&  $0.5$& $0.-202.583i$ &$211.829 - 211.798 i$     & $308.356 - 207.269 i$ &   $390.780 - 204.266 i$   \\ \cline{1-7}   \SetCell[r=6]{}$1$&$0$&$2.16302 - 1.69909 i$ & $3.84177 - 1.62618 i$  & $5.20392 - 1.53559i$&   $6.43969 - 1.45072 i$ \\   &$0.1$  &$2.17058 - 1.71095i $ &$3.84730 - 1.63091 i$    &  $5.20810 - 1.53813i$ &    $6.44299 - 1.45229i$     \\
&$0.2$ &$2.19254 - 1.74526i$  &$3.86377 - 1.64499i$     & $5.22058 - 1.54573 i$ &    $6.45286 - 1.45699 i$     \\   &$0.3$& $2.22707 - 1.79879i$ &$3.89075 - 1.66805i$     & $5.24123 - 1.55828i$ &   $6.46923 - 1.46480  i$   \\
&$0.4$& $2.27191 - 1.86765i$ &$3.92761 - 1.69952i$     & $5.26981 - 1.57567i $ &   $6.49202 - 1.47565  i$   \\    & $0.5$& $2.32497 - 1.94823i$ &$3.97359 - 1.73872i$     & $5.30603 - 1.59769i$&   $6.52107 - 1.48949 i$   \\ \cline{1-7}
\SetCell[r=6]{}$0.05$&$0$&$2.93223 - 5.39171\e{-5} i$ &$3.99493 - 4.36325\e{-6} i$ &$4.99957 - 4.53425\e{-7} i$&$5.99996 - 4.85953\e{-8} i$ \\
&$0.1$  & $2.94163 - 5.46847\e{-5} i$&$3.99999 - 4.38818\e{-6} i$&$5.00290 - 4.54977\e{-7} i$&$6.00246 - 4.87122\e{-8} i$     \\
&$0.2$ &$2.96881 - 5.69458 \e{-5} i$  &$4.01466 - 4.46281\e{-6} i$&$5.01284 - 4.59635\e{-7} i$&$6.00994 - 4.90633\e{-8}i$     \\
&$0.3$& $3.01112 - 6.05978 \e{-5} i$ &$4.03879 - 4.58669\e{-6} i$&$5.02927 - 4.67405\e{-7} i$&$6.02234 - 4.96497\e{-8}i$   \\
&$0.4$& $3.06541- 6.55234 \e{-5} i$ &$4.07167 - 4.75922\e{-6} i$&$5.05197 - 4.78298\e{-7} i$&$6.03957 - 5.04733\e{-8} i$   \\
& $0.5$& $3.12875 - 7.16258\e{-5}i$ &$4.11251 - 4.97971\e{-6} i$&$5.08069 - 4.92331\e{-7} i$&$6.06151 - 5.15367\e{-8} i$
\end{tblr}}
\caption{Fundamental modes of $\ell=1$ vector-type Proca perturbations
in $4,5,6,7$-dimensional Schwarzschild-AdS spacetime with $r_{\rm
h}=100$, $r_{\rm h}=1$ and $r_{\rm h}=0.05$, for different values of
the mass for the case of the Proca field. The case $\mu=0$ is the
fundamental modes of $\ell=1$ vector-type Maxwell field.}
\label{tab:vt_mass}
\end{table}

\twocolumngrid

%\newpage
%\centerline{}
%\newpage
%\centerline{}
%\newpage

\end{document}